\documentclass[twocolumn]{aastex631}
\usepackage{multirow}
\usepackage{color}
\usepackage{capt-of}
\usepackage{rotating}
\usepackage{longtable}
\usepackage{graphicx}
\usepackage{float}
\usepackage{amsmath,amssymb}
\usepackage{booktabs}
\usepackage{gensymb}
\usepackage{xcolor}
\definecolor{ForestGreen}{RGB}{34,139,34}
\definecolor{LinkRed}{rgb}{0.7752941176470588, 0.22078431372549023, 0.2262745098039215}
\definecolor{FireBrick}{rgb}{0.7, 0.13, 0.13}
\definecolor{DarkGold}{RGB}{184,134,11}
\newcommand{\good}{\raisebox{0pt}{\scalebox{1.75}{\textcolor{ForestGreen}{\checkmark}}}}
\newcommand{\maybe}{\raisebox{0pt}
{\scalebox{1.75}{\textcolor{DarkGold}{?}}}}
\newcommand{\bad}{\raisebox{0pt}{\scalebox{1.75}{\textcolor{FireBrick}{\ensuremath{\times}}}}}
\usepackage{natbib}
\usepackage[hang,flushmargin]{footmisc}
\usepackage{chngcntr}
\usepackage{hyperref}
\usepackage{enumitem}
\setlist{parsep=0pt,listparindent=\parindent}
\usepackage{xtab}
\usepackage{tabularx}
\usepackage{xltabular}
\usepackage{soul}
\usepackage{array}
\usepackage{mwe}
\usepackage{afterpage}
\usepackage{rotating}
\usepackage{afterpage}
\usepackage{colortbl}
\usepackage{lipsum}
\usepackage{pifont}

\usepackage{newunicodechar}
\DeclareRobustCommand{\okina}{%
  \raisebox{\dimexpr\fontcharht\font`A-\height}{%
    \scalebox{0.8}{`}%
  }%
}
\newunicodechar{ʻ}{\okina}

\renewcommand{\tablenotemark}[1]{{\normalfont\textsuperscript{\footnotesize\it\textcolor{LinkRed}{#1}}}}

\newcommand{\JHU}{Physics and Astronomy Department, Johns Hopkins University, Baltimore, MD 21218, USA}
\newcommand{\STScI}{Space Telescope Science Institute, Baltimore, MD 21218.}

\newcommand{\LCO}{Las Cumbres Observatory, 6740 Cortona Drive, Suite 102, Goleta, CA, 93117, USA}
\newcommand{\UCSB}{Physics Department, University of California, Santa Barbara, Santa Barbara, CA, 93106, USA}

\newcommand{\CfA}{Center for Astrophysics $|$ Harvard \& Smithsonian, Cambridge, MA 02138, USA}
\newcommand{\IfA}{Institute for Astronomy, University of Hawaii, 2680 Woodlawn Drive, Honolulu, HI 96822, USA}

\newcommand{\UCSC}{Department of Astronomy and Astrophysics, University of California, Santa Cruz, CA 95064, USA}
\newcommand{\QUB}{Astrophysics Research Centre, School of Mathematics and Physics, Queen's University Belfast, Belfast BT7 1NN, UK}

\newcommand{\CAPS}{Center for AstroPhysical Surveys (CAPS) Fellow}

\newcommand{\Northwestern}{Center for Interdisciplinary Exploration and Research in Astrophysics (CIERA) and Department of Physics and Astronomy, Northwestern University, Evanston, IL 60208, USA}
\newcommand{\DARK}{DARK, Niels Bohr Institute, University of Copenhagen, Jagtvej 155A, 2200 Copenhagen, Denmark}
\newcommand{\UIUC}{Department of Astronomy, University of Illinois at Urbana-Champaign, 1002 W. Green St., IL 61801, USA}

\newcommand{\TAPIR}{TAPIR, Mailcode 350-17, California Institute of Technology, Pasadena, CA 91125, USA}
\newcommand{\RESCEU}{Research Center for the Early Universe (RESCEU), School of Science, The University of Tokyo, Bunkyo-ku, Tokyo 113-0033, Japan}

\newcommand{\Melbourne}{OzGrav, School of Physics, The University of Melbourne, VIC 3010, Australia}

\newcommand{\Oxford}{Department of Physics, University of Oxford, Denys Wilkinson Building, Keble Road Oxford OX1 3RH}

\newcommand{\IAIFI}{The NSF AI Institute for Artificial Intelligence and Fundamental Interactions}
\newcommand{\IfAHilo}{Institute for Astronomy, University of Hawaiʻi, 640 N.~Aʻohoku Pl., Hilo, HI 96720, USA}
\newcommand{\MIT}{Department of Physics and Kavli Institute for Astrophysics and Space Research, Massachusetts Institute of Technology, 77 Massachusetts Avenue, Cambridge, MA 02139, USA}

\begin{document}

\title{Evidence for an Instability-Induced Binary Merger in the Double-Peaked, Helium-Rich Type~IIn Supernova 2023zkd}
\author[0000-0003-4906-8447]{A.~Gagliano}
\affiliation{\IAIFI}
\affiliation{\CfA}
\affiliation{\MIT}

\author[0000-0002-5814-4061]{V.~A.~Villar}
\affiliation{\IAIFI}
\affiliation{\CfA}

\author[0000-0002-9350-6793]{T.~Matsumoto}
\affiliation{Department of Astronomy, Kyoto University, Kitashirakawa-Oiwake-cho, Sakyo-ku, Kyoto, 606-8502, Japan}
\affiliation{Hakubi Center, Kyoto University, Yoshida-honmachi, Sakyo-ku, Kyoto, 606-8501, Japan}

\author[0000-0002-6230-0151
]{D.~O.~Jones}
\affiliation{\IfAHilo}

\author[0000-0003-4175-4960]{C.~L.~Ransome}
\affiliation{\CfA}

\author[0000-0002-2028-9329]{A.~E.~Nugent}
\affiliation{\CfA}

\author[0000-0002-1125-9187]{D.~Hiramatsu}
\affiliation{\IAIFI}
\affiliation{\CfA}

\author[0000-0002-4449-9152]{K.~Auchettl}
\affiliation{\Melbourne}
\affiliation{\UCSC}

\author[0000-0002-6347-3089]{D.~Tsuna}
\affiliation{\TAPIR}
\affiliation{\RESCEU}

\author[0000-0002-7937-6371]{Y.~Dong}
\affiliation{\CfA}

\author[0000-0001-6395-6702]{S.~Gomez}
\affiliation{\CfA}

\author[0000-0002-6298-1663]{P.~D.~Aleo}
\affiliation{\UIUC}
\affiliation{\CAPS}

\author[0000-0002-4269-7999]{C.~R.~Angus}
\affiliation{\QUB}

\author[0000-0001-5486-2747]{T.~de~Boer}
\affiliation{\IfA}

\author[0000-0002-4924-444X]{K.~A.~Bostroem}
\altaffiliation{LSST-DA Catalyst Fellow}
\affiliation{Steward Observatory, University of Arizona, 933 North Cherry Avenue, Tucson, AZ 85721-0065, USA}

\author[0000-0001-6965-7789]{K.~C.~Chambers}
\affiliation{\IfA}

\author[0000-0003-4263-2228]{D.~A.~Coulter}
\affiliation{\STScI}

\author[0000-0002-5680-4660]{K.~W.~Davis}
\affiliation{\UCSC}

\author{J.~R.~Fairlamb}
\affiliation{\IfA}

\author{J.~Farah}
\affiliation{\LCO}
\affiliation{\UCSB}

\author[0000-0002-6886-269X]{D.~Farias}
\affiliation{\DARK}

\author[0000-0002-2445-5275]{R.~J.~Foley}
\affiliation{\UCSC}

\author[0000-0002-8526-3963]{C.~Gall}
\affiliation{\DARK}

\author[0000-0003-1015-5367]{H.~Gao}
\affiliation{\IfA}

\author[0000-0003-0209-9246]{E.~P.~Gonzalez}
\affiliation{\LCO}
\affiliation{\UCSB}

\author[0000-0003-4253-656X]{D.~A.~Howell}
\affiliation{\LCO}
\affiliation{\UCSB}

\author[0000-0003-1059-9603]{M.~E.~Huber}
\affiliation{\IfA}

\author[0000-0002-5740-7747
]{C.~D.~Kilpatrick}
\affiliation{\Northwestern}

\author[0000-0002-7272-5129]{C.-C.~Lin}
\affiliation{\IfA}

\author{T.~B.~Lowe}
\affiliation{\IfA}

\author[0000-0002-1417-8024]{M.~E.~MacLeod}
\affiliation{\CfA}

\author[0000-0002-7965-2815]{E.~A.~Magnier}
\affiliation{\IfA}

\author[0000-0001-5807-7893]{C.~McCully}
\affiliation{\LCO}
\affiliation{\UCSB}

\author[0009-0003-8803-8643]{P. Mínguez}
\affiliation{\IfA}

\author[0000-0001-6022-0484]{G.~Narayan}
\affiliation{\UIUC}
\affiliation{\CAPS}

\author[0000-0001-9570-0584]{M.~Newsome}
\affiliation{\LCO}
\affiliation{\UCSB}

\author[0000-0002-1092-6806
]{K.~C.~Patra}
\affiliation{\UCSC}
\affiliation{University of California, Observatories 1156 High Street Santa Cruz, CA 95064}

\author[0000-0002-4410-5387]{A.~Rest}
\affiliation{\STScI}
\affiliation{\JHU}

\author[0000-0002-3825-0553]{S.~Rest}
\affiliation{Department of Computer Science, The Johns Hopkins University, Baltimore, MD 21218, USA}

\author[0000-0002-8229-1731]{S.~Smartt}
\affiliation{\Oxford}
\affiliation{\QUB}

\author[0000-0001-9535-3199]{K.~W.~Smith}
\affiliation{\Oxford}
\affiliation{\QUB}

\author[0000-0003-0794-5982]{G.~Terreran}
\affiliation{Adler Planetarium, 1300 S. DuSable Lake Shore Drive, Chicago, IL 60605, USA}

\author[0000-0002-1341-0952]{R.~J.~Wainscoat}
\affiliation{\IfA}

\author[0000-0001-5233-6989]{Q.~Wang}
\affiliation{\MIT}

\author[0000-0002-0840-6940]{S.~K.~Yadavalli}
\affiliation{\CfA}

\author[0000-0002-0632-8897]{Y.~Zenati}
\affiliation{\JHU}
\affiliation{\STScI}

\collaboration{1000}{(Young Supernova Experiment)}

\correspondingauthor{A.~Gagliano}
\email{alexander.gagliano@cfa.harvard.edu}

%%%%%%%%%%

\begin{abstract}
We present ultraviolet to infrared observations of the extraordinary Type~IIn supernova 2023zkd (SN~2023zkd). Photometrically, it exhibits persistent and luminous precursor emission spanning $\sim$4 years preceding discovery ($M_r\approx-15$~mag, 1,500~days in the observer frame), followed by a secondary stage of gradual brightening in its final year. Post-discovery, it exhibits two photometric peaks of comparable brightness ($M_r\lesssim-18.7$~mag and $M_r\approx-18.4$~mag, respectively) separated by 240~days. Spectroscopically, SN~2023zkd exhibits highly asymmetric and multi-component Balmer and He~I profiles that we attribute to ejecta interaction with fast-moving ($1,\!000-2,\!000\;\mathrm{km}\;\mathrm{s}^{-1}$) He-rich polar material and slow-moving ($\sim$$400\;\mathrm{km}\;\mathrm{s}^{-1}$) equatorially-distributed H-rich material. He~II features also appear during the second light curve peak and evolve rapidly. Shock-driven models fit to the multi-band photometry suggest that the event is powered by interaction with $\sim$$5-6\;M_{\odot}$ of CSM, with $2-3\;M_{\odot}$ associated with each light curve peak, expelled during mass-loss episodes $\sim$$3-4$ and $\sim$$1-2$ years prior to explosion. The observed precursor emission, combined with the extreme mass-loss rates required to power each light curve peak, favors either super-Eddington accretion onto a black hole or multiple long-lived eruptions from a massive star to luminosities that have not been previously observed. We consider multiple progenitor scenarios for SN~2023zkd, and find that the brightening optical precursor and inferred explosion properties are most consistent with a massive ($M_{\mathrm{ZAMS}}\geq30\;M_{\odot}$) and partially-stripped He star undergoing an instability-induced merger with a black hole companion.
\end{abstract}

\section{Introduction} \label{sec:intro}
The observational taxonomy continues to grow for stars that end their lives as core-collapse supernovae (CCSNe). Multi-year, wide-field imaging surveys such as the Zwicky Transient Facility \citep[ZTF;][]{2019Bellm_ZTF,2020Dekany_ZTFObs}; the Young Supernova Experiment \citep[YSE;][]{2021Jones_YSE, 2023Aleo_YSEDR1}; and the Asteroid Terrestrial-impact Last Alert System \citep[ATLAS;][]{2018Tonry_ATLAS}, have enabled detailed photometric analyses of SNe from the photospheric to the nebular phase. 

Of particular interest are CCSNe with significant photometric variability following their primary peak. These events cannot be naturally explained by the radioactive decay chain of $^{56}$Ni$\rightarrow^{56}$Co$\rightarrow^{56}$Fe, or by recombination in the ejecta. Observations of these events instead indicate the presence of circumstellar material (CSM), which can produce a broad diversity of photometric and spectroscopic signatures during SN interaction due to its efficient conversion of kinetic energy and during SN shock breakout due to its re-processing of high-energy photons (\citealt{2014ASmith_MassLoss} provides a conceptual picture of an interaction, and Figure~3 of \citealt{2024Khatami_LSNe_Lanscape} gives a sense of the variety of possible photometric signatures). 

From the recombination of the photo-ionized CSM, strongly-interacting transients may reveal narrow spectral lines of H \citep[as with SNe~IIn;][]{1990Schlegel_SNIIn,1997Filippenko_OpticalSpectra}, He \citep[SNe~Ibn;][]{2008Pastorello_SNIbn,2009Chugai_Ibn}, or C and O while lacking H/He \citep[SNe~Icn;][]{2021Galyam_Icn,2022Pellegrino_SNeIcn}. Where the density of the CSM deviates from either a smooth power-law profile with radius or spherical symmetry, interacting SNe can also re-brighten to a secondary photometric maximum (as seen in, e.g., the type~Ib SN~2005bf, \citealt{2006Folatelli_2005bf}; the candidate pair-instability SN~2016iet, \citealt{2019Gomez_2016iet}; the ambiguous stripped-envelope SN~2020acct, \citealt{2024Angus_2020acct}; the type~Ic SN~2019cad, \citealt{2021Gutierrez_2019cad}; the type Ic/Luminous SN~2019stc, \citealt{2021Gomez_2019stc}; the type~IIn SN~2021qqp, \citealt{2024Hiramatsu_21qqp}; and the type~Ic SN~2022xxf, \citealt{2023Kuncarayakti_2022xxf}, among others). Because circumstellar material can be formed from mass-loss episodes in the months to years prior to core collapse, observations of these strongly-interacting transients are of vital importance for reconstructing the extremes of late-stage stellar evolution.

A second, complementary benefit of our growing photometric datasets is the ability to identify enhanced emission from a terminal system prior to core-collapse. In some cases, these ``precursor events" are luminous enough to be confused for an explosion itself. The most well-studied of these precursor events is 2009ip, which was initially reported to be an SN until being spectroscopically confirmed as a luminous impostor with a dramatic re-brightening phase in 2012 \citep{2009Maza_2009ip,2009Berger_2009ip,2010Smith_2009ip,2011Foley_2009ip,2013Fraser_2009ip,2013Mauerhan_2009ip,2013Pastorello_2009ip,2013Prieto_2009ip,2013Smith_2009ip, 2014Graham_2009ip,2014Levesque_BalmerDecrement,2012ATel_2009ip,2014Margutti_2009ip,2015Martin_2009ip,2015Moriya_2009ip,2017Graham_2009ip,2017Reilly_2009ip,2022Smith_2009ip,2023Pessi_2009ip}. SN~2015bh is another extreme case: optical variability spanning 20 years concluded with a putative explosion peaking at $M_r<-18$~mag, although the interpretation of this final event as terminal is debated \citep{2016EliasRosa_2015bh,2017Thone_2015bh}. The population of detected CCSN precursors has grown increasingly heterogeneous in the intervening years \citep{2024Gagliano_Fuse}, and now includes the persistent and months-long emission preceding at least one normal SN~II \citep[SN~2020tlf; ][]{2022Galan_FinalMoments}; the years-long and quasi-periodic NIR variability associated with the nearby SN~II 2023ixf; \citep{2023Kilpatrick_2023ixf}; and the years-long and steadily brightening optical emission preceding the SN~Ibn 2023fyq \citep{2024Dong_2023fyq,2024Brennan_2023fyq,2024Tsuna_MergerPrecursor}. 

Systematic searches for precursors to SNe~IIn, specifically, suggest that luminous pre-explosion emission is common; estimates range from 25\% of all events for $M_r<-13$~mag precursors in the final three months preceding explosion \citep{2021Strotjohann_MonthsLong} to $>$30$\%$ of all SNe~IIn \citep{2024Reguitti_IInPrecursorSearches}. When attributed to a steady-state wind from the progenitor, the luminosity of this emission indicates mass-loss rates of $10^{-4}-1\;M_{\odot}\;\textrm{yr}^{-1}$, significantly higher than can be explained by line-driven winds alone \citep{2013Taddia_SNeIIn,2014Moriya_MassLossSNeIIn,2024Ransome_DiversityofSNeIIn}. Multiple alternative physical mechanisms have been proposed to power these brightening phases, including energy deposition in the core associated with late-stage nuclear burning \citep{2002Woosley_MassiveStars}; wave damping from vigorous stellar convection \citep{2012Quataert_WaveDriven,2014Shiode_MassLoss,2017Fuller_WaveDriven,2018Fuller_WaveDriven,2020Leung_WaveDriven,2022Wu_MassLoss}; centrifugally-induced mass-loss \citep{2000Lignieres_RapidlyRotating,2018Fuller_WaveDriven,2020Zhao_Centrifugal}; and binary interaction \citep{2014SmithArnett_Preparing,2024Ercolino_BinaryInteraction}. 

Precursor emission, in addition to characterizing the mass deposited in the terminal star's surroundings, provides insights into the nature of the progenitor itself \citep{2022Matsumoto_PrecursorModel}. The high mass-loss rates and CSM velocities associated with SN~IIn precursors have been likened to the massive eruptions of Luminous Blue Variables (LBVs), two of which have been well-documented in our own Galaxy ($\eta$ Carinae and P Cygni; \citealt{1999Israelian_Pcygni}). Though the massive outbursts of these Galactic sources are not causally connected to core collapse as SN~IIn precursors appear to be \citep[and traditional stellar theory predicts an LBV as a brief transitionary stage for stars of mass $M_{\mathrm{ZAMS}}>30\;M_{\odot}$ en route to a H-free Wolf-Rayet, instead of a stellar endpoint;][]{1994Humphreys_LBVs,2017Smith_LBVs}, an LBV-like origin for SNe~IIn has other observational support: pre-explosion photometry with \textit{HST} is consistent with an LBV-like progenitor for SN~2005gl \citep{2009Galyam_LBV}, SN~2009ip \citep{2011Foley_2009ip}, SN~2010bt \citep{2018EliasRosa_2010bt}, SN~2015bh \citep{2022Jecson_2015bh}, and SN~2010jl \citep{2011Smith_LBV2010jl}\footnote{We caution that this interpretation comes primarily from the luminosities of the pre-explosion systems, in tandem with the velocities of H-rich material measured in spectra obtained post-explosion. These properties do not unambiguously indicate an LBV (see \citealt{2011Dwarkadas_LBVProgenitors} and \citealt{2024Ransome_DiversityofSNeIIn} for related discussions).}.

This LBV interpretation is actively debated. SNe~IIn whose progenitors exhibit eruptive episodes span a wider mass range than can be easily explained by LBVs alone \citep{2008Prieto_MassProgenitor,2009Thompson_LuminousTransients}; spectropolarimetric measurements of SNe~IIn also suggest enduring CSM asymmetries not expected for an isotropic outburst in a pristine environment \citep{2024Bilinski_IInspectropolarimetry}. 

Binary interaction between an LBV-like primary and its massive companion may naturally be able to explain the additional diversity, both in SN~IIn precursors and the more ambiguous but closely-related ``SN impostors" \citep{2010Kashi_LBV,2011Smith_BinaryInteraction,2012Chevalier_BinaryInteraction,2014SmithArnett_Binary,2024Ercolino_Binary,2024Matsuoka_Binary}. Indeed, the environments of LBVs in the Milky Way and its neighboring galaxies are incompatible with the young stellar clusters where massive stars are born; ejection from their natal environment driven by the explosion of a binary companion was proposed by \citet{Smith2015_LBVEnvironments}. Binary kicks may also explain the substantial fraction of low-redshift SNe~IIn not occurring in the star-forming regions of their host galaxies \citep{2022Ransome_HalphaIIn}. An LBV in a binary system was recently suggested for the SN impostor 2000ch to explain its periodic eruptions \citep{2023Aghakhanloo_2000ch}; it has also been proposed for SN~IIn 2020ywx from an analysis of the event's long-lived CSM interaction \citep{2024BaerWay_2020ywx}.

SNe~IIn exhibiting both a luminous precursor event \textit{and} photometric variability from CSM interaction offer a unique opportunity to unify distinct lines of evidence into a unified picture of the pre-explosion system. We now undertake this task for SN~2023zkd, a helium-rich SN~IIn with two photometric peaks and a years-long precursor that steadily brightens to event discovery. As we will later argue, SN~2023zkd's unique photometric and spectroscopic evolution provide renewed evidence for the role binary interaction in type~IIn explosions.

This paper is organized as follows. We describe our photometric and spectroscopic observations in \textsection\ref{sec:data}, from the archival imaging used to construct the pre-discovery light curve to the optical and NIR spectroscopy spanning the post-discovery peaks.  We model the spectral energy distribution of the SN host galaxy in \textsection\ref{sec:host} and infer its global properties. In \textsection\ref{sec:photometric_evolution}, we describe 2023zkd's unique photometric evolution post-discovery. We characterize the spectroscopic sequence of SN~2023zkd in detail in \textsection\ref{sec:spectroscopic_evolution}: we present evidence for multiple CSM components (\textsection\ref{subsec:vel_components}), evaluate the ambiguous spectroscopic classification of the event in the context of other He-rich SNe~IIn (\textsection~\ref{subsec:spec_comparison}), and constrain the density of the interaction region using the Balmer decrement (\textsection\ref{subsec:balmer_decrement}). We infer the black-body properties of the SN from precursor to secondary maximum in \textsection\ref{sec:blackbody}. In \textsection\ref{sec:circumstellar}, we use multiple shock-luminosity models to constrain the properties of the CSM and reconstruct the mass-loss history of the progenitor system. Finally, we discuss the most likely interaction geometry for the event in \textsection\ref{subsec:disc_geometry} and consider potential progenitor systems in \textsection\ref{subsec:progenitors}. We conclude with our main findings in \textsection\ref{sec:conclusion}.

We correct all photometry and spectroscopy for Galactic extinction assuming the \citet{2023Gordon_Extinction} extinction law, where $E(B-V) = 0.038$~mag \citep{2011SFD}. We further correct all observations for a host extinction of $A_V=0.47$~mag (or $E(B-V) = 0.152$~mag, assuming a Milky-Way-like $R_V=3.1$ for the host), which we infer from the spectral energy distribution (SED) of the host galaxy derived in \textsection\ref{sec:host}.

As no spectrum has been obtained for the host galaxy, in all subsequent results we adopt an event redshift of $z\approx0.056$ from the peak of the H$\alpha$ emission in the SN classification spectrum. We note that the host galaxy lies in the footprint of the Dark Energy Spectroscopic Instrument Survey \citep[DESI;][]{2019Levi_DESI}, and a host spectrum will likely be made public in an upcoming Data Release (Target~ID~\#39628011363898921). These data will permit more precise characterization of the CSM components identified in this work and, correspondingly, the properties of the most likely progenitor system.

Throughout this paper, we assume a flat $\Lambda$CDM cosmology with $H_0=\;70\;\textrm{km}\;\rm{s}^{-1}\;\textrm{Mpc}^{-1}$, $\Omega_{\rm{M}}=0.3$, and $\Omega_{\Lambda}=0.7$. All dates are given in UTC. Unless otherwise noted, phases are presented in rest-frame~days relative to the observed maximum of the first photometric peak in ZTF-$r$, which we calculate from an uncertainty-weighted average of the dates of four comparable observations to be $\mathrm{MJD}\approx60290.6\pm 1.2$.

\begin{figure}
    \centering
    \includegraphics[width=1.1\linewidth]{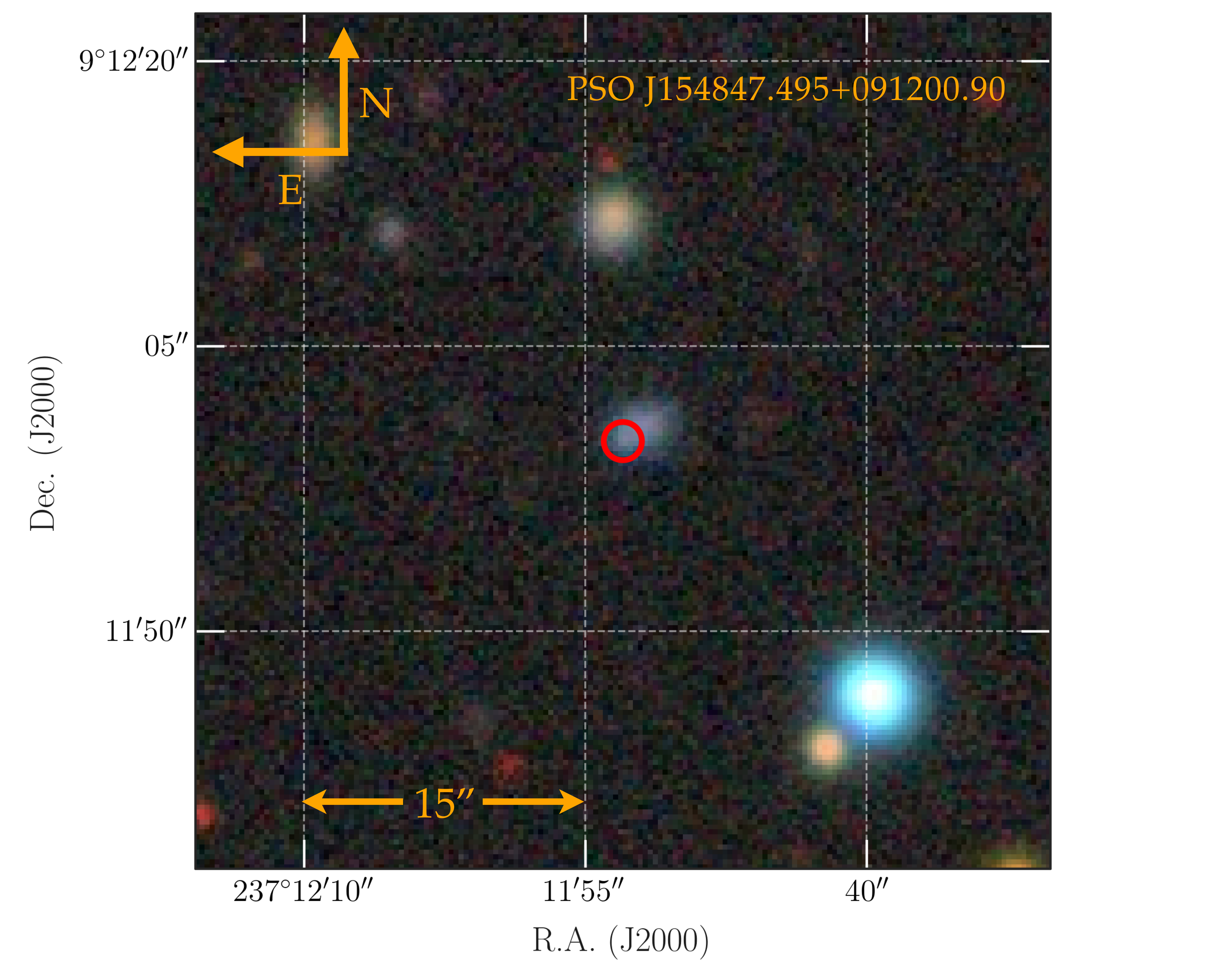}
    \caption{$grz$-band color image of the explosion site of SN~2023zkd from the Dark Energy Camera Legacy Survey (DECaLS) Data Release 9 \citep{2019Dey_DECaLS}. The SN location is denoted by the red circle with radius 1$\arcsec$, and the image scale is given bottom left. The host galaxy at image center is PS0~J154847.495+091200.90, with a redshift of $z\approx0.056$ inferred from the SN spectra.}
    \label{fig:host_image}
\end{figure}

\section{Data} \label{sec:data}
SN~2023zkd was discovered by the Zwicky Transient Facility (ZTF) on July 7th, 2023 at 04:36:28.8~UTC ($\mathrm{MJD}=60132.5$) with an apparent brightness of 20.6 in ZTF-$r$ \citep{2023ZTF_2023zkdDiscoveryReport}. The SN occurred at (J2000) $(\alpha,\delta)=(15^{\mathrm{hr}}48^{\mathrm{m}}47.536^{\mathrm{s}},\;+09\degree12'00.28\arcsec$, 0.58$\arcsec$ South and 0.70$\arcsec$ East of the center of its host galaxy (PSO~J154847.495+091200.90, shown in Figure~\ref{fig:host_image}). SN~2023zkd was later flagged on the morning of January 18th, 2024 ($\mathrm{MJD}\approx60327$) by the Light curve Anomaly Identification and Similarity Search \citep[LAISS;][]{2024Aleo_LAISS} tool running within the ANTARES Astronomical Time-domain Event Broker \citep{2021Matheson_ANTARES} and integrated into the Young Supernova Experiment via a Slack bot (details on the event's early anomalous features are provided in the Appendix). A classification spectrum was then obtained by our group (P.I. Gagliano) with the Magellan Low Dispersion Survey Spectrograph 3 \citep[LDSS3;][]{1994AllingtonSmith_LDSS3} on March 17th, 2024 (+90d). 

By comparing our +90d spectrum to templates in the GEneric cLAssification TOol (GELATO)\footnote{\url{https://gelato.tng.iac.es/}}, and noting the prominent narrow Balmer emission features, we determined an initial type~IIn classification. A public type~Ibn classification was later released on the Transient Name Server\footnote{\url{https://www.wis-tns.org/}} (TNS) on July 13, 2024 \citep{2024_ePESSTO_2023zkdClassification} (+202.2d), at which point the H$\alpha$ feature was less prominent. 

Statistically significant $>$$3\sigma$ positive flux was detected in unbinned difference imaging at the SN site in Pan-STARRS \citep{2016Chambers_PS1} GPC1-$w$ at a single epoch on March 30, 2023 ($\mathrm{MJD}\approx60033.6$, $m_w\approx21.4$), 98.9~days before discovery in the observer frame. This preliminary measurement hinted at emission above the baseline level from the pre-explosion system. We have undertaken a comprehensive analysis of archival imaging at the explosion site, and detect statistically-significant ($>$3$\sigma$) positive flux in 50-day bins spanning $\sim$1,500~days prior to discovery and in photometric filters from ZTF, ATLAS, and Pan-STARRS. 

\subsection{Optical Photometry} \label{subsec:phot}
\subsubsection{FLWO/Keplercam}\label{subsub:data_keplercam}
We observed SN~2023zkd with the KeplerCam Camera on the 1.2~meter Telescope at the at the Fred Lawrence Whipple Observatory (FLWO) from MJD 60416.23 to 60576.17 (+110.0d to +270.4d). Photometric observations were reduced using a custom point spread function (PSF) photometry routine in python. Bias and flat frames were also obtained. Multiple exposures were obtained for each observation, and observations from the same night were stacked using SWarp \citep{swarp_cite}. The PSF was constructed using a customized version of SExtractor \citep{sextractor_cite} written in python, with calibrations from the Pan-STARRS catalog \citep{2016Chambers_PS1}, where template images in the field of SN 2023zkd were obtained using PANSTAMPS \citep{panstamps_cite}. Template subtraction was done using \texttt{Pyzogy} \citep{david_guevel_2017_Pyzogy}\footnote{https://github.com/dguevel/PyZOGY}. Due to inclement weather affecting many epochs, only observations $\geq 5\sigma$ above the calculated noise background after template subtraction are reported.

\subsubsection{ATLAS}\label{subsub:data_atlas}
We query the ATLAS forced photometry web service\footnote{\url{https://fallingstar-data.com/forcedphot/}} \citep{2020Smith_ATLAS,2021Shingles_ATLAS} for photometry at the explosion site, finding observations spanning MJD 57230.3 (-2898.0d) to MJD 60561.3 (+256.3d). We use the \texttt{ATClean} code \citep{2024Rest_ATClean} to identify bad flux measurements and account for systematics in the reported $c$- and $o$-band photometry. We request 10 control light curves in an annulus of radius 15$\arcsec$ surrounding the SN site to calibrate uncertainties across phases. We further apply the quality cuts recommended in \citet{2024Rest_ATClean} and remove measurements with $\chi^2>4$, flux uncertainty $>$160~$\mu$Jy, and $\chi/N>10$ (where $N$ is the number of points used to fit the pointing PSF). 

The templates used to produce ATLAS differential forced photometry were changed on approximately $\mathrm{MJD}\approx58417$ (2018-10-26) and $\mathrm{MJD}\approx58882$ (2020-02-03). To avoid step-like artifacts in the pre-explosion flux measurements from these changes, we separately estimate and subtract baseline flux values between $58417<\mathrm{MJD}<58882$ and $\mathrm{MJD}>58882$ (we do not consider photometry prior to $\mathrm{MJD}\approx58417$). In each window, we compute the median flux from the first 200~days of observations, remove values $>$1$\sigma$ from the median, and re-compute the median from the remaining data. We have ensured that $>$20 observations remain after sigma-clipping to estimate a robust baseline. We then subtract this value from flux measurements within each window and proceed with binning using the same procedure as for ZTF data (described in detail in \textsection\ref{subsub:data_ztf}).

We bin all post-explosion photometry in 50-day bins pre-explosion and 5-day bins post-explosion if a bin contains at least two flux measurements. Due to the potential for systematics remaining after our multi-baseline binning procedure, we report only detections with $>$5$\sigma$ significance prior to discovery and $>$3$\sigma$ following discovery.

\subsubsection{Pan-STARRS}\label{subsub:data_ps1}
SN~2023zkd was observed in $grizyw$ filters with the Pan-STARRS GPC1 telescope from MJD~57179.3 (-2946.3d) to MJD~60544.3 (+240.2d). Photometry was reduced using the University of Hawaii’s PS1 Image Processing Pipeline IPP \citep[IPP;][]{2016Magnier_IPP,2020Magnier_IPP}. Pan-STARRS images were ingested, processed, and archived by IPP, then undergo template image convolution and subtraction. PS1 template images are created from stacked exposures (primarily the PS1 3$\pi$ survey), from which new nightly images are resampled and astrometrically aligned to match a skycell in the PS1 sky tessellation. A nightly image zeropoint is calculated by comparing PS1 stellar catalogs \citep{2016Chambers_PS1} to the PSF photometry. Template images are then convolved with nightly images and matched to their PSF via a three-Gaussian kernel before being subtracted with the \texttt{HOTPANTS} code \citep{2000Alard_HOTPANTS,2015Becker_HOTPANTS}.
 Photometry is then obtained using the \texttt{Photpipe} forced photometry pipeline \citep{2005Rest_Photpipe, 2014Rest_Photpipe}, where epochs flagged as potentially having bad pixels are removed. For all other epochs, a flux-weighted centroid is forced to be at the transient candidate position and a nightly zeropoint is applied to calculate the source’s brightness. Final data were then accessed through the YSE's transient management platform, {\tt YSE-PZ} \citep{2022Coulter_YSEPZ,2023PASP_YSEPZ}.

The SN was also observed in $iw$ filters with the Pan-STARRS GPC2 camera from MJD~58664.4 (-1540.0d) to MJD~60540.3 (+236.4d). As Pan-STARRS GPC2 zeropoints are not well-calibrated and the majority of pre-explosion observations could not be successfully reduced using \texttt{Photpipe}, we report only post-discovery GPC2 photometry. 

\subsubsection{Zwicky Transient Facility}\label{subsub:data_ztf}
We query the ZTF forced photometry service \citep{2019Masci_ZTF} for differential photometry at the explosion site spanning the full duration of the survey. We follow a similar procedure to that outlined in \citet{2023Masci_ForcedPhotZTF} to analyze the pre-explosion forced photometry at the explosion site. We apply a series of nominal quality cuts (ensuring $\chi^2<1.4$, removing all observations with \texttt{procstatus=56} or \texttt{seeing>4}, and excluding $g$-band photometry between MJD 58119.5 and MJD 58139.5 where images are reported to be miscalibrated). We use the reported difference image $\chi^2$ values at each epoch to re-scale reported flux uncertainties. For pre-discovery epochs, we define a reference baseline flux value as the median flux value $>$1700~days before the discovery date while ensuring that there are at least 30 epochs within the reference period for each band and field combination. We subtract all flux measurements by this value, and rescale flux uncertainties to account for potential systematics identified during the baseline subtraction. 

We place all difference flux observations on a consistent reference zeropoint and define 50-day bins spanning all pre-discovery epochs and 1-day bins spanning all post-discovery epochs. Our bin size was chosen as a trade-off between the number of detections and the statistical significance of each detection; nonetheless, we find statistically significant pre-explosion emission with bin widths of 5-200~days. We compute uncertainty-weighted averages of flux values in each bin, and convert the binned fluxes to magnitude using the reference binned zeropoint. We report statistically-significant detections ($>$3$\sigma$) at each bin center, and convert non-detections to 5$\sigma$ upper limits using the reference zeropoint.

\subsubsection{Las Cumbres Observatory}\label{subsub:data_lco}
As part of the Global Supernova Project \citep{2017Howell_GSP}, we obtained $BgVri$-band imaging data for SN~2023zkd with the Las Cumbres Observatory \citep[LCO;][]{2013Brown_LCO} network of 1m robotic telescopes with Sinistro cameras and spanning Cerro Tololo Inter-American Observatory (Región~IV, Chile), McDonald Observatory (Texas, USA), South African Astronomical Observatory (Sutherland, South Africa), and Teide Observatory (Canary Islands, Spain). The observations cover the secondary photometric peak from MJD~60508.1 to MJD~60593.1 (+206.0d to +286.5d). LCO photometry was performed with PSF fitting using \texttt{lcogtsnpipe}\footnote{
https://github.com/LCOGT/lcogtsnpipe/} \citep{2016Valenti_LCO}, a PyRAF-based photometric reduction pipeline. $BV$ and $gri$-band data were calibrated to Vega and AB magnitudes, respectively, using
the 9th Data Release of the AAVSO Photometric All Sky Survey \citep{2016Henden_AAVSO} using the 13th Data Release of the Sloan Digital Sky Survey \citep[SDSS;][]{2017Albareti_SDSS}.

\subsection{UV Photometry}\label{subsub:data_uvot}
The SN site was observed with the \textit{Swift Ultra-Violet/Optical Telescope} \citep{2005Roming_UVOT} (ObsIDs: 00016727001-00016727006) at eight epochs spanning $\mathrm{MJD}=60514.5$ (+211.9d) to $\mathrm{MJD}=60545.5$ (+241.4d). Following the analysis steps described in \citet{2014Brown_UVOT}, we used \texttt{uvotsource} from the \texttt{HEASoft v6.26} package to perform aperture photometry on UVOT pointings within a 3$\arcsec$ aperture centered on SN~2023zkd. We estimated the background flux at the SN site using an annulus with inner radius 6$\arcsec$ and outer radius 6$\arcsec$ centered on the SN. The aperture-corrected background flux was then subtracted from all observations to obtain differential photometry. 

\subsection{X-ray} \label{subsec:xray}
In addition to the \textit{Swift} UVOT data, contemporaneous X-ray observations were taken by the \textit{Neil G. Gehrels Swift X-Ray Telescope} (XRT) in photon-counting mode (ObsIDs: 00016727001-00016727008). All observations were reprocessed with the standard filter and screening criteria\footnote{\url{https://swift.gsfc.nasa.gov/analysis/xrt_swguide_v1_2.pdf}} as well as the most recent calibration files from level one XRT data using \textsc{XRTPIPELINE} version 0.13.7. Using a source region with a radius of 49$\arcsec$ centered on the location of 2023zkd and a source-free background region with a radius of 150$\arcsec$ centered at ($\alpha$,$\delta$)=(15:48:37.8212,+9:19:32.377), no significant X-ray emission associated with the source was found in the individual epochs. To increase the signal to noise of our observations, we merged all eight individual \textit{Swift} observations using \textsc{XSELECT} version 2.5b. We find no significant X-ray emission arising from SN~2023zkd. Using the source and background regions above, we derive a 3$\sigma$ upper limit to the 0.3-10.0 keV count rate of 0.004 counts/sec. Assuming a galactic column density of $3.72\times10^{20}$~cm$^{-2}$ derived from \citet{2013Willingale_HI} and adopting a thermal Brehmsstrahlung source spectrum with kT$\approx19$~keV as suggested by the strongly-interacting type~IIn SN~2010jl \citep{2015Chandra}\footnote{This event exhibited an LBV-like progenitor and multiple spectroscopic similiarities to 2023zkd; we discuss these in subsequent sections.}, we obtain a 3 sigma upper limit on the unabsorbed flux (0.3-10.0 keV) of $2.06\times10^{-13}\;\mathrm{erg}\;\mathrm{cm}^{-2}\;\mathrm{s}^{-1}$. Assuming a redshift of $z=0.056$, this corresponds to an X-ray luminosity of $1.5\times10^{42}\;\mathrm{erg}\;\mathrm{s}^{-1}$.

\subsection{Optical and Near-IR Spectroscopy} \label{subsec:spec}
In this section, we outline the spectroscopic sequence obtained for SN~2023zkd. The dates, phases, and wavelength coverage of each obtained spectrum are summarized in Table~\ref{tab:spec_observations}. All spectra will be publicly released via WISeREP\footnote{\url{https://www.wiserep.org/}} \citep{2012Yaron_wiserep} following publication.

\begin{deluxetable*}{crccccc}
\tablehead{
\colhead{Observation Date} & \colhead{MJD} & \colhead{Phase} & \colhead{Telescope} & \colhead{Spectrograph} & \colhead{Wavelength Range} & \colhead{Exposure Time} \\
 \colhead{(UT)} & \colhead{}   & \colhead{(Days)} & \colhead{} & \colhead{} & \colhead{(\AA)} & \colhead{(s)}
 }
\startdata
2024-03-17 & 68386.1 & +90.4 & Magellan & LDSS-3 & 3800 -- 10000 & 2700 \\
2024-07-13 & 60504.1 & +202.2 & NTT & EFOSC2 &  3000 -- 8800 & Unk.\tablenotemark{a} \\
2024-07-19 & 60510.9 & +208.7 & NOT & ALFOSC &  3500 -- 8500 & 3063 \\
2024-07-22 & 60514.0 & +211.5 & NOT & ALFOSC &  3500 -- 8500 & 3063 \\
2024-07-27 & 60519.0 & +216.3 & NOT & ALFOSC &  3500 -- 8500 & 3063 \\
2024-08-01 & 60523.3 & +220.4 & Gemini-North & GNIRS & 9000 -- 24000 & 6300 \\
2024-08-03 & 60525.0 & +222.0 & NOT & ALFOSC & 3500 -- 8500 & 3063 \\
2024-08-08 & 60530.1 & +226.8 & Magellan & LDSS-3 & 3800 -- 9000 & 1800 \\
2024-08-16 & 60538.2 & +234.5 & Shane & KAST & 3500 -- 9000 & 3000\tablenotemark{b} \\
2024-09-01 & 60555.0 & +250.4 & Magellan & LDSS-3 & 3800 -- 10000 & 1800 \\
2024-09-02 & 60555.1 & +250.5 & MMT & Binospec & 3500 -- 8800 & 1950 \\
2024-09-10 & 60563.1 & +258.0 & MMT & Binospec & 3500 -- 8800 & 2250 \\
\enddata
\tablenotetext{a}{Public TNS spectrum; exposure time unknown.}
\tablenotetext{b}{Exposure time for each (red/blue) side of the spectrograph.}
\caption{Spectroscopic observations of SN~2023zkd. All phases are presented in rest-frame days relative to the first ZTF-$r$ peak at MJD=60290.6.}
    \label{tab:spec_observations}
\end{deluxetable*}

\subsubsection{MMT/Binospec}\label{subsec:Binospecdata}
We obtained two spectra of SN~2023zkd with the Binospec wide-field optical spectrograph on the 6.5m MMT through a follow-up program on behalf of the Young Supernova Experiment on MJD~60555.1 (+250.5d) and MJD~60563.1 (+258.1d). Each spectra were obtained with the 270 line grating, LP3800 blocking filter, a 1$\arcsec$ slit width, and a central wavelength of 6500~Å. We reduced the spectra with standard reduction methods using the \texttt{redspec} pipeline\footnote{\url{https://github.com/gmzsebastian/redspec}} \citep{sebastian_gomez_2024_14291063}, which uses a combination of standard IRAF routines and the \texttt{twodspec} package.

\subsubsection{Magellan/LDSS3}\label{subsec:Magellandata}
We obtained three spectra of SN~2023zkd with the LDSS3 spectrograph mounted atop the Magellan 6.5m Clay telescope on MJD~60530.1 (+226.8d), MJD~60386.1 (+90.4d), and MJD~60555.0 (+250.4d). All spectra were obtained in 3 exposures using a 1$\arcsec$ slit, the LP3800 filter, and 270 grating set to a central wavelength of 6500~Å. We set 900s per exposure for the first spectrum and 600s per exposure for the final two. Spectra were reduced using the same pipeline as was used for the MMT/Binospec data. 

\subsubsection{NOT/ALFOSC}\label{subsec:ALFOSCdata}
We obtained four spectra of SN~2023zkd using the Alhambra Faint Object Spectrograph and Camera (ALFOSC) mounted on the Nordic Optical Telescope (NOT). The ALFOSC spectra were taken with a 1$\arcsec$ slit and grism 4. All spectra were bias subtracted, flat-fielded, wavelength calibrated and then extracted using standard routines within {\tt{IRAF}}. Nightly spectroscopic standard stars were used for flux calibration. 

\subsubsection{Gemini North/GNIRS}\label{subsubsec:GNIRS}
We used the Gemini North Near-IR Spectrograph (GNIRS) to obtain a single near-infrared spectrum of SN~2023zkd on August~1st, 2024 ($\mathrm{MJD}\approx60523.3$, +220.4~d), during the secondary photometric maximum. Observations were taken with the SXD prism and 32mm cross-disperser grating, in 21 exposures of 300s each. The spectrum was reduced in the standard manner with the PypeIt pipeline \citep{pypeit:joss_pub}.

\section{Host Galaxy Properties with Prospector}\label{sec:host}
SN~2023zkd's ambiguous spectroscopic classification motivates an investigation into its host properties, as systematic differences have been found between the hosts of archival SNe~IIn, SLSNe~IIn, and SNe~Ibn (see Figure~7 of \citealt{2024Qin_hosts} and Figure~12 of \citealt{2021Schulze_SNIInHosts}). The derived global properties of the host, including its stellar mass and star-formation rate, can also provide indirect evidence for a particular progenitor system \citep{2014Habergham_InteractingEnvironments,Ransome2022_SNIInHosts,2023Moriya_SNeIIn}.

An ongoing challenge in galaxy SED fitting lies in aggregating catalog-level photometry from diverse surveys, each obtained from instruments with distinct systematics and extracted using survey-specific techniques. To circumvent this issue, we use the \texttt{BLAST} web application \citep{2024Jones_BLAST,nugent2025} to retrieve postage stamps of the host galaxy of SN~2023zkd in GALEX \citep{2005Martin_GALEX}, Pan-STARRS, SDSS \citep{2000York_SDSS,2017Blanton_SDSS}, DES \citep{2019Dey_DESILegacySurveys} and 2MASS \citep{2006Skrutskie_2MASS}. We use the morphological properties of the host reported in Pan-STARRS to construct global PSF-matched elliptical apertures and extract photometry for the host in each filter. The host galaxy is detected in GALEX ($FUV$ and $NUV$), PanSTARRS ($griz$), SDSS ($ugriz$), and DES ($gz$). The source is not detected in 2MASS $J$ and $K$; given the large aperture correction required to account for the 1$\arcsec$ pixel scale, we convert the raw flux measurements to $3\sigma$ upper limits for SED fitting.

We determine the stellar population properties of the host galaxy using \texttt{Prospector} \citep{2019Leja_SFH, 2021Johnson_Prospector}. Within \texttt{Prospector}, we apply a nested sampling fitting routine through \texttt{dynesty} \citep{2020Speagle_Dynesty} to fit the observed host photometry and upper limits and obtain posterior distributions on the stellar population properties of interest, including redshift, mass formed ($M_F$), stellar metallicity ($Z_*$), and dust attenuation from old ($\tau_{v,2}$) and young ($\tau_{v,1}$) stellar light. We construct a model SED with \texttt{FSPS} and \texttt{python-FSPS} \citep{2009Conroy_FSPS, 2010Conroy_FSPS}, which internally uses the \texttt{MIST} stellar isochrones \citep{2011Paxton_MIST3,2013Paxton_MIST4,2015Paxton_MIST5,2016Choi_MIST2,2016Dotter_MIST1} and \texttt{MILES} spectral library \citep{2006Sanchez_MILES,2011Falcon_MILES}. 

For the \texttt{Prospector} SED fit, we employ a \citet{2003Chabrier_IMF} initial mass function, the \citet{2013Kriek_DustLaw} dust attenuation model, which measures an offset from the \citet{2000Calzetti} attenuation curves and determines the fraction of light attenuated from old to young stellar light, a nebular emission model, and the \citet{2005GCB_MassMetallicity} mass-metallicity relation to probe realistic $M_F$ and $Z_*$ combinations. Given that we do not have a spectroscopic redshift for the host, we allow redshift to be a sampled parameter, with a broad prior of $0.02 < z < 0.07$ covering the redshift indicated by the SN spectra. We use a parametric delayed-$\tau$ star formation history model, SFH $\propto t\times \exp{-t/\tau}$, where $t=t_\textrm{age}$ is a sampled lookback time, constrained to be less than the age of the Universe at the sampled redshift, and $\tau$ is an additional sampled parameter representing the e-folding time. To compare to commonly used values in the literature, we convert $M_F$ to a stellar mass $M_*$, $\tau_{v,2}$ and $\tau_{v,1}$ to the total dust attenuation ($A_V$) in mag, and determine a mass-weighted age ($t_m$) and present-day star formation rate (SFR) from $M_F$, $\tau$, and $t_\textrm{age}$. We refer the reader to \citet{2020Nugent_ClusterEnv} and \citet{2022Nugent_GRBs} for descriptions on these conversions.

\begin{figure*}
    \centering
    \includegraphics[width=\linewidth]{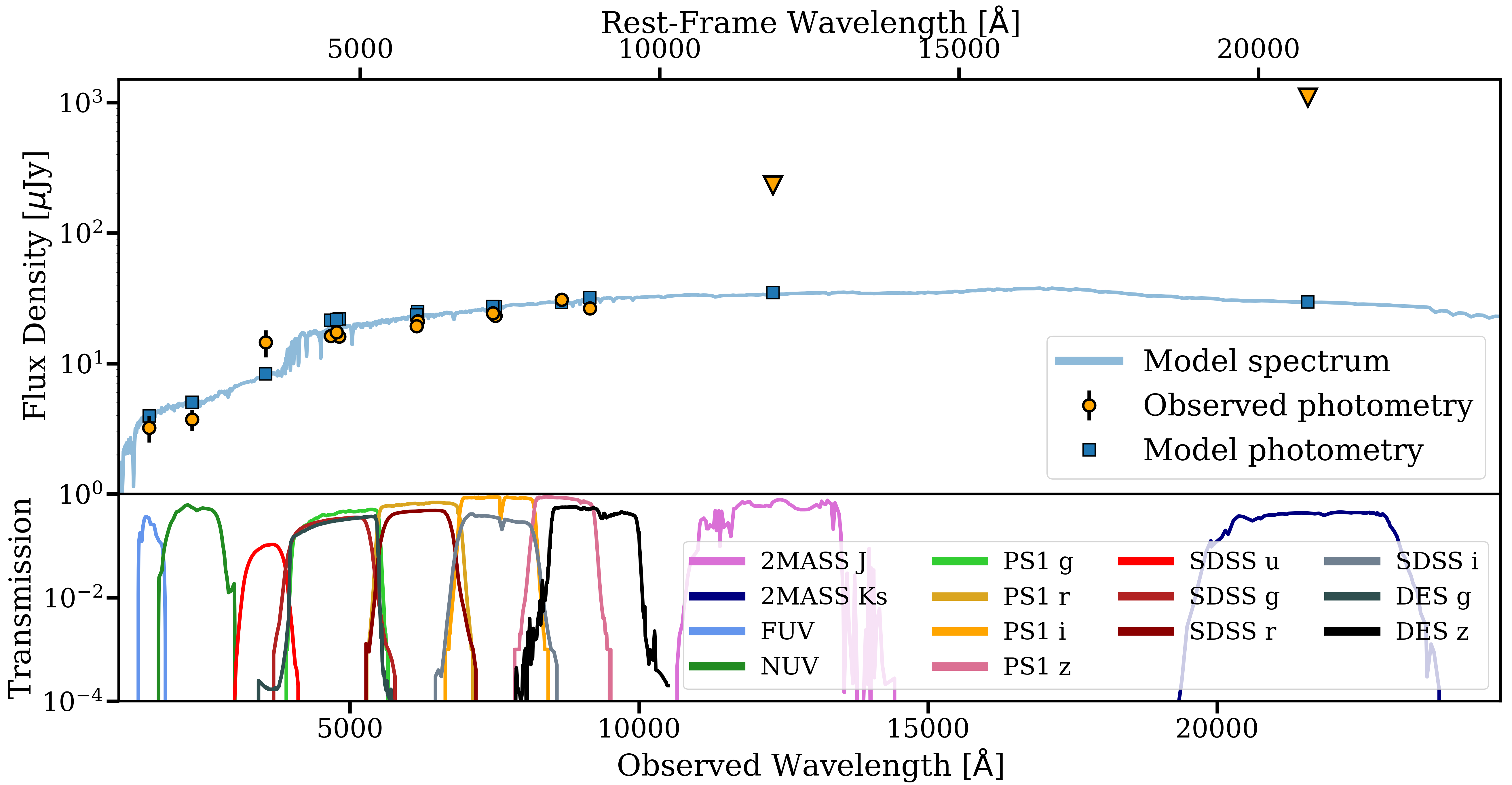}
    \caption{The Prospector model spectra (blue line) and photometry (blue squares) derived at the median of the posterior distributions of each stellar population property compared to the observed photometry (orange circles). 2MASS NIR observations are S/N$<$3, and converted to upper limits (shown as orange triangles). Transmission curves for UV, optical, and NIR photometry are shown in the bottom panel. The model underestimates the observed SDSS-$u$ photometry but is otherwise consistent with a Balmer break at $\sim$4000~Å for $z~0.05$, indicating minimal star-formation.}
    \label{fig:prospector_sed}
\end{figure*}

A corner plot of the posteriors is provided in the Appendix, and the spectral energy distribution associated with the median posteriors is plotted along with the host photometry in Figure~\ref{fig:prospector_sed}. The model underestimates the galaxy flux in SDSS-$u$ but is otherwise consistent with the measured photometry; the median posterior redshift of $z=0.04^{+0.01}_{-0.01}$ is broadly consistent with the value of $z=0.056$ obtained from the SN but less tightly constrained.  

\begin{deluxetable}{cc}
\tablehead{
\colhead{Parameter} & \colhead{Posterior Median and 1$\sigma$ Range}
 }
\startdata
$z$ & $0.04^{+0.01}_{-0.01}$\\
$A_V$ & $0.47^{+0.38}_{-0.23}$\\
$t_m$ & $2.61^{+1.12}_{-1.21}$\\
$\tau$ & $8.05^{+11.00}_{-5.66}$\\
SFR [$M_\odot$~yr$^{-1}$] & $0.02^{+0.02}_{-0.01}$\\
log(M$_*$/M$_{\odot}$) & $7.61^{+0.24}_{-0.41}$\\
log(Z$_*$/Z$_{\odot}$) & $-0.73^{+0.32}_{-0.18}$
\enddata
\caption{Median and 1$\sigma$ values for the parameter posteriors of the \texttt{Prospector} stellar population model fit to the host galaxy photometry.}
\label{tab:host_properties}
\end{deluxetable}

\begin{figure}
    \centering
    \includegraphics[width=\linewidth]{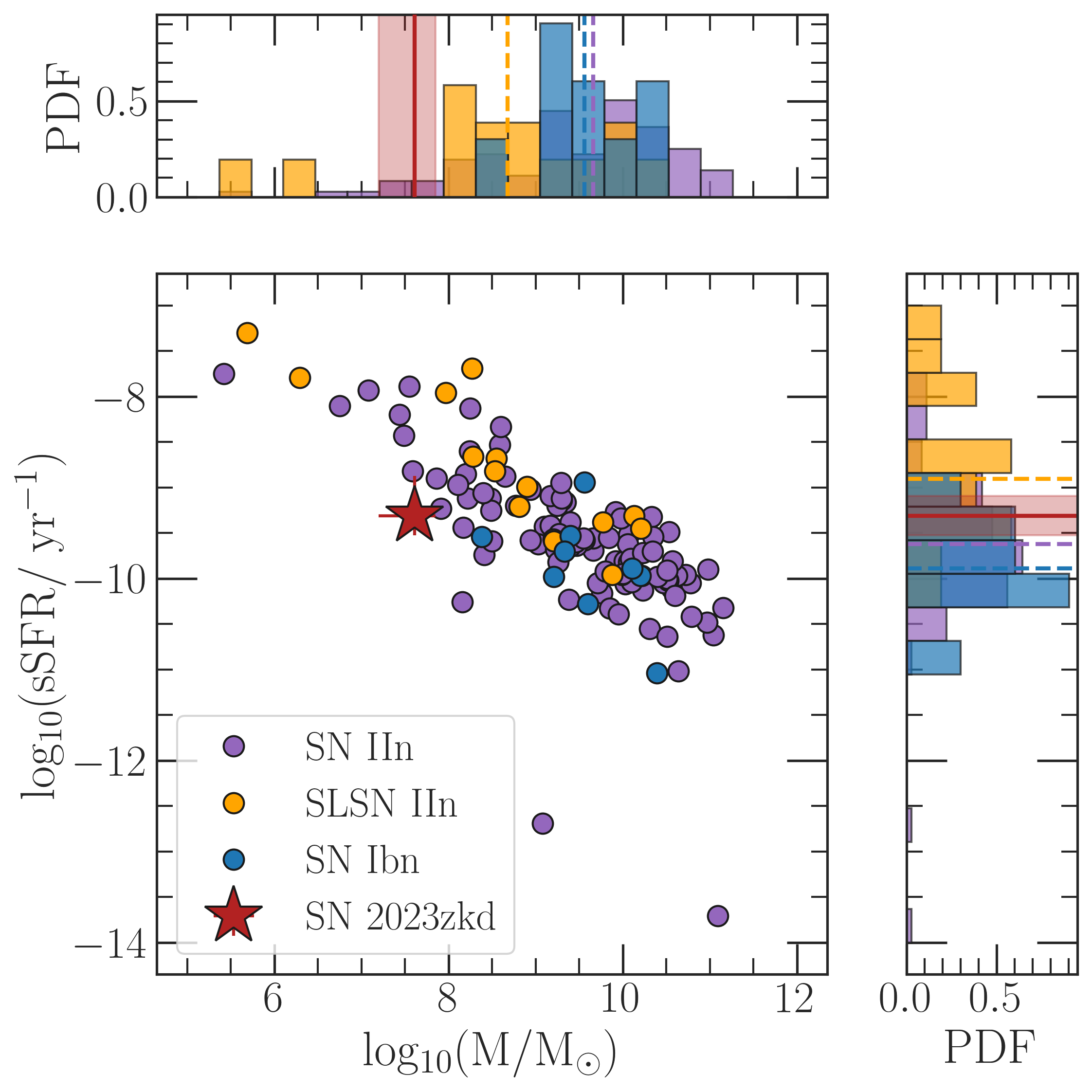}
    \caption{Comparison between the \texttt{Prospector}-derived global stellar mass and star-formation rate for the host of SN~2023zkd (red star) and those reported for the hosts of interaction-powered transients from \citet{2021Schulze_SNIInHosts} (classes given in legend). Probability density functions are given at top and right, with red line and shaded region corresponding to the median and 1$\sigma$ uncertainties for the properties of the SN~2023zkd host. Dashed lines indicate the median values for the reference samples. The host galaxy mass is lower than most comparison objects, but the specific star-formation rate is comparable.}
    \label{fig:host_prop_comparison}
\end{figure}

The posterior median values are presented in Table~\ref{tab:host_properties}. We compare them to the properties of archival SN~Ibn, SN~IIn, and SLSN~IIn host galaxies from \citet{2021Schulze_SNIInHosts} in Figure~\ref{fig:host_prop_comparison}. The inferred stellar mass of log(M$_*$/M$_{\odot}$)$=7.61^{+0.24}_{-0.41}$ is in the bottom 7\% of the 97 SN~IIn hosts from \citet{2021Schulze_SNIInHosts}, and substantially lower than the value of log(M/M$_{\odot}$)$=10.2$ reported for SN~2021qqp (an SN~IIn with similar photometric properties to SN~2023zkd, as we discuss in \textsection\ref{sec:photometric_evolution}). It is also lower than the majority of reported SN~Ibn and SLSN~IIn hosts.

We constrain the global star-formation rate to $0.02^{+0.02}_{-0.01}$~M$_{\odot}$~yr$^{-1}$. Only four hosts in the type~IIn sample (PTF09bcl, PTF10dk, PTF10uiz, and iPTF14ajx) exhibit lower values, though this can be predominantly explained by the low stellar mass: the specific star-formation rate of $\mathrm{log}_{10}(sSFR)=-9.3^{+0.43}_{-0.22}$ is marginally ($<$1$\sigma$) lower than means of the SN~IIn and SN~Ibn samples (and higher than the mean for SLSNe-IIn). The mass of the host is consistent with the reported distribution of SLSNe~IIn, a class of strongly-interacting H-rich SNe defined by a brightness at peak of $M_r<-21$~mag \citep{2012GalYam_LSNe}. A sample median of $8.89^{+0.38}_{-0.37}\;M/M_{\odot}$ is reported for SLSN~IIn hosts versus $9.63^{+0.12}_{-0.12}\;M/M_{\odot}$ for SN~IIn hosts, though we caution that sample sizes are small. 

Finally, we note the inferred metallicity of the host galaxy is 0.5~dex lower than was inferred for 2021qqp using \texttt{Prospector} for the same assumed star-formation model \citep{2024Hiramatsu_21qqp}; and lower than all values reported for SN~IIn host galaxies in \citet{2016Galbany_Metallicity}, where it was estimated from the O3N2 tracer in spaxel-integrated host spectra \citep{1979Alloin_O3N2}.

\section{Photometric Evolution}\label{sec:photometric_evolution}

\begin{figure*}
    \centering
    \includegraphics[width=\linewidth]{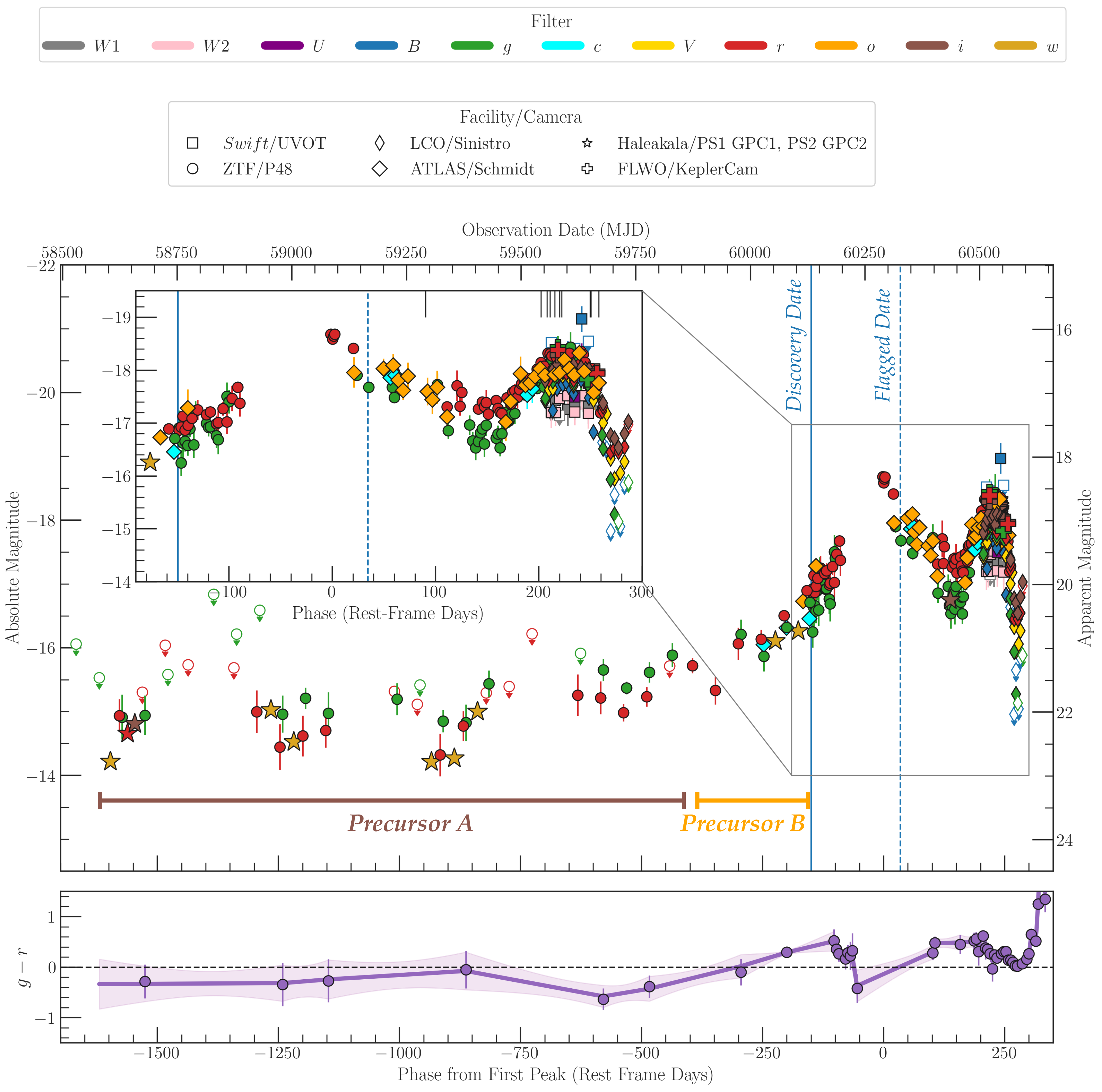}
    \caption{Multi-band photometry (top panel) and observed $g-r$ color evolution (bottom panel) for SN~2023zkd. All data have been corrected for Galactic and host extinction ($A_{V, \mathrm{host}}\approx0.47$). Photometry prior to discovery at MJD~60132.5 is shown in 50-day bins from ZTF and Pan-STARRS; post-discovery photometry is binned in 1-day bins for ZTF photometry, 5-day bins for ATLAS photometry, and unbinned otherwise. Filled points indicate detections with $>$3$\sigma$ significance, and open markers indicate 5$\sigma$ upper limits. The inset plot highlights the post-discovery photometry for the event, with vertical black lines at top indicating the phases of obtained spectra. The blue solid line marks the discovery phase, and the blue dashed line marks the date the transient was flagged by LAISS. The gap during the first maximum is due to Sun constraints. Precursor emission is detected to $>$3$\sigma$ at multiple epochs in ZTF, ATLAS, and Pan-STARRS1 data $\sim$4.5 years before the first ZTF-$r$-band maximum, and $\sim$4 years before SN discovery. The $g-r$ color is estimated with weighted averages of photometry in 100-day bins pre-discovery and 5-day bins post-discovery; the purple line and shaded region correspond to the mean and 1$\sigma$ standard deviation, respectively, of a univariate spline fit to the binned measurements.}
    \label{fig:all_phot}
\end{figure*}

We plot the full photometric sequence obtained for SN~2023zkd in Figure~\ref{fig:all_phot}. The data reveal both long-lived precursor emission and two pronounced post-explosion peaks. Prior to a seasonal gap between MJD 60196 and 60289 (93~days in the observer frame, 88~days in the rest frame), the post-discovery light curve is characterized by a 62-day rise at a median rate of 24~mmag~day$^{-1}$. A peak brightness of $M_r=-18.7$~mag was observed after re-appearing from behind the sun, followed by a 1.3~magnitude decline over the following 170~days. The event then underwent a dramatic re-brightening period, reaching a secondary maximum of $M_r=-18.4$~mag in ZTF-$r$ 85~days after first minimum. It finally exhibited a secondary dimming period, dropping by $\sim$2 magnitudes in 22~days before passing behind the Sun.

\begin{figure*}
    \centering
    \includegraphics[width=\linewidth]{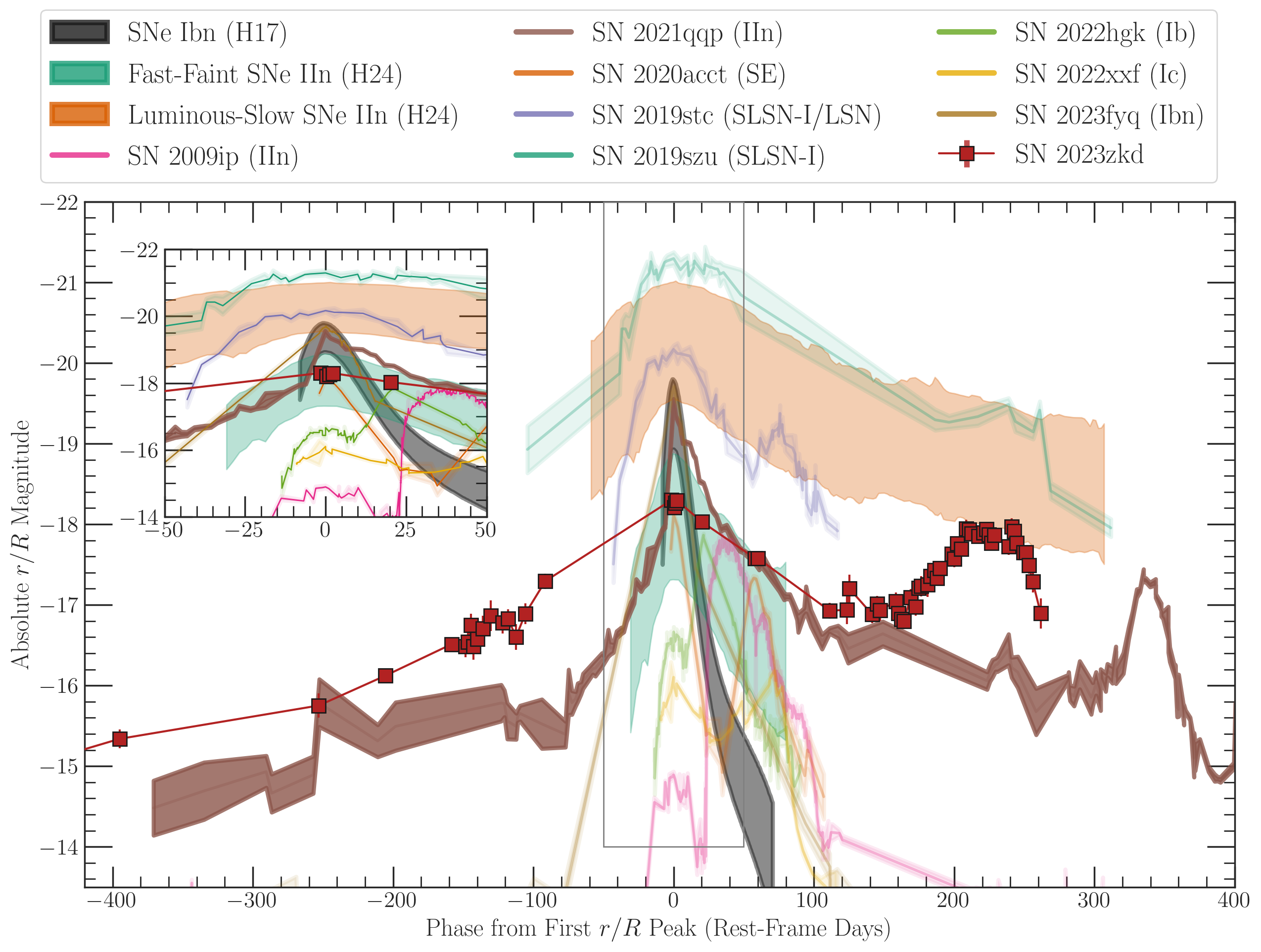}
    \caption{Comparison between the photometric evolution of SN~2023zkd in ZTF-$r$ and other interacting SNe, including the double-peaked sample from \citet{2024Angus_2020acct} (see references therein) and 2021qqp \citep[the closest photometric analog to 2023zkd, shown in brown;][]{2024Hiramatsu_21qqp}. The 95\% range (1.96$\sigma$) for the SN~Ibn template from \citet{2017Hosseinzadeh_SNeIbn}is shown as the black shaded region, as is done for the ``fast-faint" and ``luminous-slow" SN~IIn templates from \citet{2024Hiramatsu_IInSample} shown as green and orange shaded regions, respectively. SN~2023zkd evolves more slowly than expected for an SN~Ibn and is more consistent with the ``fast-faint" SN~IIn template, reinforcing its type~IIn classification in this work. Transient names and classes are given in legend (the alternative classification of ``Luminous SN" (LSN) is suggested by \citealt{2022Gomez_LSNe} for the double-peaked SN 2019stc).}
    \label{fig:photometric_comparison}
\end{figure*}

The binned pre-discovery photometry is characterized by emission at a median $M_r\approx-15$~mag for $\sim$1,500~days. This may represent roughly persistent emission across this period or multiple long-lived eruptive episodes; the detections ostensibly suggest two distinct periods spanning $\sim$500~days between MJD~58623 and MJD~59100 (-1,601d to -1,127d) and $\sim$550~days between MJD~59300 and MJD~59850 (-938d to -417d). This secondary interpretation is suggestive of the multiple eruptive mass-loss episodes associated with SNe~IIn reported in \citet{2021Strotjohann_MonthsLong}, although the maxima of these proposed episodes are not detected in 50-day bins and additional analysis (described in the Appendix) indicate that the observed variability is statistically insignificant. We therefore define this phase of roughly persistent emission, spanning $\mathrm{MJD}=58623$ (-1579~d, or -1429~d relative to discovery) to $\mathrm{MJD}=59873$ (-417~d, or -259~d relative to discovery) as highlighted by the brown line in Figure~\ref{fig:all_phot}, as `Precursor A'.

In contrast with the plateau-like emission in Precursor A, the precursor emission in the $\sim$300~days prior to discovery (in the rest frame) exhibits a persistent brightening accompanied by a monotonic increase in the observed $g-r$ color (the color temperature decreases to $\sim$8,000~K over this period, unlike any period during Precursor A). We define this phase of the pre-explosion emission as `Precursor B' (spanning the yellow line in Figure~\ref{fig:all_phot}). Precursor B,
coupled with a concave-up rise in the post-discovery photometry, is unlike the \citet{2021Strotjohann_MonthsLong} sample of SN~IIn precursors and, instead, reminiscent of the inferred observational signatures of runaway binary accretion leading to an SN or merger event \citep[e.g., ][]{2018MacLeod_RunawayCoalescence,2020MacLeod_MassLoss,2020Schroder_MergerPrecursor,2024Tsuna_MergerPrecursor}. We discuss potential progenitors for SN~2023zkd in the context of Precursor B in \textsection\ref{subsec:progenitors}.

Considering only the photometry prior to discovery ($\leq-149.7\;\mathrm{d}$), the median precursor brightness of $M_r\approx-15.2$~mag and $M_g\approx-15.5$~mag is on the luminous end of the sample consolidated by \citet{2021Strotjohann_MonthsLong}; if all photometry prior to the first seasonal gap is included ($<$0$\;\mathrm{d}$), this median increases to $M_r\approx-16.1$~mag. 

Precursor B brightens at a median rate of 5~mmag~day$^{-1}$ in ZTF-$r$. This is significantly slower than the median post-discovery rise rate of 24~mmag~day$^{-1}$. If the post-discovery photometry is associated with the SN explosion, we can infer a rise time from discovery to $r$-band maximum of 60-148~days (with the spread corresponding to the duration of the seasonal gap). This lower limit is $\sim$40 days longer than the maximum rise time for the SNe~IIn with precursors reported in \citet{2021Strotjohann_MonthsLong}, but comparable to the inferred rise time of Type IIn SN~2021qqp \citep{2024Hiramatsu_21qqp}. We caution, however, that the rise times reported in \citet{2021Strotjohann_MonthsLong} are defined as the number of days to $r$-band peak from 1.086~mag below peak; this timescale will be lower than the period reported for SN~2023zkd, but this pre-peak phase occurs during the seasonal gap for 2023zkd.

Our lower limit on the rise time is also substantially higher than the average for the population of SNe~IIn modeled by \cite{2024Ransome_DiversityofSNeIIn}, who calculate $r$-band rise times of 39$^{+16}_{-23}$ days in the rest frame from discovery across their sample (the longest rise of 71.2~days attributed to SN~2020jhs is more consistent). In interaction-dominated SNe, the diffusion timescale of the CSM sets a lower limit on the SN rise. SN~2023zkd's long rise suggests that either the mass of the CSM whose interaction dominates the first peak is higher than most archival events (a median value of $\sim1.2;M_{\odot}$ is inferred for the sample in \citealt{2024Ransome_DiversityofSNeIIn}), or the early post-discovery photometry is emission from the pre-explosion system and not the SN. As potential evidence for a later explosion date, ZTF-$r$ observations starting at $\mathrm{MJD}\approx60180$ (-124d) indicate an increased brightening rate of $\sim$35~mmag~day$^{-1}$, with an even higher brightening rate of $\sim$49~mmag~day$^{-1}$ observed in ZTF-$g$ over this period; however, few observations exist at this rate prior to the seasonal gap.

Post-discovery, SN~2023zkd brightens to a maximum of $M_r\leq-18.7$~mag before dimming for $sim$170~days. Excess flux is still detected from the system at this post-peak minimum. The SN then begins a secondary re-brightening period, and rises to a comparable secondary r-band maximum of $M_r\approx-18.4$~mag 240 days after the first. 

Following the secondary peak, SN~2023zkd dims at a rate of 48~mmag~day$^{-1}$ in ZTF-$r$, six times faster than the decline rate during the first peak. This behavior is emblematic of the ejecta reaching the outer edge of the CSM, leading to a rapid drop in energy input from SN-CSM interaction. Similar behavior has been observed in other SNe~IIn \citep[e.g., 1994W;][]{2009Dessart_1994W}, and may be also suggestive of substantial CSM asymmetry \citep[see \textsection4.3 of][]{2014Moriya_SNeIInProgenitors}.

Figure~\ref{fig:all_phot} also shows the $g-r$ color of the transient in ZTF filters in 100-day bins pre-explosion and 5-day bins thereafter. The color of Precursor~A is consistent with $g-r=0$, whereas we observe a sustained reddening period from $g-r\approx-0.5$ to $g-r\approx0.5$ throughout Precursor B. We then observe a decrease in $g-r$ of $\sim$1 magnitude from discovery to first $r$-band maximum, followed by a more gradual reddening in the rise to secondary maximum at $g-r\approx0$. The rapid final dimming period is associated with a monotonic increase in $g-r$, and concludes at $g-r>1$ on $\mathrm{MJD}\approx60640$ (+331d). When the cold-dense shell formed in an SN interaction reaches the edge of the CSM, a flood of suppressed redward-scattered photons can be released from behind the interaction front, potentially explaining the rapid reddening \citep{2017Smith_InteractingSNeIIn}. Prompt dust formation may also be possible \citep{2014Gall_2010jl}; late-time NIR follow-up observations would be valuable for confirming this scenario.

Double-peaked SNe~IIn are relatively rare in the literature: of a sample of 119 SNe~IIn analyzed by \citet{2024Ransome_DiversityofSNeIIn}, five featured prominent re-brightening phases. Among other SN classes, double-peaked events span the diversity of stripped-envelope explosions: a sample of nine presented in \citet{2024Angus_2020acct} include SNe~Ib/c and SLSNe-I, with each suggesting ongoing SN interaction with material stripped from the progenitor prior to explosion. We plot the photometric evolution of SN~2023zkd in ZTF-$r$ in Figure~\ref{fig:photometric_comparison}, along with the double-peaked sample from \citet{2024Angus_2020acct}. 

To more directly compare events, we exclude SNe where the first peak is not temporally resolved in $r$ and do not correct 2023zkd's light curve for host extinction. The photometric peaks of SN~2023zkd are separated by 223.5~days (236~days in the observer frame), 104.5~days (in the rest frame) higher than the event with the widest separation for the sample reported by \citet{2024Angus_2020acct} \citep[119~days for SN~2023aew, an SN~IIb/Ic hybrid with no spectroscopic signatures of CSM interaction;][]{2024Kangas_2023aew,2024Sharma_2023aew}. Of the events in this reference sample, only SN~2022xxf also features a secondary peak with maximum brightness within 1~mag of the first peak.

In Figure~\ref{fig:photometric_comparison}, we also show $r$-band templates for ``fast-faint" and ``luminous-slow" SNe~IIn from \citet{2024Hiramatsu_IInSample} (green and orange shaded regions, respectively) and for SNe~Ibn (gray shaded region) from \citet{2017Hosseinzadeh_SNeIbn}. SN~2023zkd's evolution across both peaks is significantly slower than the SN~Ibn template, and most closely matched to the fast-faint SN~IIn population (if the pre-discovery detections are not considered). Considering the full evolution across both peaks, the rise and decline rates are more well-matched to the luminous-slow population of SNe~IIn (although the observed first peak is $\sim$2~mag fainter than the typical luminous-slow SN~IIn peak). These comparisons further reinforce SN~2023zkd's classification as an SN~IIn.

The missing photometry for the event, coupled with the host mass more closely-matched to SLSN-IIn hosts, raises the possibility of a luminous unobserved peak. Given the similar luminosity of the observed first peak relative to the SN~IIn population from \citet{2024Hiramatsu_IInSample}, we can place loose constraints on the behavior of the event during the seasonal gap. For the events in the sample, an SN~IIn with a decline time to 20\% $r$-band flux of $\sim$150~days exhibits a rise time from 20\% flux before peak of $\sim$40~days, versus the $\sim$160~days that we measure for SN~2023zkd (the phase of pre-peak 20\% flux is comparable to the discovery phase). This suggests that either the intrinsic event peak occurred during the seasonal gap, or the rise time of SN~2023zkd is atypically long among SNe~IIn. 

If we shift the fast-faint and slow-luminous templates to the phase of the last 2023zkd observation prior to the seasonal gap (to mimic an unobserved SN~IIn peak), the observed post-peak luminosity falls squarely between these two populations. This suggests an intrinsic peak $r$-band brightness brighter than the faint population but fainter than the bright population ($-18.3\;\mathrm{mag}<M_r<-20.2\;\mathrm{mag}$); linear fits extrapolated from the final pre-gap observations and the earliest post-gap observations intersect at $-18.5\;\mathrm{mag}<M_r<-19\;\mathrm{mag}$, or $-18.9\;\mathrm{mag}<M_r<-19.4\;\mathrm{mag}$ after correcting for host extinction. This is substantially lower than required for a SLSN-IIn ($M_{r/R}<-21\;\mathrm{mag}$). If a much higher brightening rate began during the seasonal gap, a rise time associated with SLSNe~IIn to $M_r\approx-21\;\mathrm{mag}$ would require a decline rate of $\geq60\;\mathrm{mmag}\;\mathrm{day}^{-1}$ to be observed at $M_{r}=-18.7$~mag after re-appearing behind the sun, three times higher than the decline rate observed for the SLSN~IIn PTF09uy \citep{2020Nyholm_SLSNIIn}. We conclude that an unobserved peak much more luminous than the observed maximum is unlikely. 

Qualitatively, the photometric evolution of SN~2023zkd appears remarkably similar to that of the strongly-interacting and double-peaked SN~IIn 2021qqp \citep[][ also shown in Figure~\ref{fig:photometric_comparison}]{2024Hiramatsu_21qqp}. A long-lived precursor was also observed in SN~2021qqp at $M_r\approx-14$~mag and lasting $>$300~days in the frame of the SN, with marginal ($\sim$3$\sigma$) detections reported $\sim$2,500 days prior to first $r$-band maximum. The separation between observed $r$-band peaks in SN~2023zkd is $\sim$112~days smaller than the separation for SN~2021qqp (though shifting the phase of 2023zkd to account for an unobserved peak brings the difference down to $\sim$50 days). Though both events exhibited increases in brightening rate leading to first $r$-band maximum, the observed rise rates are substantially different: the fastest rise observed for SN~2023zkd of $\sim$50~mmag~day$^{-1}$ is still substantially slower than the 70~mmag~day$^{-1}$ associated with the explosion first light for SN~2021qqp. The lack of a clear distinction between precursor and SN provides additional evidence that the first emission from the SN~2023zkd ejecta occurred just prior to or during the first seasonal gap. In this case, all post-discovery observations prior to MJD 60183-60188 would be associated with the SN precursor.

\section{Spectroscopic Evolution}\label{sec:spectroscopic_evolution}
\subsection{Optical Spectra}\label{subsec:optical_spec_evolution}

\begin{figure*}[!b]
    \centering
    \includegraphics[width=\linewidth]{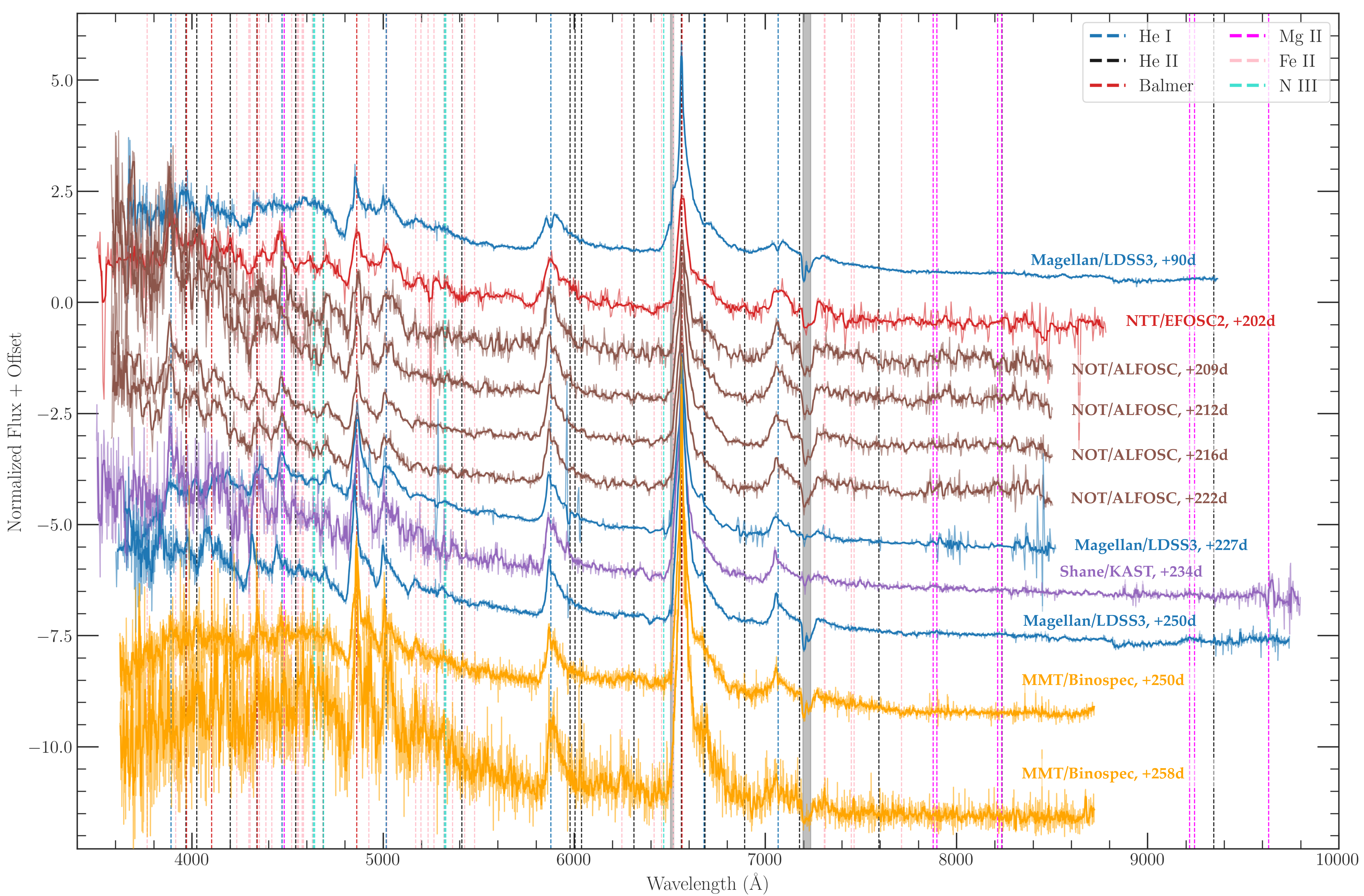}
    \caption{Optical spectroscopic sequence obtained for SN~2023zkd. Spectra are colored by spectrograph and relevant lines are annotated. Spectra have been normalized within 5500 - 7500 Å and corrected for Galactic and host extinction. Multi-component and asymmetric Balmer and He~I features are visible throughout its evolution, while a forest of Fe~II lines appear alongside candidate He~II and N~III features during the second light curve peak.}
    \label{fig:spec_sequence}
\end{figure*}

\begin{figure*}
    \centering
    \includegraphics[width=0.85\linewidth]{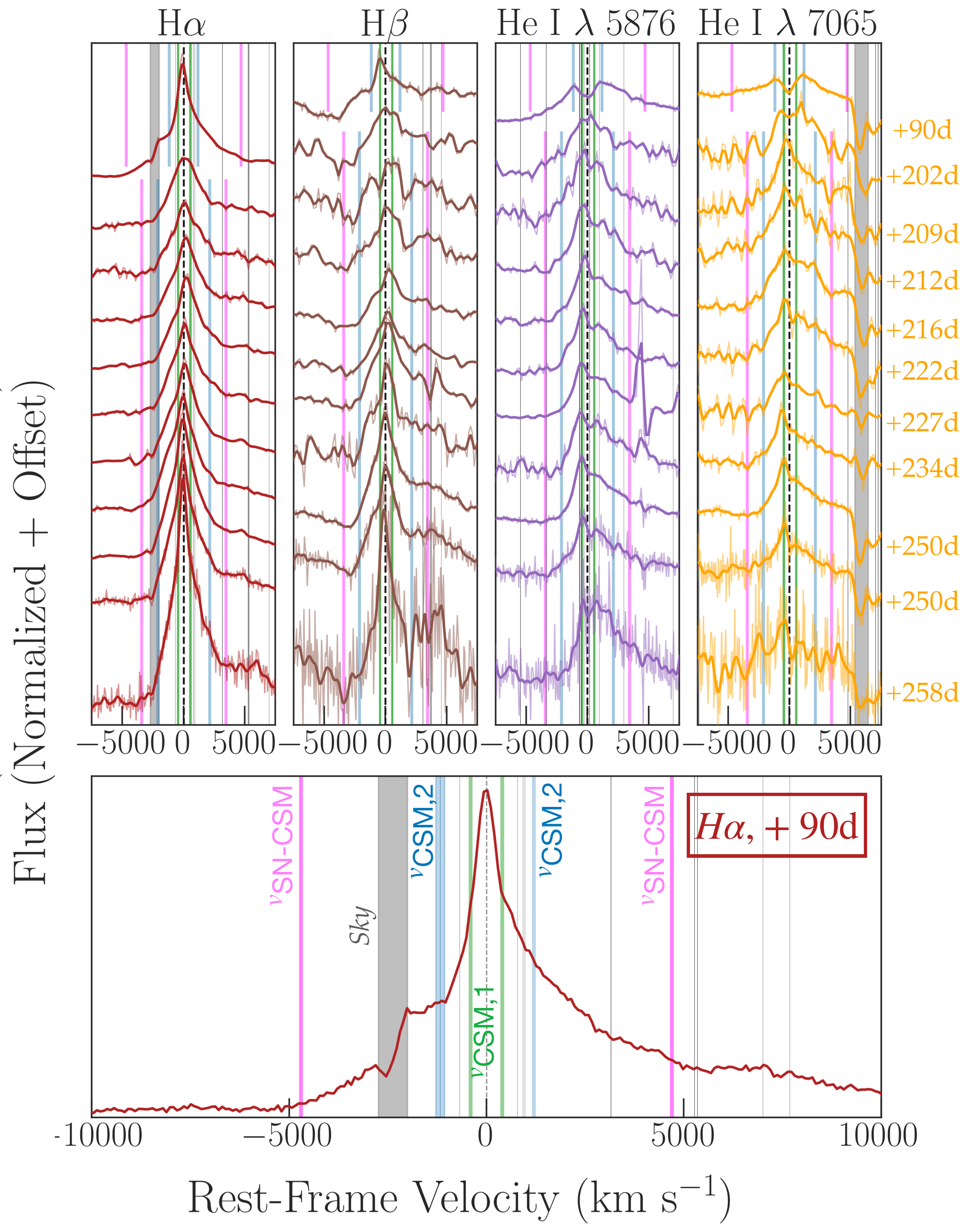}
    \caption{\textit{Top Panels:} Line profiles for H$\alpha$, H$\beta$, He~I~$\lambda5876$ and He~I~$\lambda7065$ spanning +90d to +258d. All spectra have been smoothed to a consistent resolution of $R=750$. Vertical lines indicate the range of the CSM (blue) and SN-CSM interaction shell velocities (magenta) derived in \textsection\ref{subsec:spec_fitting}; for clarity, only the mean velocities are shown for spectra obtained during the second light curve peak (+202d onward). Additional emission near the rest wavelength is suggestive of a secondary CSM component at $\pm400\;\mathrm{km}\;\mathrm{s}^{-1}$ (green lines), and the +90d spectrum shows two He~I emission components at the SN-CSM shell velocity. The low-velocity components are confirmed in the GNIRS spectrum in Figure~\ref{fig:GNIRS_spec}. The broad base observed in the +90d spectrum persists throughout the final spectrum at +258d. \textit{Bottom Panel:} H$\alpha$ profile at +90d, with the same velocity components as above. The unresolved blue shoulder is consistent with a velocity of $\sim1,\!170\;\mathrm{km}\;\mathrm{s}^{-1}$ (shaded magenta region) as measured from the FWHM of the Lorentzian profile fit to the emission (\textsection\ref{subsec:spec_comparison}), but a narrower core can also be seen.}
    \label{fig:line_profiles}
\end{figure*}

We plot our full spectroscopic sequence for SN~2023zkd, spanning 168~days, in Fig~\ref{fig:spec_sequence}. The spectra are dominated by multi-component Balmer and He~I emission features that evolve slowly from +90d to +258d. We plot the emission lines of H$\alpha$, H$\beta$, He~I $\lambda$5876, and He~I $\lambda$7065 in Figure~\ref{fig:line_profiles}. These spectral profiles are highly complex and grow increasingly asymmetric throughout the second peak, with prominent red wings most likely the product of electron scattering from the ionized CSM (similar to e.g., SN~2013L, \citealt{2020Taddia_redwing}; and SN~2021adxl, \citealt{2024Brennan_adxl}). The H$\alpha$ profiles at all epochs show a narrow core and a wide base (with the base spanning $\sim$$6,\!000\;\mathrm{km}\;\mathrm{s}^{-1}$).

SN~2023zkd is unique among SNe~IIn for its persistent He~I emission features. The Magellan/LDSS3 spectrum obtained at +90d exhibits double-peaked He~I profiles at $\lambda$5876, $\lambda$5016, and $\lambda$7065. In these profiles, the minimum between the peaks is consistent with the rest frame wavelength for all lines. In addition, a distinct `notch' is also seen near $\sim$$5,\!000\;\mathrm{km}\;\mathrm{s}^{-1}$ relative to H$\alpha$, and is likely to be the same complex feature at He~I~$\lambda$6678. 

In the second spectrum at +202d (during the rise of the second peak), He~I $\lambda$4471 and the H$\alpha$ notch are clearly observed in emission, and the He~I features at 5876~Å and 7065~Å exhibit a more flat-topped profile. The double-peaked profile re-appears by +212d, and the redward peak fades until only a single emission feature blueward of the rest wavelength dominates the profile in the Magellan/LDSS3 spectrum at +250d. 

The double-peaked He~I profiles at +90d are uncommon among even SNe~Ibn \citep{2017Hosseinzadeh_SNeIbn}. These could alternatively be two He~I features in emission and symmetric about the rest wavelength, or absorption from material with negligible velocity relative to the observer superimposed atop a broad emission feature. A detailed analysis of the velocity components in \textsection\ref{subsec:vel_components} favors the former interpretation.

A narrow P-Cygni feature also appears at $\sim$4700~Å in the NTT/EFOSC2 spectrum at +202d. It grows stronger by the phase of the second NOT/ALFOSC spectrum at +212d, then decreases again but persists through the final spectrum at +258d. Given the similar evolution of a likely He~II profile near $\lambda$8237 (potentially blended with Mg~II), we associate this with He~II~$\lambda$4686. Additional features are identified as He~II at 5977~Å, 6004~Å, and 6037~Å. 

Late-time He~II features are extremely rare among interacting SNe \citep[He~II features from ultraviolet photons mediating the SN shock are typically observed only in the few days following explosion;][]{2020Gangopadhyay_2019uo,2022Gangopadhyay_InteractingSNe,2023Chugai_HeII,2024JacobsonGalan_ggi,2024Pessi_FlashIonization2024ggi}. The collision between the SN ejecta and a secondary CSM overdensity provides a natural explanation for renewed ionizing photons at these phases (as has been observed during the first peak of SN~2020acct, \citealt{2024Angus_2020acct}; and in the peculiar type~Ib, \citealt{2008Smith_2006jc}). The P-Cygni profile of He~II~$\lambda$4686 has its absorption trough at $\sim$$1,\!000\;\mathrm{km}\;\mathrm{s}^{-1}$ in all spectra, similar to the velocity of the fast CSM we derive in \textsection\ref{subsec:spec_fitting}. 

2023zkd also shows a forest of additional narrow emission components in all spectra covering the second light-curve peak ($>$202d). The majority of these are identified as Fe~II, as shown by the pink lines in Figure~\ref{fig:spec_comparison} \citep[we adopt the Fe~II line list reported by][]{2004Chugai_1994W}. Additional blends near 7850~Å and 8200~Å are partially explained by Mg~II, although additional unknown elements may be suggested by the blends near 8000~Å and 8400~Å. An emission line observed from +209d to +250d is most easily explained as N~III at 4634~Å and 4641~Å, as seen in the SN~IIn 1998S \citep{2015Shivvers_1998S}. 

\subsection{Optical Spectral Line Fitting}\label{subsec:spec_fitting}
To better characterize the spectroscopic evolution of SN~2023zkd in the optical, we model the spectral features of H$\alpha$, H$\beta$, and He~I at $\lambda5876$ (given its strength relative to the other He~I lines). For each observed feature, we fit a combination of Gaussian and Lorentzian profiles. Each fitted component is parameterized by the amplitude, the center wavelength in the rest frame, and the profile spread ($\sigma$, the standard deviation of the Gaussian profile; and FWHM, the full-width-half-maximum for a Lorentzian profile). 

\begin{figure*}
    \centering
    \includegraphics[width=\linewidth]{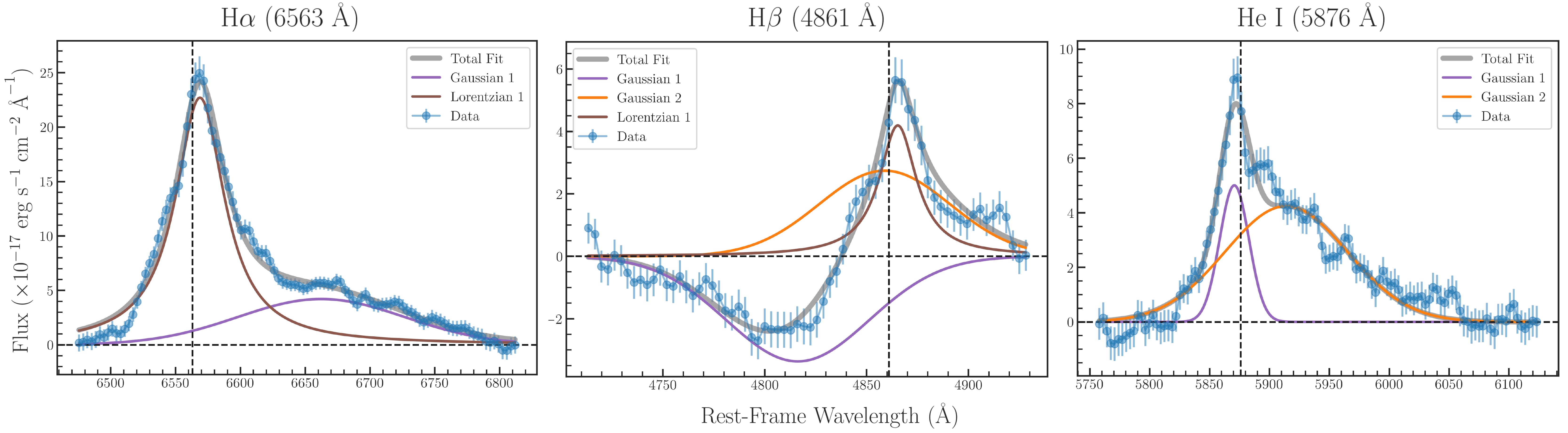}
    \caption{Line fits to the H$\alpha$ (left), H$\beta$ (middle), and He~I$\lambda$5876 (right) profiles of the NOT/ALFOSC spectrum obtained on MJD 60519 (+216.3d). Data are shown in blue, and the combined model for each profile is shown as a shaded gray line.}
    \label{fig:line_fits}
\end{figure*}

We first fit a low-order polynomial to the spectral region of interest (masking the emission profile) and subtract it from the line region. We then visually determine the line width for fitting. We define uniform priors for each parameter and compare the wavelength and flux values $(\lambda, f)$ with associated flux uncertainties $\sigma_{f}$ to the combined model fit $M(\lambda)$ using a standard log-likelihood under assumptions of Gaussianity: 
\begin{equation}
    \mathrm{ln}\;\mathcal{L} \propto -\frac{1}{2}  \sum \left[ \frac{\left(f - M(\lambda) \right)^{2}}{\sigma_{f}^{2}} +\mathrm{ln}\left(\sigma_{f}^{2}\right) \right]
\end{equation}
For each spectral feature, we initialize 64 walkers at randomized positions about an initial guess for the parameter vector with noise scale of $10^{-4}$. We then run an MCMC simulation using the \texttt{emcee} package for 10,000 total steps and a burn-in period of 1,000 steps, with the log-likelihood as our update criterion. We verify convergence by examining the parameter posteriors with the \texttt{corner} package while ensuring that the final fits all have reduced $\chi^2\approx1$. We show an example fit to our H${\alpha}$, H${\beta}$, and He~I~$\lambda$5876 profiles at MJD=60519 (+216d) in Figure~\ref{fig:line_fits}. 

We fit two profiles per line for each optical spectrum of SN~2023zkd, and integrate the combined fit from each chain to obtain the total line flux with associated uncertainties for H$\alpha$, H$\beta$, and He~I $\lambda$5876. Due to the strong asymmetries observed in the profiles, we find our best results by adopting one narrow Lorentzian component (if resolved) and one broad Gaussian component to the emission for H$\alpha$ and H$\beta$, and two Gaussian components for He~I~$\lambda$5876. For H$\beta$, we additionally fit a third Gaussian profile in absorption to the trough of the P-Cygni profile. We caution that the red wing of H$\alpha$ is completely blended with He~I at 6678~Å, which contributes additional flux to the profile. This can be seen as an absorption at this wavelength transitions to emission in co-evolution with He~I $\lambda$7065 and $\lambda$5876.

We measure the FWHM of the Lorentzian component for H$\alpha$ in velocity space, which we associate with the CSM velocity; we measure the minimum of the P-Cygni trough for H$\beta$ in velocity space and associate it with the velocity of the SN-CSM interaction shell. The minimum of the H$\beta$ trough decreases from $4,\!698^{+190}_{-150}$~km~s$^{-1}$ at $\mathrm{MJD}\approx60386$ (+90d) to $2,\!797^{+700}_{-700}$~km~s$^{-1}$ at $\mathrm{MJD}\approx60504$ (+202d), where it remains roughly constant through the second photometric peak (the highest measured value during the second peak is $3,\!622^{+670}_{-480}$~km~s$^{-1}$ at $\mathrm{MJD}\approx60530$, or +227d). The FWHM of the narrow Lorentzian component, in contrast, increases from $1,\!176^{+110}_{-110}$~km$^{-1}$ at $\mathrm{MJD}\approx60386$ (+90d) to a maximum of $2,\!228^{+92}_{-92}$~km~s$^{-1}$ at $\mathrm{MJD}\approx60525$ (+222d), nearing to the SN-CSM shell velocity\footnote{We caution that the resolution of the NOT spectrum ($\sim$400~km~s$^{-1}$) introduces additional uncertainty to the measured CSM velocity at +222d.}. At the decline of the secondary peak, the measured CSM velocity marginally decreases to $2,\!032^{+86}_{-86}$~km~s$^{-1}$ ($\mathrm{MJD}\approx60555$, +250d) and then to $1,\!824^{+46}_{-46}$~km~s$^{-1}$ ($\mathrm{MJD}\approx60563$, +258d). 

The evolution of the CSM velocity determined by our Lorentzian fits is similar to that reported for SN~2021qqp (1,300~km~s$^{-1}$ for the first peak and 2,500~km~s$^{-1}$ for the second peak), and similarly suggests that the CSM associated with the second photometric peak was ejected at higher velocities (though our measured SN-CSM velocity is substantially lower than their values of 8,500~km~s$^{-1}$ and 5,600~km~s$^{-1}$ during the first and second peak, respectively). All measured CSM velocities are on the highest end of both the SN~IIn samples associated with eruptive precursors \citep{2021Strotjohann_MonthsLong} and of the broader SN~IIn population \citep{2024Ransome_DiversityofSNeIIn}.

\subsection{Evidence for High- and Low-Velocity CSM Components}\label{subsec:vel_components}
Having identified the dominant components in the emission lines of H$\alpha$, H$\beta$, He~I $\lambda$5876, and He~I $\lambda$7065, we now investigate their unusual structure.

In addition to the CSM and SN-CSM components fit by our spectral modeling, each profile exhibits additional and marginally-resolved features that cannot be easily attributed to other transition lines (Figure~\ref{fig:line_profiles}). The blue shoulder and the primary H$\alpha$ emission at +90d both point to an unshocked CSM velocity of $\sim$1,200$\;\mathrm{km}\;\mathrm{s}^{-1}$; to probe the symmetry and composition of this CSM, we have added vertical lines to search for symmetric velocity components associated with the SN-CSM interaction shell (magenta) and the CSM (blue). We find, surprisingly, that the CSM velocities align precisely with the peaks of the He~I profiles at 5876~Å and 7065~Å at +90d. These velocities are also consistent with blueward and redward dips in the He~I $\lambda$5876 and the  He~I $\lambda$5876 profiles at +202d, and peaks in the H$\beta$ profile at multiple epochs during the second light curve peak. 

Furthermore, we find evidence for a secondary, lower-velocity CSM component: the peaks in the He~I $\lambda$5876 profile starting during the second peak at +202d are more consistent with velocities of $\pm400\;\mathrm{km}\;\mathrm{s}^{-1}$. These lower velocities are also suggested by a shift in the emission peaks of H$\beta$ toward the approaching component. While the peaks in the H$\alpha$ and H$\beta$ profiles at +90d are Lorentzian and suggestive of Thomson scattering through a high-density plasma, the double-peaked He~I profiles at this phase instead appear Doppler-broadened. This strongly suggests two distinct emitting regions. Finally, we observe that, when the two He~I profiles transition to a single dominant emission feature from +202d onward (along with a broader rightward feature), the peak of the emission lies occurs near $-400\;\mathrm{km}\;\mathrm{s}^{-1}$. The limited resolution of our spectra ($\sim$$250\;\mathrm{km}\;\mathrm{s}^{-1}$) prevents us from precisely constraining the velocity of the slow-moving material, but the components are persistent and distinct from the higher-velocity CSM at all phases after +202d.

These features suggest two distinct and axisymmetric CSM components: one lower-velocity component where narrow H emission is observed at +90d, and another higher-velocity component in which the peak of the He emission occurs during the second photometric peak. The enhancement of the blue emission component during the second light curve peak is probably the result of electron scattering, with redward photons behind the interaction shell preferentially scattered out of sight.

Fast-moving and low-moving CSM components have been identified in multiple SNe~IIn, including 2016jbu, with a lower-velocity component at $\sim$$250\;\mathrm{km}\;\mathrm{s}^{-1}$ attributed to a stellar wind in the terminal progenitor; and in ASASSN-15ua, 2013L, and 2010jl, where a spectral component near $\sim$$100\;\mathrm{km}\;\mathrm{s}^{-1}$ was reported for each event \citep{2014Gall_2010jl,2014Fransson_2010jl,2018Huang_ElectronScattering,2020Taddia_2013L,2024Dickinson_ASASSN15ua}. In 2010jl, the narrow component was also suggested to originate in material separate from the scattering medium \citep{2018Huang_ElectronScattering}. 

We favor a similar interpretation for the narrow H components at +90d. In this scenario, emission from low-velocity material scatters through the more extended, faster-moving material dominating the photometric emission, leading to the observed emission at each of the velocities through the second brightening phase. While our spectral resolution at optical wavelengths is comparable to the inferred low velocities ($\sim400\;\mathrm{km}\;\mathrm{s}^{-1}$), this view is further supported by lower-intensity He~I profiles in our higher-resolution NIR spectrum. We discuss these data in \textsection\ref{sec:nir_spectrum}. 

\subsection{Spectroscopic Comparison to other Interaction-Powered SNe and Classification as a Type IIn Supernova}\label{subsec:spec_comparison}
We next consider the ambiguous spectroscopic classification of SN~2023zkd. The event was publicly classified as an SN~Ibn due to its narrow H and He profiles, but He has been also observed in the spectra of some SNe~IIn; notable examples include SN~2016jbu \citep[an SN~IIn which also exhibited both precursor emission and a post-explosion re-brightening event 250~days after first peak;][]{2022Brennan_2016jbu}; and SN~1996al \citep[a transitional object that first appeared as a spectroscopically-normal SN~II, then later evolved to show narrow H$\alpha$ features reflective of an SN~IIn; ][]{2016Benetti_1996al}.

To understand the 2023zkd system within the broader context of interacting SNe, we query WiSeREP for the spectra of other SNe~IIn with prominent He features at similar phases to 2023zkd. We find similar spectra for SNe~2016jbu \citep{2022Brennan_2016jbu} and 1996al \citep{2016Benetti_1996al}; along with 2008iy (which showed an unprecedented rise time of $\sim$400~days, though photometric coverage of the event was poor; \citealt{2010Miller_2008iy}); and 2021adou \citep{2021JacobsonGalan_adouClassification}\footnote{Despite photometric similarities, only marginal He emission was observed in the SN~IIn 2021qqp \citep{2024Hiramatsu_21qqp}.}. We also compare the spectra of the SNe~Ibn/IIn 2021foa \citep{2024Farias_2021foa,2024Gangopadhyay_2021foa} and 2005la \citep{2008Pastorello_HeRich}.

\begin{figure*}[!htbp]
    \centering
    \includegraphics[width=\linewidth]{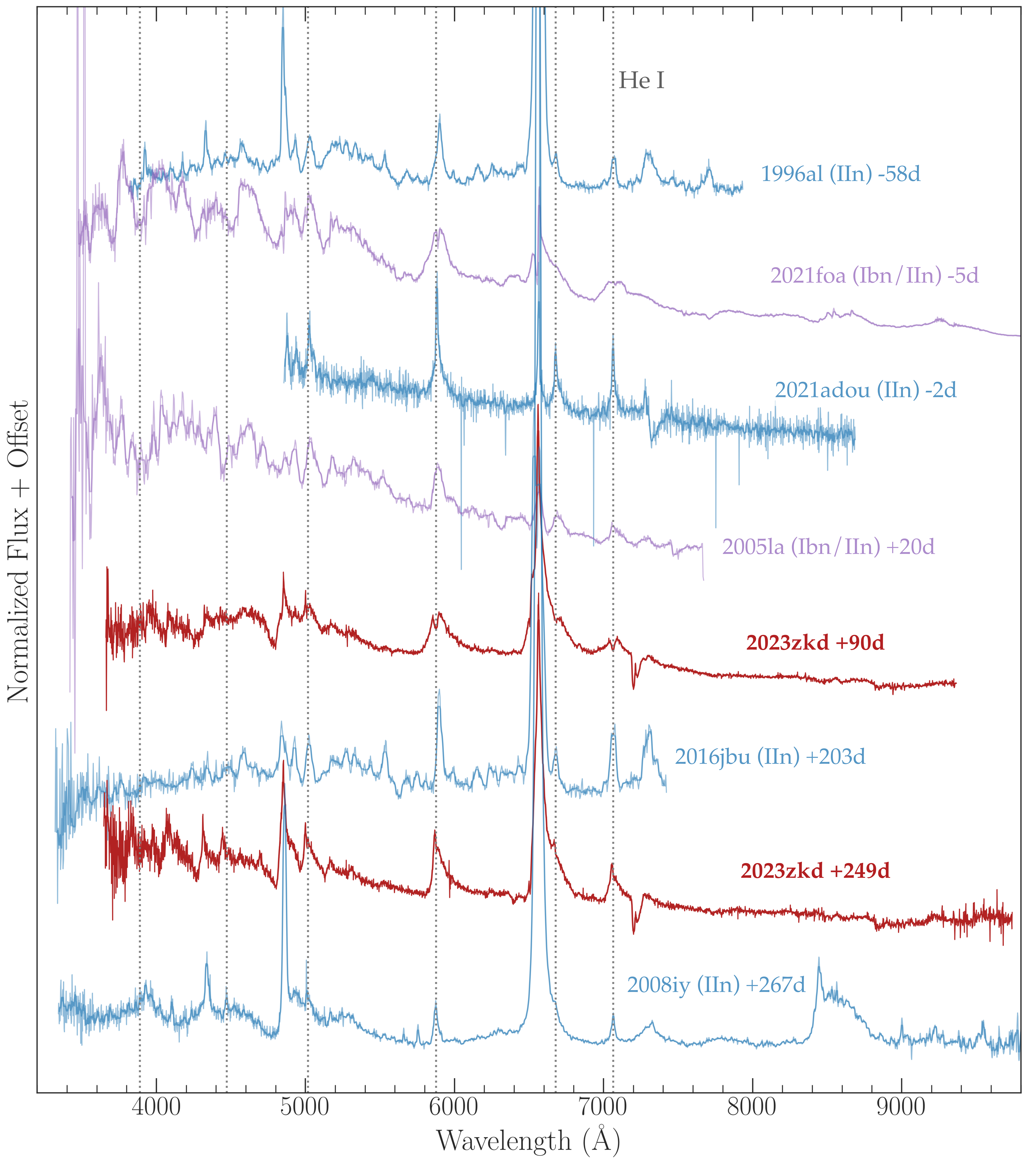}
    \caption{Comparison of 2023zkd spectra during the primary and secondary peaks to strongly-interacting SNe from literature with prominent H and He emission features. All spectra have only been corrected for Galactic extinction. Comparison spectra are colored by spectroscopic class, and vertical dashed lines indicate strong He~I lines. The emission profiles for SN~2023zkd are highly asymmetric, with narrow cores and a broad redward tail not observed in the features of other SNe~IIn but potentially similar to the emission from the SN~Ibn/IIn 2021foa.}
    \label{fig:spec_comparison}
\end{figure*}

 We plot the most similar spectra for these strongly-interacting SNe in Figure~\ref{fig:spec_comparison}. We do not observe a red wing in the Balmer features of any of our comparison SNe~IIn (whose spectra span -58 to +203~days relative to peak), suggesting more extreme electron scattering in 2023zkd. Interestingly, the earliest spectrum of 2021foa (-5d from $r$-band maximum) shows similar double-peaked emission profiles of He~I $\lambda$7065 and He~I $\lambda$5876, and with some line profile asymmetries; \citet{2024Farias_2021foa} interpret the double-peaked He~I emission as a blend with the Na~I doublet at $\lambda$5876, and the secondary component at $\lambda$7065 to the presence of an unknown element. 

We repeat the process described in \textsection\ref{subsec:spec_fitting} to measure the integrated line flux of H$\alpha$ and He~I $\lambda5876$ for the comparison SNe in Figure~\ref{fig:spec_comparison}. We consider only optical spectra 50-365~days after $r$-band maximum (which we have either estimated photometrically or taken from the text of the cited publications, if available). The same combination of line fitting components was adopted as in 2023zkd, with the exception of SN~1996al: the prominent blue shoulder of its H$\alpha$ profile at later phases required an additional Gaussian emission profile to accurately reconstruct its total flux.

\begin{figure}
    \centering
    \includegraphics[width=\linewidth]{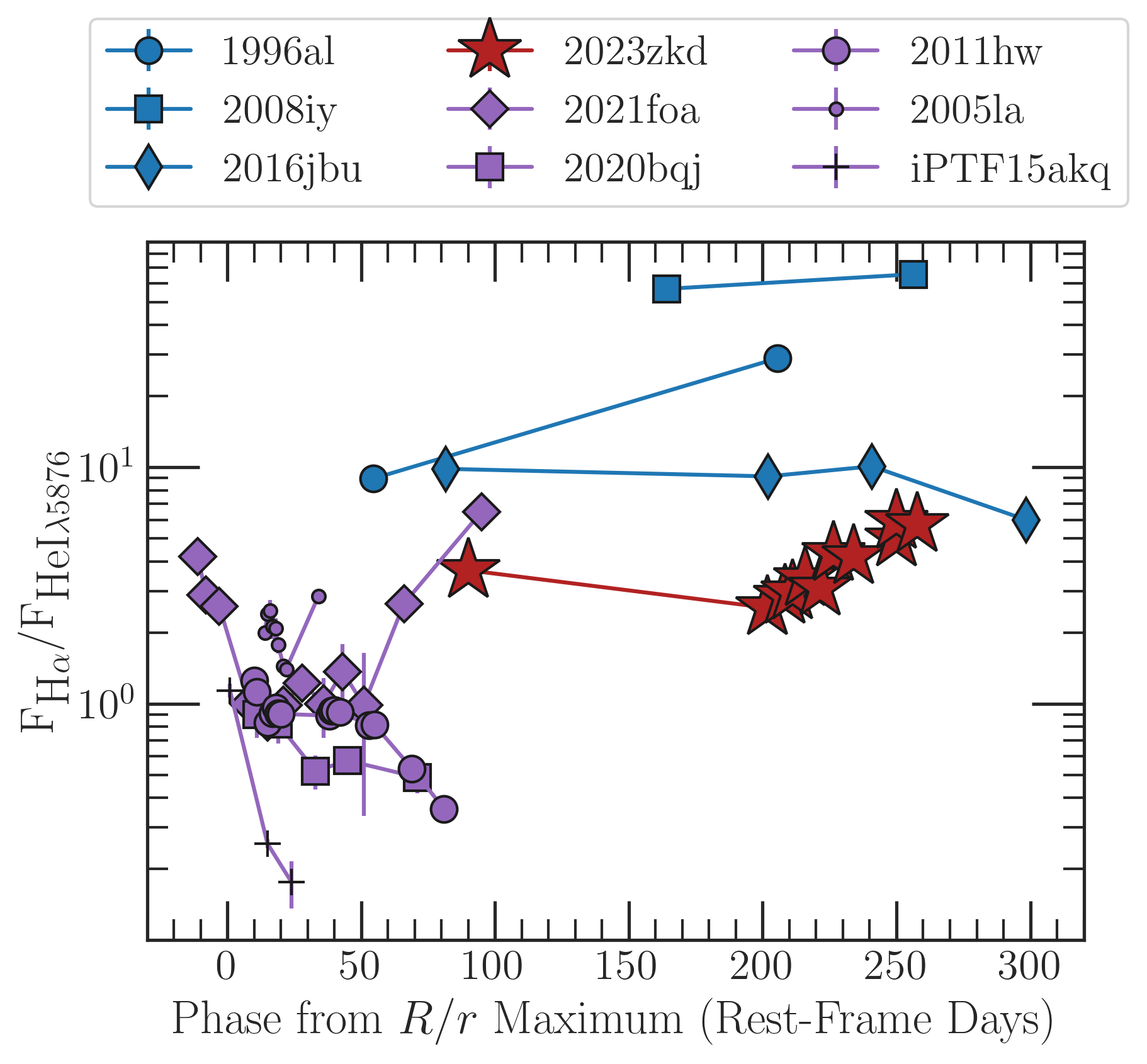}
    \caption{F$_{\textrm{H}\alpha}$/F$_{\textrm{HeI}\lambda5876}$ line ratios for SN~2023zkd (red), He-rich SNe~IIn between 50~days and 1~year from $R/r$-band maximum (blue), and for transitional SNe~IIn/Ibn (purple, from \citealt{2024Farias_2021foa}). SN~2023zkd lies at the boundary between spectroscopically-normal SNe~IIn and transitional SNe~Ibn/IIn events.}
    \label{fig:line_ratios}
\end{figure}

We show the resulting F$_{\mathrm{H}\alpha}$/F$_{\mathrm{HeI}\;\lambda5876}$ flux ratios for the sample in Figure~\ref{fig:line_ratios}, along with the values calculated by \citet{2024Farias_2021foa} for the transitional SNe~Ibn/IIn 2020bqj \citep{2021Kool_2020bqj}, 2011hw \citep{2012Smith_2011hw}, 2005la \citep{2008Pastorello_2005la}, and iPTF15akq \citep{2017Hosseinzadeh_SNeIbn}. The flux ratio for SN~2023zkd is systematically lower than is observed in all comparison He-rich SNe~IIn. SN~2016jbu has the lowest average F$_{\mathrm{H}\alpha}$/F$_{\mathrm{HeI}\;\lambda5876}$ ratio among SNe~IIn, from 10.2 at +241d to 6.0 at +299d relative to $r$-band maximum, and only reaches a similar value to 2023zkd at a single epoch. In addition, we observe a gradual but monotonic increase in the ratio over the second light curve peak: the value increases from 2.2 in the second spectrum at +202d, to 5.8-6.0 in the final two spectra at +250d and +258d. In comparison, the line ratios for the ``flip-flopping" SN~2021foa reach those of 2023zkd's later epochs during its ``SN~IIn" phases, from 4.2 at -11d to 6.5 at +95d. The transitional SN~IIn/Ibn sample has a mean line flux ratio of F$_{\mathrm{H}_{\alpha}}$/F$_{\mathrm{HeI}\lambda5876}\approx1$ \citep[see Figure~12 of ][]{2024Farias_2021foa}, suggesting that the spectral evolution of SN~2023zkd is still more akin to a He-rich SN~IIn than a transitional SN~Ibn (though we note that the dearth of late-time spectroscopy for these populations limits a more comprehensive analysis). 

Nonetheless, it is interesting that SN~2023zkd has the lowest F$_{\mathrm{H}\alpha}$/F$_{\mathrm{HeI}\lambda5876}$ ratio of the SNe~IIn analyzed. Blending of the H$\alpha$ profile with He~I at 6678~Å likely leads to an overestimation of the integrated flux we report, and so the intrinsic F$_{\mathrm{H}\alpha}$/F$_{\mathrm{HeI}\lambda5876}$ ratio for SN~2023zkd may fall even closer to the `transitional' SN~Ibn/IIn events. While the photometric evolution and host-galaxy properties of 2023zkd strongly suggest a type~IIn, these line ratios indicate an explosion filling the gap between SNe~Ibn and He-rich SNe~IIn. 

\subsection{Evolution of the Balmer Decrement}\label{subsec:balmer_decrement}
Next, we consider the evolution of the H$\alpha$ and H$\beta$ line strengths and the implications for the circumstellar emission and geometry. The Balmer decrement is the ratio between the intensities of these profiles. Balmer decrements have been well-constrained for photo-ionized nebulae that absorb Ly$\alpha$ photons but are optically thin to H$\alpha$; collisionally-excited material produce systematically lower Balmer decrements that also correlate with the properties of the emitting region \citep{1980Drake_LineIntensity}. This makes the measurement a valuable (and underutilized in SN science) probe of the ionization physics of both high-density and low-density material.

Because of the strong blending observed in H$\alpha$ (both with multiple-velocity components and with He~I at 6678~Å), we compute the Balmer decrement at each spectral epoch using the peak flux values of each profile (F$_{\mathrm{peak,H}\alpha}$ and F$_{\mathrm{peak,H}\beta}$) instead of their integrated flux. The peak flux values are expected to more reliably trace the low-velocity component of the CSM across the spectral sequence than the integrated flux, particularly when multiple velocity components are blended \citep{2014Levesque_BalmerDecrement}. 

\begin{figure}
    \centering
\includegraphics[width=\linewidth]{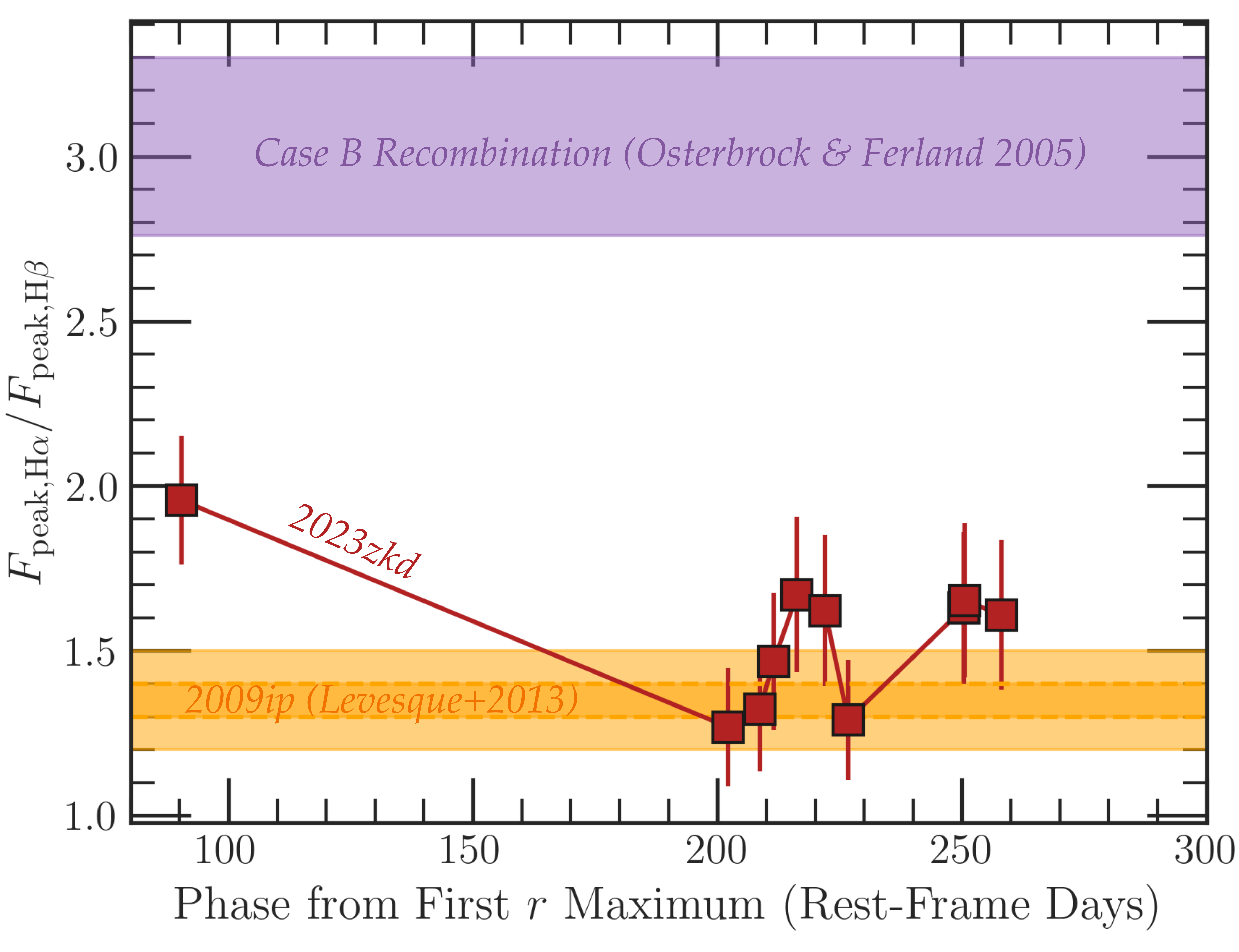}
\caption{F$_{\mathrm{peak,H}\alpha}$/F$_{\mathrm{peak,H}\beta}$ values for SN~2023zkd across the spectral sequence (red squares), compared to the range expected for Case B recombination (purple shaded region) and those reported for SN~2009ip during its 2012-B event (orange dashed lines, with shaded regions corresponding to 1$\sigma$ uncertainties; \citealt{2014Levesque_BalmerDecrement}).}\label{fig:balmerDecrement}
\end{figure}

We plot the resulting F$_{\mathrm{peak,H}\alpha}$/F$_{\mathrm{peak,H}\beta}$ ratio in Figure~\ref{fig:balmerDecrement}. We also compare our results to the Balmer decrement expected for Case B recombination \citep{2006Osterbrock_Ferland} across a broad range of gas temperatures (2,500~K - 20,000~K) in purple, and to those measured during the luminous `2012-B' event from the LBV SN~2009ip \citep[orange;][]{2014Levesque_BalmerDecrement}, which was also calculated using the profile peaks. 

We measure a low Balmer decrement spanning the full spectral sequence of SN~2023zkd, from a value of $2.0\pm0.2$ at +90d and spanning a consistently low 1.2-1.7 across the second $r$-band peak (with a minimum value of $1.2\pm0.2$ calculated from the NTT/EFOSC2 spectrum at +202d). 

There are two interesting properties to note in the evolution of our observed Balmer decrement. The first is its systematic decrease from the first spectrum to all subsequent ones tracing the second peak. The Balmer decrements measured at those later phases (+202d to +258d) are comparable to the values observed in SN~2009ip \citep{2014Levesque_BalmerDecrement}, where they were argued as evidence for emission in an extremely high-density plasma ($n_{e} > 10^{13}\;\mathrm{cm}^{-3}$) in comparison to the line intensity calculations from \citet{1980Drake_LineIntensity}. In 2023zkd, the systematic decrease in the Balmer decrement leading into the secondary peak is suggestive of a secondary high-density emitting region (or the same region with increased density, potentially from compression by the SN ejecta). 

We also observe marginal evolution in the Balmer decrement during the secondary peak. Rapidly-expanding CSM could provide a plausible explanation: the expanding CSM decreases in density until the photosphere recedes inward and probes a higher-density interaction region at +250d, although the scatter in our blackbody measurements around this phase prevents us from confirming this scenario.

\subsection{NIR Spectroscopy at Secondary Maximum}\label{sec:nir_spectrum}
\begin{figure*}
    \centering
    \includegraphics[width=\linewidth]{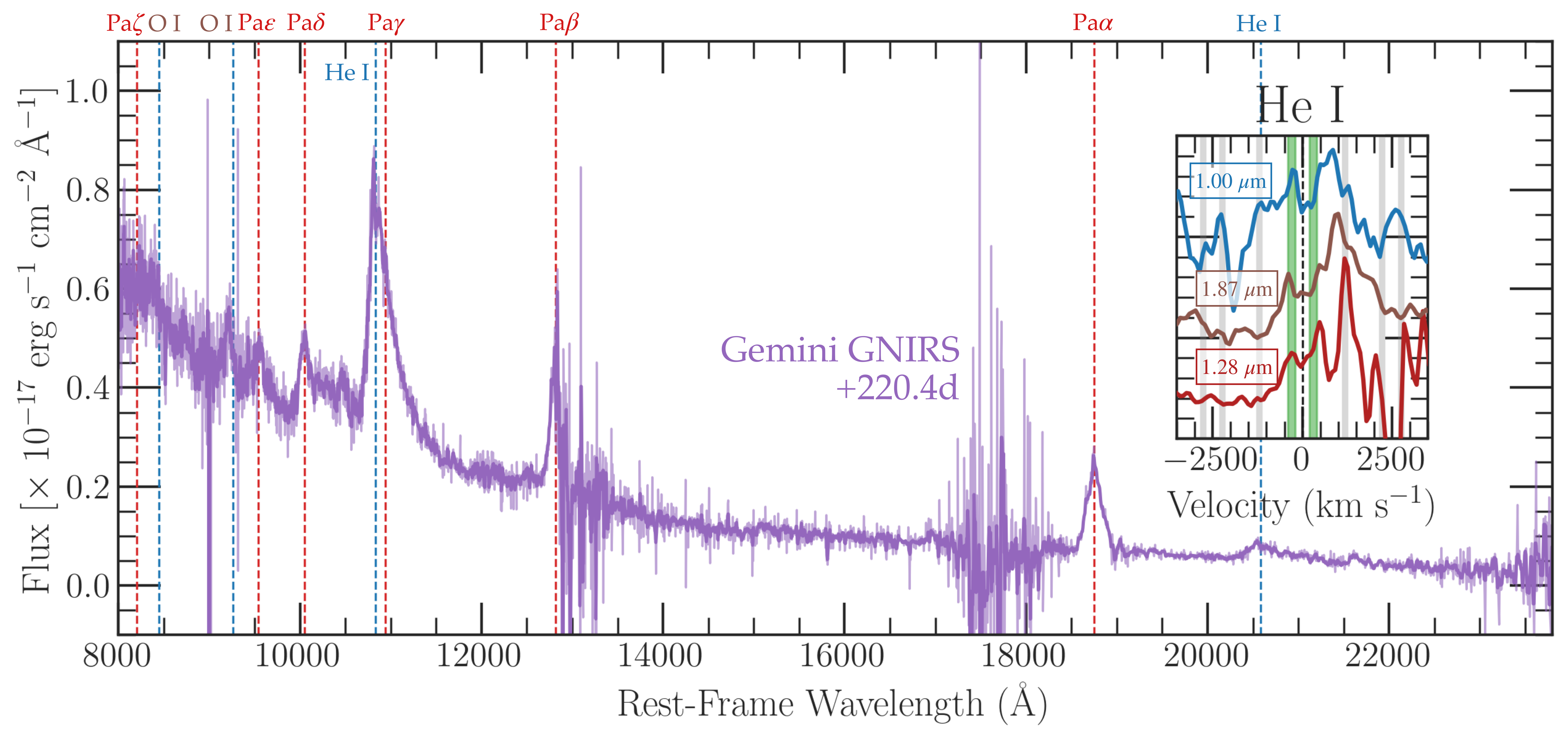}
    \caption{NIR spectrum of 2023zkd at secondary $r$-band maximum obtained with GNIRS, showing prominent H and He features superimposed on a blue continuum. Spectrum has been corrected for Galactic and host extinction. The He~I profiles at 12785~Å, 10031~Å, and 18685~Å are shown in velocity space in the inset top right, with the CSM velocity components identified in optical spectra annotated (green for $\pm400\;\mathrm{km}\;\mathrm{s}^{-1}$ and gray for the velocities of the faster CSM and the SN-CSM shell during the two peaks). All profiles show low-velocity CSM components consistent with those identified at optical wavelengths.}
    \label{fig:GNIRS_spec}
\end{figure*}

In Figure~\ref{fig:GNIRS_spec}, we plot the NIR spectrum of the SN obtained near secondary $r$-band maximum at +220d. The spectrum shows a blue continuum and prominent Paschen features, and the absorption trough to the Pa$\delta$ P-Cygni profile at 4,000-5,000~km~s$^{-1}$ is consistent with the velocity of the SN-CSM shell inferred from H$\beta$ at +222d (\textsection\ref{subsec:spec_fitting}). He~I emission features are also observed at 1.08~$\mu$m and 2.06~$\mu$m, and the profile of He~I at 1.08~$\mu$m exhibits the same asymmetric and double-peaked profile as observed in the optical at +222d. 

To further investigate a multi-component origin for our H/He-rich CSM, we search for He~I features that are NIR-bright and minimally contaminated by Paschen/Brackett or other strong transition lines (there are very few strong and uncontaminated H lines within this wavelength range). We select the lines at 10030.6~Å, 18685.3~Å, and 12785.0~Å from the NIST Database of Atomic Lines \citep{ralchenko2020development}. We estimate and subtract a flux continuum near each line, and normalize each profile within $\pm$2,500$\;\mathrm{km}\;\mathrm{s}^{-1}$ to a maximum flux of unity.

We plot these three spectral regions in velocity space in the inset of Figure~\ref{fig:GNIRS_spec}, along with the inferred velocities of the CSM and SN-CSM components from \textsection\ref{subsec:spec_fitting} (grey lines). We observe correlated excess flux symmetric about the rest wavelength at the same velocities suggested by the additional components in our optical spectra (shown in green) to within the spectral resolution of our data, confirming the presence of confined, slow-moving CSM. The emission at the positive component is stronger than at the negative component; this may be a viewing angle effect.

O~I is also detected in the NIR spectrum at 8446~Å and at 9263~Å, as in the SNe~IIn 1987F \citep{1989Flippenko_1987F,1996Wegner_1987F} and 2013L \citep{2020Taddia_2013L}. The blue continuum suggests that no substantial dust has formed, as has been observed with other SNe~IIn at these or earlier phases relative to peak (e.g., 2005ip, \citealt{2018Nielsen_Dust}; 2010jl, \citealt{2012Smith_2010jl,2014Gall_2010jl}); this is consistent with the high temperatures maintained during the secondary CSM interaction ($>$6$\times10^3$~K is suggested by the black-body fits in Figure~\ref{fig:blackbody_evolution}), as dust cannot exist or form at $\geq3,\!000\;$K \citep{2011Gall_Dust,2018Nielsen_Dust}. 

\section{Black-Body Fitting and Bolometric Luminosity}\label{sec:blackbody}
Next, we infer the bolometric properties of SN~2023zkd.

We define a set of 100-day bins spanning the pre-discovery photometry (from the earliest $>$3$\sigma$ detection) and 5-day bins after discovery. In each bin, we calculate an uncertainty-weighted average of the available photometry in each filter and fit a black-body spectral energy distribution to the resulting flux values. We require at least two photometric filters in a bin for a fit pre-discovery, and three filters after discovery. We then calculate a pseudo-bolometric luminosity in each bin by trapezoidal integration of the available filters, and a bolometric luminosity from the Stefan-Boltzmann law with the best-fit black-body parameters. 

\begin{figure}
    \centering
    \includegraphics[width=\linewidth]{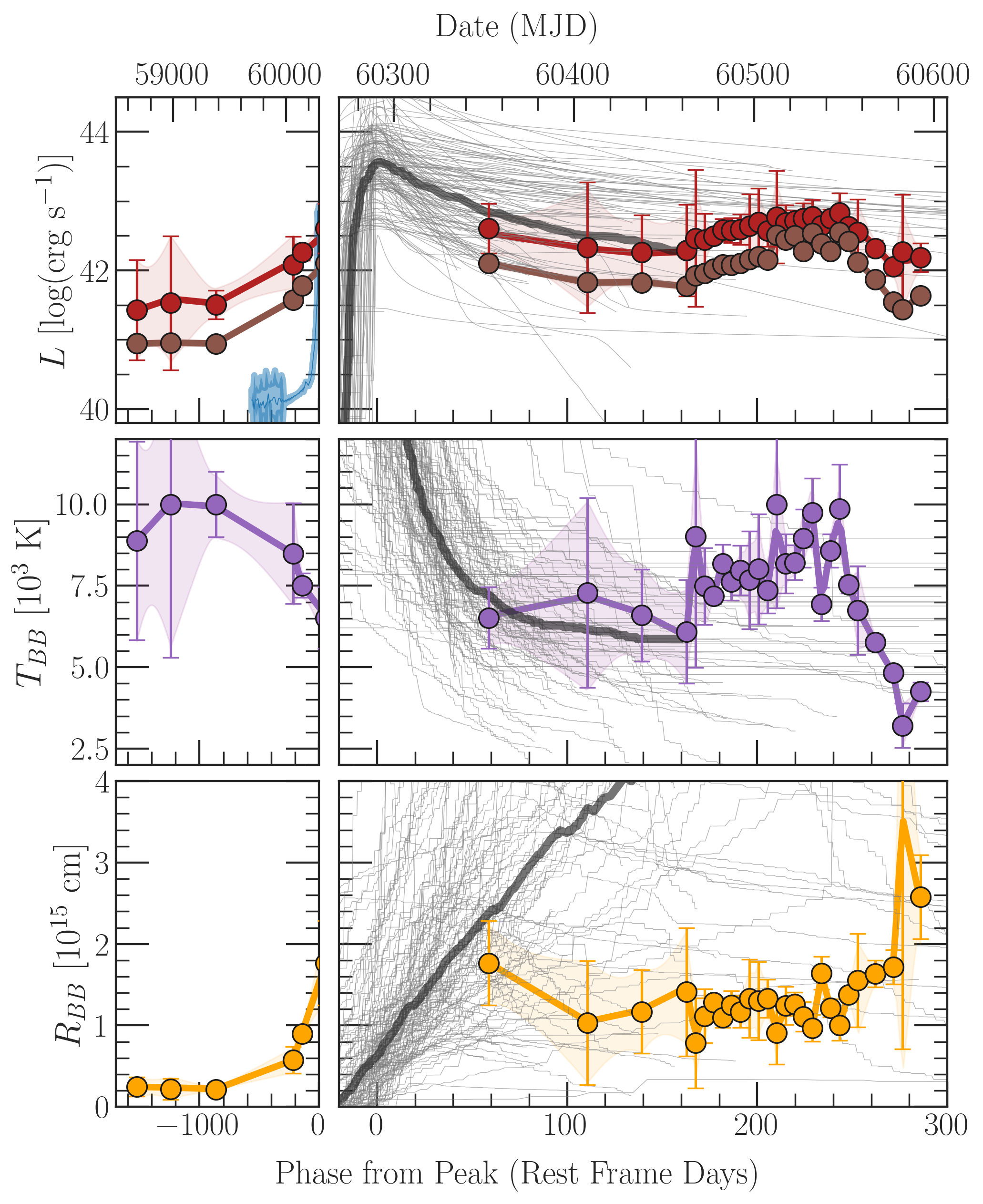}
    \caption{Black-body properties of SN~2023zkd in 100-day bins pre-discovery (left column) and 5-day bins after discovery (right column). Spline fits are shown for all properties (colored lines) with associated 1$\sigma$ uncertainties (shaded regions). The event's pseudo-bolometric luminosity (top panel, brown points) is estimated by trapezoidal integration of available photometry in each bin; the bolometric luminosity (red points) is calculated by integrating the best-fit black-body curves for each bin. The black-body temperature (middle panel, purple) and radius (bottom panel, gold) is also shown. The precursor emission associated with the SN~Ibn 2023fyq \citep{2024Dong_2023fyq} is shown in blue in the top left panel for comparison, as are the median (thick black line) and individual (thin gray lines) blackbody properties for archival SNe~IIn derived from the fits in \citet{2024Ransome_DiversityofSNeIIn} (grey lines, obtained via private comm.).}
    \label{fig:blackbody_evolution}
\end{figure}

We plot the resulting black-body temperature, radius, and luminosity for bins corresponding to converged fits in Figure~\ref{fig:blackbody_evolution}. To guide the eye, we also plot first-order spline fits of the blackbody properties along with their associated 1$\sigma$ uncertainties. We show the bolometric luminosity of the precursor associated with the SN~Ibn 2023fyq (top left panel, blue line; \citealt{2024Dong_2023fyq}) and the \texttt{MOSFiT}-inferred blackbody properties for the SN~IIn sample modeled by \citet{2024Ransome_DiversityofSNeIIn} (thin grey lines for individual events and thick black line from a spline fit to the population; values have been obtained by personal communication). 

Precursor~A has a pseudo-bolometric luminosity of $\sim$$3\times10^{40}$~erg~s$^{-1}$ and a bolometric luminosity of $\sim$$1.4\times10^{41}$~erg~s$^{-1}$, roughly an order of magnitude higher than the flat portion of the precursor detected in 2023fyq but comparable to the bolometric luminosities reported for the precursors to the type~IIn SNe~2016bdu \citep{2018Pastorello_2016bdu}, 2015bh \citep[which exhibited long-lived eruptions reminiscent of SN~2009ip;][]{2016EliasRosa_2015bh,2017Thone_2015bh,2018Kilpatrick_2015bh}, 2021qqp \citep{2024Hiramatsu_21qqp}, and SN~2010mc \citep{2013Ofek_2010mc}. Precursor A is consistent with a $\sim$$10^{4}\;\mathrm{K}$ blackbody of radius $0.2\pm0.1\times10^{15}\;\mathrm{cm}$, where the small radius $\sim10\;\mathrm{AU}$ is suggests emission extremely close to the progenitor system. 

Interestingly, the blackbody properties of Precursor B are distinct from those of Precursor A. In Precursor B, we observe a lower blackbody temperature of $\sim$$8,\!000\;K$ (also suggested by the $g-r$ color evolution in Figure~\ref{fig:all_phot}) and a marginally higher blackbody radius of $0.7\pm0.2\times10^{15}\;\mathrm{cm}$. The luminosity is also higher, approaching $\sim$$10^{42}\;\mathrm{erg}\;\mathrm{s}^{-1}$ and aligns with the steady brightening seen in the multi-band photometry.

After each of the photometric peaks the bolometric luminosity reaches a comparable observed maximum of $\sim$$4\times10^{42}$~erg~s$^{-1}$. The black-body temperature increases from $5\times10^{3}$~K to $7\times10^{3}$~K at the start of the seasonal gap, and then increases again to a maximum of $\sim$$8,\!000$~K at the second photometric peak. It then decreases through the decline of the second peak, reaching a minimum of 3,000-4,000~K. The second peak is also accompanied by a decrease in the blackbody radius to a minimum of $\sim$10$^{15}$~cm. 

The median bolometric properties of the SN~IIn sample from \citet{2024Ransome_DiversityofSNeIIn} align closely with the first-peak evolution of 2023zkd, further confirming its classification and suggesting an unobserved first peak bolometric luminosity near $\sim$$3\times10^{43}\;\mathrm{erg}\;\mathrm{s}^{-1}$. The blackbody temperature near the end of the first peak is comparable to the SN~IIn population, but the blackbody radius is systematically lower at all phases. This may suggest emission from a more compact interaction zone, or may be an artifact of the temperature floor (the minimum temperature the photosphere can cool to before it must recede) set in those models. 

Finally, we note the potential presence of additional structure in the blackbody evolution during the second light curve peak. The blackbody radius appears to decrease near +240d, then begins increasing again along with a decrease in blackbody temperature. This may be the sign of multiple emitting regions or some physical change in the interaction.

\section{Inferring the Properties of the Circumstellar Medium}\label{sec:circumstellar}

\subsection{Constraints on the CSM Mass with MOSFiT}\label{subsec:mosfit}
We now constrain the properties of the CSM using the Modular Open Source Fitter for Transients \citep[\texttt{MOSFiT};][]{2017Guillochon_MOSFiT}. \texttt{MOSFiT} takes user-provided multiband photometric data and fits a model to the data using a chosen sampling method (we use nested sampling via \texttt{dynesty}). We use the CSM interaction models of \citet{Chatzopoulos_2012}, which were originally based on the work of \citet{Chevalier_1982}, and then extended by \citet{Villar_2017} and \citet{2020Jiang_SS_CSM}. Most models in \texttt{MOSFiT}, including the CSM interaction model, take the form of a blackbody SED, and an expanding and cooling photosphere which recedes when a minimum temperature is reached. We select a set of physically informed priors on the parameters, motivated by previous works and presented in Table~\ref{tab:mosfit_params}. The parameters that are used in the fitting of the CSM model are:

\begin{enumerate}
    \item $s$, the CSM profile geometry parameter, where $\rho\,\propto\,r^{-s}$. Unlike most other work, $s$ is left as a free parameter (between 0 and 2), but we note that smaller values of $s$ indicate a more shell-like CSM geometry. Conversely, when $s$ approaches 2, the CSM can be considered wind-like with a steady mass-loss rate.
    \item $n$, describing the SN ejecta density profile set by the polytropic index of the stellar envelope. $n$ is varied between 7 and 12.
    \item $R_0$, the inner CSM radius. 
    \item $\rho_0$, the density of the CSM at $R_0$, i.e. the inner CSM density.
    \item M$_\mathrm{CSM}$, the total CSM mass.
    \item $v_{ej}$, the characteristic velocity of the ejecta (which scales with the ejecta mass and explosion energy as $v_{ej} = \sqrt{\frac{10\;E_{SN}}{3\;M_{ej}}}$)
    \item $T_{min}$, the temperature at which the photosphere begins to recede.
    \item Nuisance parameters such as $\sigma$ which assess how under-estimated the uncertainties may be.
\end{enumerate}

The ejecta mass is set to $M_{ej}=10\;M_{\odot}$. Using our set of physically informed priors, we fit the photometry of SN~2023zkd during only the secondary peak (we exclude first-peak photometry due to the lack of temporal coverage and \texttt{MOSFiT}'s inability to model CSM structure deviating from the expected $\rho\,\propto\,r^{-s}$ profile). We construct posterior distributions of the parameters listed above, and estimate the CSM properties powering SN~2023zkd's secondary peak from each posterior's median and 1$\sigma$ standard deviation. 

We present the median of the posteriors obtained from our modeling in Table~\ref{tab:mosfit_params}. We find a reasonably well-constrained CSM mass of 2-2.2$\;M_{\odot}$ whose interaction with the SN ejecta powers the secondary peak. $s$, which parameterizes the density profile of the CSM, is constrained to $s=0.68^{+0.24}_{-0.18}$. This suggests a ``shell-like" density profile, and is lower than the median value of $s=1.27$ inferred for the population SNe~IIn fit with the \texttt{MOSFiT} model in \cite{2024Ransome_DiversityofSNeIIn}. A corner plot of the joint posteriors across all realizations is shown in the Appendix. 

The median value for the posterior corresponding to $n$, the parameter characterizing the ejecta density profile, is $9.46^{+0.64}_{-0.42}$. $n$ can be used to distinguish between red supergiant-like progenitors \citep[whose indices are typically considered to be $n = 12$;][]{1999Matzner_CCSNe,2013Moriya_IIn} and luminous blue variable or Wolf-Rayet-like progenitors for SNe~IIn \citep[having $7<n<10$;][]{1969Colgate_SNe}; the range inferred for SN~2023zkd suggests a density structure more aligned with LBV-like progenitors, though the range is large. The inner CSM density required to power the second maximum, $\rho \sim$$10^{-12.4}$~g$\;$cm$^{-3}$ is well-constrained and consistent with archival SNe~IIn \citep{2024Ransome_DiversityofSNeIIn}. 

The inferred ejecta velocity associated with the secondary peak is 2,340-2,460~km~$\rm s^{-1}$, substantially lower than the median value of $4,\!810^{+3,454}_{-2,082}$~km~$\rm s^{-1}$ associated with the SN~IIn sample from \cite{2024Ransome_DiversityofSNeIIn}. These low velocities could be due to substantial deceleration of the SN ejecta from collision with the CSM powering the first light curve peak.

\setlength{\tabcolsep}{8pt} 
\begin{deluxetable*}{lccc}
\tablehead{
\colhead{Parameter} & \colhead{Units} & \colhead{Prior} & \colhead{Posterior Median and 1$\sigma$ Range}
 }
\startdata
log$_{10}$(M$_{\mathrm{CSM}})$ & log$_{10}$(M$_{\odot}$) & log$U(0.1, 100)$ & $0.32^{+0.02}_{-0.02}$\\
$n$ & -- & U(7, 12) & $9.46^{+0.64}_{-0.42}$\\ 
log$_{10}$($n_{\mathrm{H,host}}$) & log$_{10}($cm$^{-3}$) & log$U(10^{16}, 10^{23})$ & $17.49^{+0.95}_{-1.02}$\\ 
log$_{10}$($R_0$) & log$_{10}$(AU) & log$U(1, 100)$ & $0.93^{+0.35}_{-0.33}$\\ 
log$_{10}$($\rho_0$) & log$_{10}$(g$\;$cm$^{-3}$) & log$U(10^{-15}, 10^{-11}$) & $-12.42^{+0.60}_{-0.33}$\\ 
$s$ & -- & U(0,2) & $0.68^{+0.24}_{-0.18}$\\
log$_{10}$($T_{\mathrm{min}}$) & log$_{10}$(K) & log$U(1, 10^4$) & $1.39^{+2.20}_{-0.93}$\\ 
$t_{exp}$ & days & U(-20, 0) & $-19.17^{+1.03}_{-0.59}$\\ 
log$_{10}$($\sigma$) &  log$_{10}$(mag) & log$U(10^{-5}, 1)$ & $-0.53^{+0.01}_{-0.01}$\\ 
log$_{10}$($v_{\mathrm{ej}}$) & log$_{10}$($\mathrm{km}\;\mathrm{s}^{-1}$) & log$U(10^{3}, 10^{5}$) & $3.38^{+0.01}_{-0.01}$\\ 
\enddata
\caption{Median and 1$\sigma$ confidence intervals for the parameters of the circumstellar interaction model in \texttt{MOSFiT}. Parameter definitions are given in the text.}
    \label{tab:mosfit_params}
\end{deluxetable*}

We have chosen to keep the ejecta mass fixed at $M_{ej}=10\;M_{\odot}$ in our modeling. In separate \texttt{MOSFiT} runs where we fit this parameter, two solutions are degenerate and consistent with the photometry: one with lower ejecta mass ($10\;M_{\odot}<M_{ej}<15\;M_{\odot}$, comparable to our assumed value) and higher CSM mass ($1.5\;M_{\odot}<M_{\mathrm{CSM}}<2.5\;M_{\odot}$); and another with higher ejecta mass ($40\;M_{\odot}<M_{ej}<45\;M_{\odot}$) and lower CSM mass ($0.5\;M_{\odot}<M_{\mathrm{CSM}}<1\;M_{\odot}$). The solution with higher ejecta mass can be ruled out from the explosion energy most consistent with the SN-CSM shell velocities, as we discuss in \textsection\ref{subsec:shock_model}. We conclude that an ejecta mass of $10\;M_{\odot}<M_{ej}<15\;M_{\odot}$ is most likely.

Finally, we can use the properties of the model fit to infer an average mass-loss rate associated with the secondary peak. We calculate $R_{CSM}$, the characteristic CSM radius probed at these epochs under the assumption of a constant CSM velocity, as \begin{equation}
    R_{CSM} = (A \cdot M_{CSM} + B)^C
\end{equation}
with
\begin{equation}
A = \frac{3-s}{4\pi \rho R_0^s}
\end{equation}
\begin{equation}
    B = R_0^{3-s}
\end{equation}
\begin{equation}
C = \frac{1}{3-s}
\end{equation}

We find $R_{CSM}\approx0.4\times10^{15}$~cm, comparable to the results from our black-body fits (bottom panel of Figure~\ref{fig:blackbody_evolution}). Then, we estimate the average mass-loss rate as 
\begin{equation}
\textlangle\dot{M}\textrangle \approx \frac{M_{CSM}\cdot v_{CSM}}{R_{CSM}}
\end{equation} where we adopt a median CSM velocity during the second peak of $v_{CSM}\approx2,\!200$~km$\;$s$^{-1}$ from our spectroscopic fits in \textsection\ref{sec:spectroscopic_evolution}. Substituting values, we find $\dot{M}\approx20\;M_{\odot}\;\mathrm{yr}^{-1}$. This is likely an overestimate of the true mass-loss rate associated with the second peak due to MOSFiT's underestimation of the CSM extent $R_{CSM}$ \citep[see ][ for a discussion]{2024Ransome_DiversityofSNeIIn}, as well as the possibility of mechanical acceleration of the CSM probed by our spectral modeling. If we allow for the possibility that the CSM from the second peak is accelerated by collision and adopt the lower CSM velocity of $v_{CSM}\approx1,\!200$~km$\;$s$^{-1}$ from the +90d spectrum, we calculate a value of $\dot{M}\approx11\;M_{\odot}\;\mathrm{yr}^{-1}$. This value is more consistent with the mass-loss rate of $\sim$$7\;M_{\odot}\;\mathrm{yr}^{-1}$ we infer from the shock model in the following section.

\subsection{Shock Modeling}\label{subsec:shock_model}
We can also reconstruct a non-parametric CSM density profile and associated mass-loss history for 2023zkd using the same shock-luminosity formalism that was adopted for SN~2021qqp \citep{2024Hiramatsu_21qqp}. Since both the photometric and spectroscopic evolution of SN~2023zkd suggest an interaction-dominated transient, the model's assumption that the transient is exclusively powered by the interaction between the expanding fast-moving SN ejecta and surrounding CSM (with kinetic energy at the forward and the reverse shocks dissipating energy to optical photons) is well-justified. 

We adopt a fiducial value of $M_{ej} = 10\;M_{\odot}$ for the explosion ejecta mass (as was used for SN~2021qqp and suggested by our earliest \texttt{MOSFiT} modeling in \textsection\ref{subsec:mosfit}; we discuss the validity of this assumption below), along with $\delta=0$ and $n=12$ for the indices of the broken power-law density profile describing the SN ejecta (consistent with red supergiant progenitors; though our \texttt{MOSFiT} fits from \textsection\ref{subsec:mosfit} suggest an index closer to 10, the posterior is broad). We define a constant radiative efficiency for both the forward and the reverse shock of $\varepsilon=0.5$, as is assumed in our \texttt{MOSFiT} models.

Given the lack of spectroscopic constraints through the first light curve peak, we adopt an initial SN-CSM shell velocity of 10,000~km~s$^{-1}$. We set the initial CSM velocity to 1,176~km~s$^{-1}$, as calculated from our H$\alpha$ emission line fits; though we find evidence for lower-velocity CSM in \textsection\ref{subsec:vel_components}, the broad line profiles indicate that the interaction with the faster CSM dominates the observed emission. The total energy of the explosion, $E_{\mathrm{SN}}$, is constrained by comparing the model's predicted shell velocities to those we measure in \textsection\ref{subsec:spec_fitting}. 

\begin{figure*}
 \includegraphics[width=\linewidth]{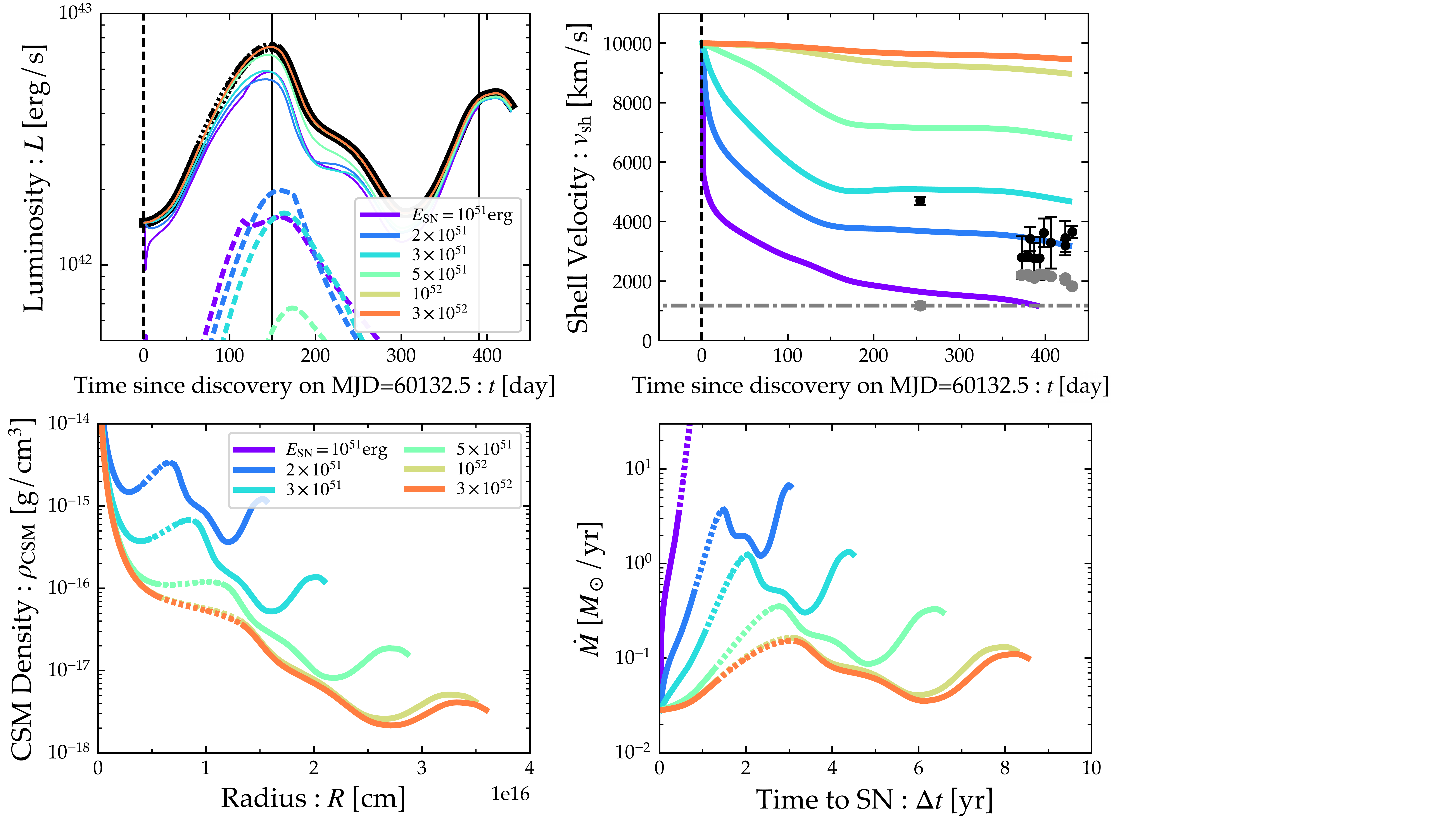}
\caption{\textit{Top Left:} Luminosity evolution of the forward (solid colored lines) and reverse (dashed colored lines) shocks for the best-fitting models to the interpolated 2023zkd photometry (black curve). The dashed section of the black curve denotes the seasonal gap in observations. \textit{Top Right:} Evolution of the associated SN-CSM interaction shell velocity for the models shown at left. Black points indicate the SN-CSM velocities measured from the absorption trough of the H$\beta$ profile, and grey points indicate the CSM velocities measured from the narrow component of the H$\alpha$ profile. The SN-CSM measurements are consistent with an explosion of energy $E_{\mathrm{SN}}=2\times10^{51}\;\mathrm{erg}\;\mathrm{s}^{-1}$ (dark blue line). \textit{Bottom Left:} Inferred density profiles for the models at top, which suggest highly structured CSM with two distinct overdensities. \textit{Bottom Right:} Inferred mass-loss rates for the SN~2023zkd progenitor. With an assumed efficiency of $\epsilon=0.5$ and ejecta mass of $10~M_{\odot}$, the total CSM mass of the system is estimated to be $5-6\;M_{\odot}$.}
\label{fig:shock_models}
\end{figure*}

We plot the results of our shock modeling in Figure~\ref{fig:shock_models}. The upper left panel compares the luminosity contributions from the forward and reverse shocks to the bolometric luminosity, while the upper right panel shows the temporal evolution of the SN-CSM shell velocities for different assumed $E_{\mathrm{SN}}$. The bottom panels show the associated CSM density profile (left panel) and mass-loss rates as a function of time to explosion (right panel), assumed to be the discovery date of $\mathrm{MJD}=60132.5$.  The measured SN-CSM shell velocities are consistent with an explosion of energy $E_{\mathrm{SN}}=2\times10^{51}$~erg (the observed scatter is a result of the blueward skew of the P-Cygni absorption feature in the later spectra, which leads to degeneracies in the emission and absorption Gaussian profiles used to fit the H$\beta$ line). 

For an assumed explosion energy of $E_{\mathrm{SN}}=2\times10^{51}$~erg, we infer a mass-loss history with two distinct peaks at $\sim$1-2 and $\sim$3 years prior to explosion (blue line), with the progenitor ejecting material at a maximum rate of $\sim$$4\;M_{\odot}\;$yr$^{-1}$ and $\sim$$7\;M_{\odot}\;$yr$^{-1}$ at the first and second episodes, respectively.  Integrating the full mass-loss history leads to a total CSM mass of $5-6\;M_{\odot}$, with $M_{CSM}\approx2\;M_{\odot}$ inferred from the second peak alone. This second-peak estimate is compatible with the mass suggested by \texttt{MOSFiT}, despite different assumptions about the ejecta density profile $n$.

We caution that this approach has assumed that the SN explosion began at the discovery date, whereas the photometry prior to the first seasonal gap could be entirely precursor emission due to the slow rise times (see \textsection\ref{sec:photometric_evolution}). If this is the case, it will not be well-characterized under the assumption of SN-CSM interaction. For this reason, we consider the mass-loss episode $\sim$3$\;$yr before explosion to be more tightly constrained. We note, however, that if Precursor A is taken to be comprised of two distinct mass-loss episodes, the first set of detections spans 2.7-4.0 years prior to discovery while the second set spans 0.7-2.2 years prior to discovery. These are well-matched to the mass-loss phases suggested by our shock modeling assuming an explosion time of the discovery epoch, although at the level of our Precursor A detections we cannot distinguish between persistent emission and two eruptive episodes with statistical significance.

Finally, we comment on the validity of the assumed $M_{ej}=10\;M_{\odot}$. The adopted shock model gives a rough scaling relation between the event's explosion energy $E_{SN}$ and the ejecta mass $M_{ej}$. Re-arranging Equation~14 from \citet{2024Hiramatsu_21qqp}, we find that for explosion energies of $10^{51}\;\mathrm{erg}\;\mathrm{s}^{-1}<E_{SN}<5\times 10^{51}\;\mathrm{erg}\;\mathrm{s}^{-1}$ (higher values are incompatible with our shell velocity evolution, shown in the top right panel of Figure~\ref{fig:shock_models}), 2023zkd requires $2\;M_{\odot}<M_{ej}<15\;M_{\odot}$. This range is consistent with the lower-mass solution found by \texttt{MOSFiT} ($10\;M_{\odot}<M_{ej}\;<15\;\;M_{\odot}$) and comparable to the ejecta mass inferred for 2021qqp \citep[they find a lower total CSM mass of $2-4\;M_{\odot}$ due to their higher shell velocities/inferred $E_{SN}$; ][]{2024Hiramatsu_21qqp}.

\subsection{Pre-Explosion Mass-Loss Limits from Swift/XRT Non-Detections}\label{sec:xray_massloss}
We follow the prescription outlined in \citet{2006Immler_2023zkd} to convert the unabsorbed X-ray flux upper limit derived in \textsection\ref{subsec:xray} during the secondary maximum to a maximum mass-loss rate for the system. We assume that the SN luminosity during the second peak is fully shock-powered, with the reverse shock $L_{r}$ contributing $30$x as much as the forward shock $L_{f}$ to the total observed luminosity (due to the higher CSM density at the reverse shock; relative ratios of 10-100 are suggested in \citealt{1982Chevalier_CSMXrays}, which change our upper limit by at most a factor of two). We adopt a CSM power-law index of $s=0.6$ from our \texttt{MOSFiT} results in \textsection\ref{subsec:mosfit}, and an ejecta profile index of $n=12$. We assume our CSM has a mean mass per particle of $\approx2.1\times10^{-24}$g as appropriate for H+He plasma \citep[e.g., ][]{2010Miller_2008iy}; and adopt a value for the cooling function of $\Lambda=3\times10^{-23}\;$erg$\;$s$^{-1}$ cm$^3$, appropriate for a plasma with $T=10^9$~K. 

Substituting values, our X-ray non-detection corresponds to a 3$\sigma$ upper limit of \begin{equation}\frac{\dot{M}}{M{\odot}\;\mathrm{yr}^{-1}}\leq2.2\times10^{-7}\left(\frac{v_{sh}}{\mathrm{km}\;\mathrm{s}^{-1}}\right)^{0.61}\left(\frac{t}{\;\mathrm{day}}\right)^{0.61}\left(\frac{v_w}{\mathrm{km}\;\mathrm{s}^{-1}}\right)\end{equation} Assuming a shock velocity at secondary maximum of $\sim$3,300~km$\;$s$^{-1}$ (from the median of the velocities of the H$\beta$ trough in \textsection\ref{subsec:spec_comparison}), a CSM velocity of $\sim2,\!200\;$km$\;$s$^{-1}$ from the median FWHM of the narrow component of H$\alpha$, and a phase of $t\approx74$~days from the start of the secondary peak in the rest-frame, we infer a mass-loss rate of $\dot{M}<1.0\;M_{\odot}\;$yr$^{-1}$. Adopting the initial shock velocity of $10,\!000\;\mathrm{km}\;\mathrm{s}^{-1}$from \textsection\ref{subsec:shock_model} increases the limit by a factor of two, while assuming $n=10$ from our \texttt{MOSFiT} results leads to a tighter limit of $\dot{M}<0.1\;M_{\odot}\;$yr$^{-1}$. All of these estimates are lower than the mass-loss rate derived from the shock modeling in the following section for an SN explosion energy of $2\times10^{51}\;$erg$\;\mathrm{s}^{-1}$, but only take into account absorption by the Galactic ISM. Photoelectric absorption by a thick neutral medium, as well as an asymmetric explosion or CSM structure \citep[as has been invoked for SN~2023ixf as an explanation for \textit{Swift} X-ray non-detections; ][]{2024Panjkov_Xray23ixf}, could suppress X-ray emission from photons produced during the SN-CSM interaction.

\section{Discussion}\label{sec:discussion}
\subsection{Emission Phases and CSM Geometry}\label{subsec:disc_geometry}
The observational sequence obtained for SN~2023zkd presents multiple peculiarities. The distinct phases of precursor emission (persistent for years during Precursor A followed by a brightening and reddening to discovery during Precursor B) are unlike those observed in other SNe~IIn \citep[e.g.,][]{2021Strotjohann_MonthsLong} and qualitatively similar to that observed in 2023fyq \citep{2024Dong_2023fyq,2024Brennan_2023fyq}, where a brightening precursor was associated with runaway accretion onto a binary companion \citep{2024Tsuna_MergerPrecursor}. Nevertheless, important differences also exist with the 2023fyq precursor: Precursor A is an order of magnitude more luminous than the plateau phase of 2023fyq (with an integrated pseudo-bolometric luminosity of $\sim$$10^{41}\;\mathrm{erg}\;\mathrm{s}^{-1}$ for 2023zkd), while the brightening rate during Precursor B is substantially lower. Further, the progenitor system proposed for 2023fyq was a low-mass ($\sim$3$\;M_{\odot}$) helium star with a neutron star companion, whereas the ejecta and CSM masses inferred for 2023zkd ($M_{ej}\approx10\;M_{\odot}$ and $M_{CSM}\approx5\;M_{\odot}$) favor a substantially more massive progenitor. 

Across all phases, 2023zkd also exhibits H/He line flux ratios lower than He-rich SNe~IIn but higher than transitional SNe~Ibn/IIn. Further, the multi-peaked H and He~I profiles suggest axisymmetric CSM components associated with both lower velocities ($200-400\;\mathrm{km}\;\mathrm{s}^{-1}$ and higher velocities ($1,\!200-2,\!200\;\mathrm{km}\;\mathrm{s}^{-1}$). The higher-velocity components are most pronounced in the He~I profiles at +90d, while the lower-velocity components dominate the peaks of the H profiles at the same epoch. Throughout the second light curve peak, multiple components are simultaneously observed (though many are unresolved) across both H and He profiles. 

These observations strongly suggest interaction with two axisymmetric CSM components, one H-rich and the other He-rich. The appearance of multiple components in both H and He from the +90d to the +258d spectrum suggests these components have distinct axes of symmetry, such that both are exposed to the initial SN shock and subsequent ejecta.
 
 A configuration involving an outer, slow-moving, and toroidal/ring-like outflow and an inner, fast-moving bipolar outflow provides a natural explanation. This geometry is regularly observed in the circumstellar environments of Galactic stars, and equatorially-distributed CSM has been invoked to explain the observations of numerous other interacting SNe: for the re-brightening of 2009ip \citep{2014Levesque_BalmerDecrement}, a detached, disk-like CSM was favored over a spherically-symmetric CSM shell due to the low Balmer decrement and absence of P-Cygni profiles\footnote{The latter are expected in the strong velocity gradient established by an expanding spherical shell, whereas narrow features are more suggestive of a narrow line-emitting region.}. A toroidal CSM geometry was similarly proposed for the type~Ib SN~2014C (that later transitioned into an SN~IIn) due to the persistence of a $2,\!000\;\mathrm{km}\;\mathrm{s}^{-1}$ feature observed in H$\alpha$ and no other lines \citep[though, in the proposed collision between the SN ejecta and the toroidal CSM, the source of the H-rich material moving at this velocity was ambiguous; ][]{2022Thomas_2014C}. High polarization in the emission from the SN~IIn 1998S, coupled with double-peaked emission profiles, were similarly explained as CSM having a dense disk- or ring-like morphology \citep{2000Leonard_1998S,2012Mauerhan_1998S}; while the early double-peaked He~I profiles for SN~IIn 1996al were argued to be the result of an explosion into stratified equatorial CSM \citep[see Figure~22 of][]{2016Benetti_1996al}. 

The mostly distinct peaks of H (low-velocity) and He~I (high-velocity) at +90d (during the decline of the first peak) can be explained by CSM components that are also partially compositionally distinct, with the bulk of the H confined to the lower-velocity CSM and the He associated with the higher-velocity CSM (though fainter high-velocity shoulders are marginally detected in e.g., the H$\beta$ component at this epoch, indicating that the components may be mixed during the interaction). The He~I profiles are significantly broader than the H profiles at this phase, suggesting that the He~I profiles may be Doppler broadened at this phase.

By +209d at the rise of the second peak, both H and He are emitted from the high-velocity CSM: much narrower double-peaked He~I profiles are observed and the H$\alpha$ peak has broadened. This may be suggestive of a secondary collision with CSM more enriched in H than the first collision. This interaction dominates the second photometric peak.

A progenitor scenario for the explosion must reconcile these observational properties. We consider the viability of multiple progenitor systems and their implied mass-loss histories below. 

\subsection{Progenitor Scenarios for SN~2023zkd}\label{subsec:progenitors}
\subsubsection{Core Collapse of A Single Massive Star}\label{subsubsec:single_star}
We first consider a single massive star as the progenitor to 2023zkd. LBVs and Wolf-Rayets (WRs) have each been considered as the progenitors of strongly-interacting SNe, with multiple lines of evidence in favor of each \citep{2009Nature_2005gl,2011Dwarkadas_LBVProgenitors,2015Pastorello_MassiveStars}. 

The high-velocity CSM component observed in 2023zkd is faster than what is typically inferred for LBV winds, but is well within the range of wind velocities proposed for WRs \citep{2017Smith_InteractingSNeIIn}. This, coupled with the detection of He~II and tentatively N~III during the secondary peak, ostensibly suggest a Nitrogen-rich WR (WN) wind. A young WN has been proposed as the progenitor of multiple strongly-interacting SNe, including the SN~Ibn 2006jc \citep{2007Foley_2006jc} and the transitional type~IIn/Ibn SN~2005la \citep{2008Pastorello_2005la}.

The pre-explosion photometry, on the other hand, supports the LBV scenario: the luminosity and decreasing temperature of Precursor A ($\sim$$8,\!000\;\mathrm{K}$ to $10,\!000$~K) aligns well with the luminosities and temperature evolution reported for LBV outbursts \citep{1994Humphreys_LBVs,2018Jiang_hotLBV} and observed in e.g., 2009ip \citep{2011Smith_BinaryInteraction}. Finally, the timescale of Precursor A is suggestive of an LBV undergoing S-Dor eruptions, whereas both the increasing bolometric luminosity of Precursor B and its magnitude ($\sim$$3\times10^{41}\;\mathrm{erg}\;\mathrm{s}^{-1}$) are consistent with observed giant outbursts (\citealt{1994Humphreys_LBVs}; see also the evolution of the 2012a outburst of 2009ip in, e.g., Figure~1 of \citealt{2013Soker_2009ip}). 

Both progenitor systems may naturally give rise to axisymmetric CSM structures with distinct compositions. \citep{2001Dwarkadas_LBVNebulae,2012Akras_3D}. In the Generalized Interacting Stellar Wind (GISW) model, a fast-moving and isotropic wind emitted from an LBV or WR collides with a slow-moving and dense toroidal structure \citep{1995Frank_GISW}. The interaction between components produces bipolar lobes of fast-moving material, as observed in WR planetary nebulae \citep{Danehkar_2022} and nearly all Galactic LBVs \citep[e.g., $\eta$ Car, HG Car, and A Car;][]{2009Groh_LBVRotation}. Hydrodynamic simulations of massive LBV outbursts \citep{1997GarciaSegura_LBVBipolar,1997Cassinelli_LBVs} suggest that near-critical rotation can also give rise to \textit{both} bipolar outflows and dense equatorial material during a single outburst \citep[see also ][]{2001Dwarkadas_LBVNebulae}. Further, the majority of Galactic eruptive LBVs are also near-critical rotators: \citet{2009Groh_LBVRotation} suggests that this rapid rotation may also prevent the loss of angular momentum required to transition to a WN, potentially leading to an explosion as an LBV. 

Despite the compositional similarities, a single WR alone seems extremely unlikely to have produced 2023zkd. The mass-loss rates inferred for WNs ($10^{-6}-10^{-4}\;M_{\odot}\;\textrm{yr}^{-1}$) are many orders of magnitude lower than suggested by 2023zkd's two photometric peaks \citep{2017Smith_InteractingSNeIIn} (though this can be partially reconciled by proposing short-lived WR phases, which can rapidly sweep up CSM and produce dense shells despite low intrinsic mass-loss rates; see \citealt{2011Dwarkadas_LBVProgenitors}, for a discussion). An obvious additional problem is that the WR stage of stellar evolution, canonically, denotes core-He burning with the absence of H, while we observe prominent and multi-component H features at all phases of emission. Intermediate velocities, as we have observed for 2023zkd, are also not expected beyond the short-lived early WN stage \citep{2005vanMarle_Constraints}.

There are also multiple challenges with a single LBV origin for 2023zkd. Precursor A ($M_r\approx-15$~mag)
is three magnitudes brighter than the brightest LBV S-Dor outburst observed to date \citep{1994Humphreys_LBVs}, and the steady brightening of Precursor B is more suggestive of a continuous than an eruptive emission mechanism\footnote{As a note of caution, the limited number of statistically-significant detections pre-discovery make it challenging to distinguish between a monotonic increase in brightness and a long-lived eruption blending with the early SN rise near discovery; the statistical analysis of the rise in Appendix A (Figure~15, left) suggests that an eruption may also be also consistent with the photometry, and the giant outburst in the LBV HD~5980 was of comparable duration \citep{2017Smith_InteractingSNeIIn}.}. Finally, the LBV stage for stars of initial mass $>$30$\;M_{\odot}$ is believed to be short-lived \citep[$10^4$~yr;][]{2005vanMarle_Constraints}. The explosion of an LBV in the midst of transitioning to an H-free WN may require excessive fine-tuning.

In conclusion, while 2023zkd exhibits signs of both massive LBV-like and WN progenitors, neither is a strong match. There has been steadily mounting evidence for alternative evolutionary channels for a massive star that include an LBV stage; for massive H-rich systems, in particular, the WN stage could \textit{precede} the LBV stage instead of following it. The CSM velocities and compositions that suggest WN-like winds, coupled with the LBV-like Precusor B, suggest that this rare evolutionary pathway is worth considering.

\subsubsection{Transition of an H-rich WN to an LBV}\label{subsec:2023zkd}
The observational taxonomy of WNs has recently expanded to include H-\textit{rich} WNs \citep[designated ``WNhs";][]{2008Smith_WRFeedback,2023Martins_WNh,2024Gormaz_WNh}, stars whose compositions are not predicted by traditional evolutionary theory (e.g., \citealt{1994Humphreys_LBVs}). These systems are produced at systematically higher masses than H-free WNs, and might provide an explanation for the H-rich ejecta in 2023zkd. As 2023zkd occurred in a low-metallicity host ($Z\approx0.1\;Z_{\odot}$), it may be also possible to minimize mass-loss rates during the WNh stage such that an LBV-like successor achieves near-critical rotation. This could reproduce the assumed multi-component circumstellar medium (CSM) geometry and the narrow, double-peaked profiles of H and He.

While the mass-distribution of H-\textit{poor} WNs peaks closer to the inferred $M_{ej}+M_{CSM}\approx15\;M_{\odot}$ value for 2023zkd (10-20$\;M_{\odot}$, compared to $>$$30\;M_{\odot}$ for WNhs; \citealt{2008Smith_WRFeedback}), this could be taken as a lower limit on the true progenitor mass given the likelihood of sustained enhanced mass-loss (assuming a characteristic mass-loss rate of $10^{-5}\;M_{\odot}\;\mathrm{yr}^{-1}$ from Figure~3 of \citet{2008Smith_WRFeedback} over a typical lifetime of $\sim$3~Myr introduces 30$\;M_{\odot}$ of additional mass lost during the MNH stage, placing the progenitor mass near the peak of the distribution of spectroscopic masses measured for Galactic WNhs; see Figure~1 of \citealt{2008Smith_WRFeedback}). Due to the observed scaling between the WNh luminosity and H mass fraction in Figure~2 of \citet{2008Smith_WRFeedback}, a WNh progenitor luminous enough to power Precusor A would also give rise to the H-rich CSM that dominates the spectral sequence for 2023zkd. 

Nonetheless, the WNh stage of a single massive progenitor can be ruled out for this system based on theoretical grounds. The observed luminosity of Precursor A, if attributed to an WNh wind, surpasses even the Eddington luminosity of a 100$\;M_{\odot}$ star. Though luminosity phases approaching the Eddington limit are expected for massive stars given their myriad instabilities \citep{1967Rakavy_Instabilities,1997Langer_LBVs,2009Saio_StrangeModes,2020Davidson_LBVs}, a star with mass close to 2023zkd's ejecta mass of 10$\;M_{\odot}$ surpassing its Eddington limit by multiple orders of magnitude for years seems exceedingly unlikely.

Sustained super-Eddington luminosities can be naturally achieved with additional energy input from binary interaction \citep[e.g.,][]{2002Begelman_ApJL}. Binary interaction affects the majority of massive stars \citep{2012Sana_BinaryInteraction} and is expected to play a sizable role in the evolution of LBVs \citep{2011Smith_BinaryInteraction}. A binary progenitor also provides an intuitive explanation for equatorially-distributed CSM \citep[e.g., ][]{2020ASchroder_MergerSN}; $\eta$~Carinae, the well-studied Galactic system whose LBV-like outbursts produced the bi-polar lobes and equatorial ``skirt" of the Homunculus nebula \citep{1997Davidson_EtaCar,2001Hillier,2012Groh_EtaCarBinary,2022Zapata_Homunculus}, is known to contain \textit{at least} two massive stars \citep{2011Bednarek_EtaCarBinary,2018Panagiotou_EtaCar}. we investigate whether a binary progenitor can reproduce the observational properties of 2023zkd below.

\subsection{Binary-Driven Merger-Supernova}\label{subsubsec:binary_merger}
Motivated by the above considerations, we now consider a binary progenitor system for SN~2023zkd. 

A massive stellar companion has been suggested for systems with similar CSM geometries to $\eta$~Car., but existing theoretical studies suggest that the energy budget provided by colliding winds from these stars is insufficient to explain Precursor A \citep[e.g.,][]{2023Abaroa_CollidingWinds}. Super-Eddington accretion onto a compact companion provides a more favorable energy budget, but luminosities of $\sim$$10^{41}\;\mathrm{erg}\;\mathrm{s}^{-1}$ favor systems involving a black hole accretor \citep[assuming spherical symmetry, the Eddington limit for accretion onto a companion of mass $M_*$ is $\sim$$10^{38}\;(M_*/M_{\odot})\;\mathrm{erg}\;\mathrm{s}^{-1}$][]{2019Brightman_SuperEddington}. As a result, we discuss the feasibility of a binary progenitor between a massive primary and a black hole companion below.

After leaving the main sequence, an inflated primary star orbiting a black hole companion could overflow its Roche-Lobe and undergo stable mass-transfer of its envelope \citep{2017Pavlovskii_MassTransfer,2021Gallegos_BBHFormation}. The companion could detach before the star is fully stripped, leaving the primary with a residual H envelope. The mass of H-rich material remaining after this stable mass-transfer phase is highly dependent on metallicity, with a lower-metallicity primary producing a less-inflated envelope and more massive H-rich material remaining after stable mass transfer has ceased \citep{2020Klencki_WR}. Extrapolation of the results in \citet{2020Laplace_SESNe} suggests that a progenitor of initial mass $M_{\mathrm{ZAMS}}>30\;M_{\odot}$ may be able to end this initial stage of mass transfer with a $\sim$$10\;M_{\odot}$ He core and a residual envelope of $\sim$$1\;M_{\odot}$ H/He\footnote{This estimate of the companion mass is also supported by the $\sim$linear scaling between the CSM and the accretor's mass suggested by the binary mass-transfer simulations of \citet{2020MacLeod_MassLoss}.}. 

Following core-He burning, the re-expansion of the progenitor can restart mass transfer onto the companion \citep{2020Laplace_SESNe}. Mass transfer is expected to be unstable at this secondary stage due to the progenitor's convective envelope \citep{2021Marchant}, causing the majority of stripped material to become unbound and distributed equatorially as a circumbinary outflow \citep[CBO, as in e.g., ][]{2006Morris_MassLoss}. The loss of angular momentum from this material can cause a tightening of the binary in a runaway interaction that drives a monotonic increase in luminosity from disk winds and ends in a merger \citep{2018MacLeod_RunawayCoalescence,2020MacLeod_RunawayCoalescence}. 

The luminosity and the years-long timescale of Precursor~A can be produced by super-Eddington accretion during this unstable mass-transfer stage \citep{2011Weng_AccretionDiskEvolution}. The brightening seen during Precursor B is a strong indication of a shrinking binary separation during unstable accretion and is predicted by multiple models \citep{2013Soker_Mergerburst, 2024Tsuna_MergerPrecursor}. 
In the merger-SN formalism modeled by \citet{2024Tsuna_MergerPrecursor}, a low-velocity CSM component arises from the CBO of stripped H/He (moving at a few hundred $\mathrm{km}\;\mathrm{s}^{-1}$) and a faster-velocity component originates from polar outflows of the He-rich circumbinary disk (moving at $1,\!000-10,\!000\;\mathrm{km}\;\mathrm{s}^{-1}$); both predictions are consistent with the components observed in 2023zkd. Binary interaction can also produce CSM components with distinct geometries, compositions, and velocities according to the interaction phases at which they were ejected, with CSM ejected during runaway interaction probing He-rich material in the primary and collimated along the poles from interaction with pre-existing toroidal CSM from an earlier mass-phase \citep{2018MacLeod_BoundOutflows}. Further, models for a merger-driven SN suggest an initial emission signature from radiative cooling of a confined inner envelope heated by the SN shock; substituting the properties of 2023zkd into equations 40 and 41 of \citet{2024Tsuna_MergerPrecursor} suggests that this early emission can span 20-30~days, comparable to the timescale of a small light curve feature observed immediately after discovery.

It is difficult to distinguish between a rapidly-rotating massive star and one undergoing runaway mass-loss onto a binary companion by geometric arguments alone; both can produce slow and H-rich toroidal CSM ($\sim$400$\;\mathrm{km}\;\mathrm{s}^{-1}$) and a faster and He-rich bipolar outflow ($\sim$1,200$\;\mathrm{km}\;\mathrm{s}^{-1}$). In the case of unstable mass-transfer, however, the disk wind and CBO outflow expand along orthogonal axes of symmetry and are not expected to interact prior to explosion. This leads to an observational puzzle: if interaction with the toroidal material powers the second peak, how is emission delayed for over 250~days? 

Interaction between the SN ejecta and the CBO can occur soon after core-collapse, and prior to the interaction with the fast-moving disk wind; however, the rapidly expanding SN ejecta may quickly overrun the interaction region and instead trace the collision with the fast-moving CSM. If this is the case, photons from the CBO interaction may be scattered through the secondary interaction region as they escape the photosphere and contribute to the total luminosity of the event. A reverse shock produced from the ejecta's collision with polar material should also propagate inward and drive additional collisions with the toroidal material, further powering a long-lived event. The complex inner interaction may only be revealed once the photosphere fully recedes. This scenario has already been proposed to explain the late-time interaction observed in some SNe~IIn \citep[e.g.,][]{2011Smith_BinaryInteraction}. 

The reconstructed mass-loss history and high CSM masses ($\sim$$2\;M_{\odot}$ associated with each peak) suggest eruptive events rather than the steady disk winds proposed in the low-mass binary merger model developed by \citet{2024Tsuna_MergerPrecursor}. It is plausible that two long-lived pre-explosion outbursts, potentially arising from instabilities associated with the mass transfer, ejected fast-moving ($1,\!200\;\mathrm{km}\;\mathrm{s}^{-1}-2,\!200\;\mathrm{km}\;\mathrm{s}^{-1}$) polar material. The SN ejecta's interaction with each shell-like structure along the polar axis then powers the two distinct light curve peaks, and the SN-CSM shell velocity is lowered during the second peak by the ejecta's first collision. An inner H-rich disk or CBO may contribute additional luminosity throughout the two peaks.

If any CSM from either the CBO or the first polar shell is accelerated by the SN ejecta, it might also collide with the polar material. Assuming the SN-CBO interaction occurs immediately after explosion and the material is accelerated to $\sim$$2,\!200\;\mathrm{km}\;\mathrm{s}^{-1}$ (the faster CSM velocity calculated during the second peak), it would overrun the interaction photosphere at the first polar shell (which is expanding at $\sim$$1,\!200\;\mathrm{km}\;\mathrm{s}^{-1}$ at a distance of $\sim$$2\times10^{15}\;\mathrm{cm}$) $\sim$230~days after discovery in the rest frame. This is suspiciously close to the start of the second light curve peak, particularly given the uncertainty in the time of explosion. These ancillary interactions could compress the outer CSM to higher densities, and increase the energy budget available to power the second peak. For starting CSM shells of identical mass, this would naturally lead to both the decrease in the observed Balmer decrement and the appearance of He~II features from the extremely high-density ($n_e>10^{13}\;\mathrm{cm}^{-3}$) interaction region during the second peak. This possibility reveals a potential limitation of our shock modeling: interaction between shells might systematically increase both the presumed CSM density at the time of explosion and the associated mass-loss rates inferred for the progenitor during the earlier outburst.

Finally, we cannot exclude an LBV-like star in a binary system as the progenitor to 2023zkd. If a pre-existing H-rich disk from stable mass transfer is coupled with an LBV outburst of the primary funneled by the disk into bipolar outflows during Precursor B, it could also produce two CSM interaction regions with distinct densities, compositions and velocities. This scenario, however, must also explain the spectral evolution observed in 2023zkd. The H and He~I features at +90d suggest the fast-moving CSM is more He-rich than the slow-moving CSM, contrary to expectations for an H-rich LBV outburst. The post-LBV primary could launch fast and He-rich winds as a WN after the outburst, but it is not easy to reconcile how the WN alone could enrich the polar regions with enough He-rich CSM to power interaction signatures over 200~days from mass-loss in the single year between the start of Precursor B and the system's discovery. The spectral sequence may be alternatively be reconciled if the LBV outburst sweeps up and accelerates He-rich winds from the binary disk, but this would require an inversion in composition (the He-rich material must be distributed along the poles before the H-rich LBV outburst occurs) and additional assumptions to be made about the evolutionary history of the companion. 

\subsection{Final Remarks on the 2023zkd Progenitor}\label{subsec:final_progenitor}
We summarize the key observational properties of 2023zkd and the evidence for and against the leading progenitor scenarios in Table~\ref{tab:prog_summary} below.

\renewcommand{\arraystretch}{1.25}
\begin{deluxetable*}{p{0.17\textwidth}@{\extracolsep{10pt}}p{0.25\textwidth}
                    @{\extracolsep{10pt}}p{0.25\textwidth}
                    @{\extracolsep{15pt}}p{0.25\textwidth}}
\tablewidth{0pt}
\tabletypesize{\footnotesize}
\setlength{\tabcolsep}{10pt}
\tablecaption{Summary of Candidate Progenitor Scenarios for SN~2023zkd (see text for details).}
\tablehead{
\colhead{\textbf{Observation}} &
\colhead{\textbf{Single LBV-like or WN Star}} &
\colhead{\textbf{Single WNh Star}} &
\colhead{\textbf{Interacting Binary}}
}
\startdata
\textbf{Precursor \newline Luminosity} & 
\bad\;$M_r\!\approx-15\!$\,mag is $\sim$3~mag above known giant outbursts\tablenotemark{a}.\newline\newline
\good\;Blackbody properties of Precursor B are consistent with an LBV outburst\tablenotemark{g}.
 &
\bad\;Requires extreme, super-Eddington mass-loss for $>$100$\,M_{\odot}$ star\tablenotemark{d}.
&
\good\;Super-Eddington accretion onto BH naturally yields $L>$$10^{39}$\,erg\,s$^{-1}$\tablenotemark{i,l}.
\\[5pt] \hline
\textbf{Precursor Timescales} &
\maybe\;LBV outbursts are diverse\tablenotemark{g}.
&
\good\;Explains Precursor B's blackbody properties as terminal LBV-like outburst\tablenotemark{c}. \newline \newline
\maybe\;No known mechanism for sustaining super-Eddington outflows for years.
&
\good\;Runaway mass-transfer increasing luminosity spanning years\tablenotemark{l}.
 \\[5pt] \hline
\textbf{Ejecta \& CSM Masses} &
\maybe\;WR wind rates ($10^{-6}$--$10^{-4}\,M_{\odot}\,\mathrm{yr}^{-1}$) require $\gtrsim$$10^4\,$yr to produce $\sim$5$\,M_{\odot}$ of CSM; giant outbursts are possible\tablenotemark{f,m}.
 &
\maybe\;Similar challenges to the LBV/WN scenario; fine tuning needed to reach $M_{CSM}\!\approx\!5\,M_{\odot}$.
&
\good\;Binary mass-transfer can easily eject multiple $M_{\odot}$ of material\tablenotemark{h,l}.
\\[5pt] \hline
\textbf{CSM Geometry,\newline Composition} &
\good\;Can produce bipolar \& toroidal CSM for near-critical rotators, even in a single outburst\tablenotemark{b,c}.
 &
\good\;Produces fast, He-rich CSM and slow H-rich CSM.
 &
\good\;Naturally explains slow (200--400\,km\,s$^{-1}$) H-rich torus \& fast (1,200--2,200\,km\,s$^{-1}$) He-rich polar outflow\tablenotemark{h,l}.
\\[5pt] \hline
\textbf{Spectral Evolution} &
\bad\;Persistent, multi-component H and He lines require strong mixing of CSM from LBV outburst \& WN wind.
 &
\maybe\;May explain persistent H \& He if near-critical rotation leads to explosion as a post-WN LBV.
&
\good\;Matches double-peaked profiles\tablenotemark{j}. \newline\newline
\good\;Stable RLOF phase can leave residual H envelope in low-$Z$ systems\tablenotemark{k}.\newline\newline
\maybe\;Exact composition sensitive to evolutionary state at the time of final CCSN/merger.
\enddata
\tablerefs{ (\textit{a}) \citet{1994Humphreys_LBVs}; (\textit{b}) \citet{1997Cassinelli_LBVs}; (\textit{c}) \citet{1997GarciaSegura_LBVBipolar}; (\textit{d}) \citet{2008Smith_WRFeedback}; (\textit{e}) \citet{2011Smith_BinaryInteraction}; (\textit{f}) \citet{2017Smith_InteractingSNeIIn}; (\textit{g}) \citet{2017Smith_LBVs}; (\textit{h}) \citet{2018MacLeod_RunawayCoalescence}; (\textit{i}) \citet{2019Brightman_SuperEddington}; (\textit{j}) \citet{2019Suzuki_DoublePeaked}; (\textit{k}) \citet{2020Laplace_SESNe}; (\textit{l}) \citet{2024Tsuna_MergerPrecursor}; (\textit{m}) \citet{2024Ransome_DiversityofSNeIIn}}
\label{tab:prog_summary}
\end{deluxetable*}

\begin{figure*}
    \centering
    \includegraphics[width=\linewidth]{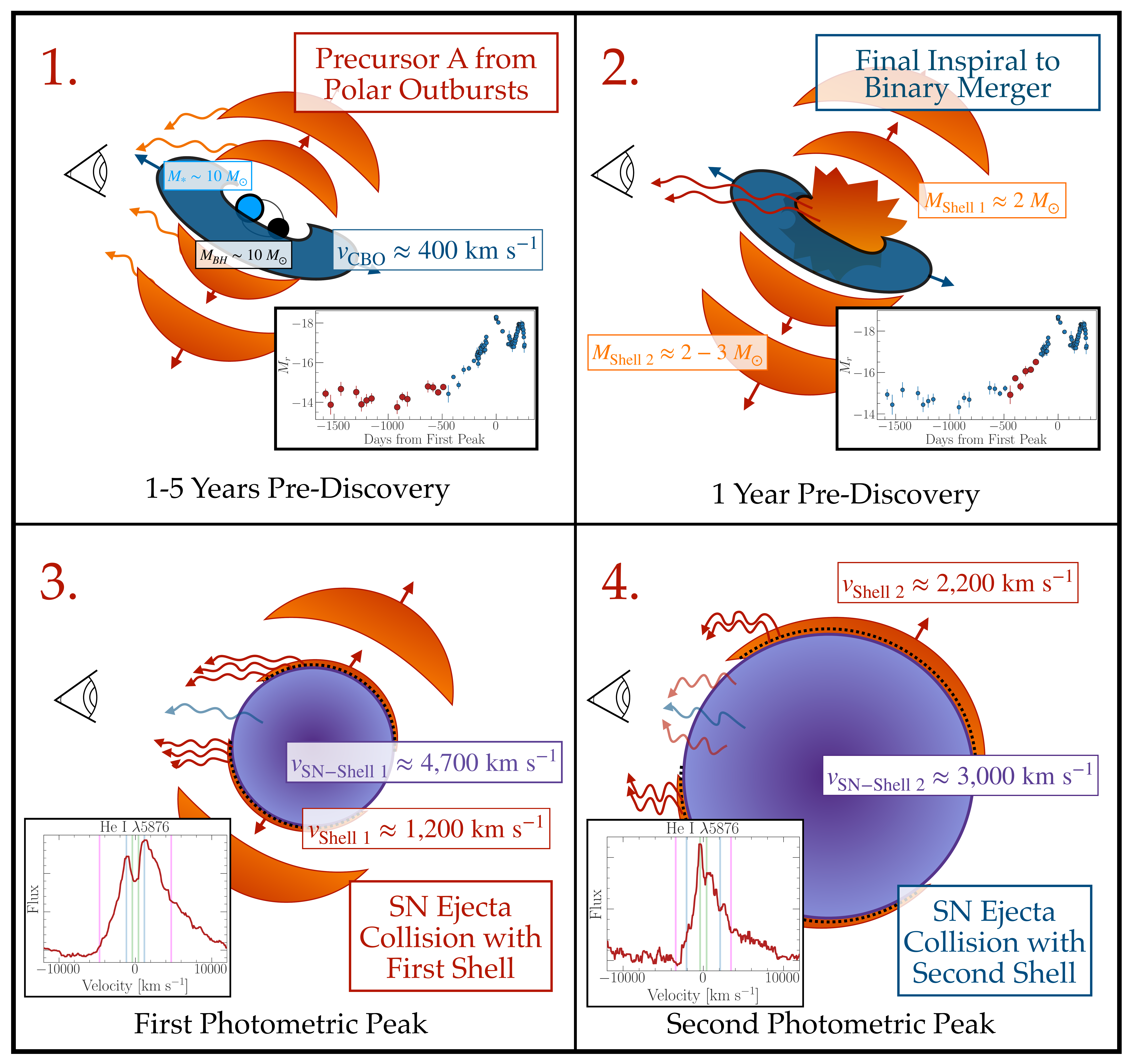}
    \caption{The proposed evolution of the putative binary system associated with SN~2023zkd. \textbf{1.} 1-5 years before discovery, photometric variability (red points) is detected from two polar pre-merger outbursts that each eject $\sim$2$\;M_{\odot}$ from the system. \textbf{2.} During the final year pre-explosion, runaway accretion leads to a tightening of the binary. The early photometric rise (red points) is associated either with unstable and enhanced accretion onto the black hole following a tightening of the binary, or an LBV-like outburst from the primary star. \textbf{3.} The primary merges with the black hole companion and explodes. The dense equatorial CBO funnels the ejecta along the polar axis and into the fast-moving, He-rich polar shells. Interaction with the first shell powers the first light curve peak, and He~I profiles are seen in double-peaked emission at the shell velocity (magenta). \textbf{4.} Interaction with the second polar shell powers the second light curve peak, with potential additional luminosity provided by CBO-interaction photons that scatter through the cooling shell (causing a peak in the He~I profiles at the CBO velocity in green). The transient rapidly dims when the SN interaction shell reaches the edge of the outer polar material.}
    \label{fig:progenitor}
\end{figure*}

From the above considerations, we conclude that a binary system with two main eruptive episodes that produce structured polar CSM and a slow equatorial outflow before merger is the most likely explanation. The binary companion is likely to be a black hole, or involve colliding stellar winds between binary components at luminosities without strong theoretical or observational support. We present a conceptual picture for this scenario at key phases in the pre- and post-explosion system in Figure~\ref{fig:progenitor}, and note its strong similarity to the progenitor systems suggested by \citet{2017Smith_InteractingSNeIIn}, \citet{2020ASchroder_MergerSN} (though our event suggests more structured polar outflows), and \citet{2022Metzger_IbnSNe}. 

In the model proposed by \citet{2022Metzger_IbnSNe}, the tidal disruption and hyper-accretion of a He core onto a black hole companion drives a luminous transient akin to an SN~Ibn or SN~IIn \textit{without} requiring the core collapse of the primary. A massive, H-rich, slowly-expanding disk surrounds the binary as a relic of the binary's common-envelope evolution stage (with additional toroidal enrichment from a CBO) and the merger is preceded by H-depleted outflows from the primary which expand unimpeded along the poles at $\sim$1,000$\;\mathrm{km}\;\mathrm{s}^{-1}$. The observation of SN~2023zkd as an SN~IIn with spectral similarities to a SN~Ibn, together with evidence for equatorially-distributed H, would then suggest that SN~2023zkd occurred within $10^3-10^4$~yr of the initial mass transfer phase that tightens the binary. 

Though the rise of Precursor~B suggests a connection to the final mass-transfer phase, the density profile derived in \textsection\ref{subsec:shock_model} indicate that the major mass-loss episodes occurred during Precursor A. Multiple mass-loss mechanisms associated could power this emission. One is periastron passage of the stellar companion along an eccentric orbit \citep{1998Layton_MassTransfer,2005Regos_MassTransfer}, which can accelerate accretion as the companion sweeps through material surrounding the primary \cite[with the orbit's effects on the disk geometry able to drive additional mass-loss; ][]{2016Munoz_EccentricBinary}. Given the likely polar geometry of the emission, however, it is reasonable to consider outbursts not just from the progenitor star but from the disk itself.

Accretion disk outflows will naturally be oriented toward the polar direction \citep{2015Fernandez_NSMergers}. Local temperature changes in a hydrogen-rich and convective accretion disk can also drive transitions between ionization modes (from ionized to neutral or neutral to ionized). The transitionary stage is well-known to be both thermally and viscously unstable (see \citealt{2020Hameury_DiskInstabilities} for a recent review, and their Figure~1 for the stability criteria associated with each mode), and can power luminous outbursts exhibiting distinct timescales and color temperatures. These modes may explain Precursor A and Precursor B in one self-consistent framework, particularly if the enhanced accretion rates associated with one mode are causally connected to the merger that drives the SN. These thermal-viscous instabilities have been explored for H-rich disks around low-mass binaries \citep[to explain the primary outburst modes of dwarf novae, e.g., ][]{1986Mineshige_DiskInstabilities,2018Bollimpalli_DiskInstabilities}, but little work has been done to investigate their relevance for high-mass binaries such as the putative progenitor to 2023zkd. Very recent work has shown that angular momentum transport from polar disk-driven outflows, in particular, can dramatically impact the orbital separation of a binary and the stability of its associated mass transfer \citep{2023Willcox_BinaryMassTransfer,2024GallegosGarcia_DiskWinds}. Additional effort on the simulation front is sorely needed to explore whether these effects can reconcile the properties of Precursor A and B (and the geometry of the CSM at the time of explosion).

We have refrained from speculating about the primary component of the progenitor system in the above analysis. Nonetheless, here we note a potential analog to the proposed progenitor system in high-mass X-ray binaries (HMXBs). HMXBs consist of a $>$$10\;M_{\odot}$ primary and a black hole or a neutron star companion. Galactic HMXBs undergo accretion-induced instabilities that can produce CSM distributed in a disk around the primary or along the polar axis around the accreting companion \citep{2023Bahramian_XrayBinary}. The circumstellar disk of a Be star can also exhibit double-peaked emission profiles in He~I and He~II from the star's rotational velocity \citep[as in, e.g., MWC~656; optical line profiles associated with the system are shown in Figure~1 from ][]{2022Zamanov_OpticalSpectroscopy_BeBH}. Further, outburst modes from the system (actively debated as originating from thermal disk instabilities, periastron passages, X-ray heating, or interaction of accreting material with the compact object's magnetosphere in the case of a neutron star; \citealt{2023Bahramian_XrayBinary}) have been observed with emission varying over both month and $\sim$year-long timescales (e.g., the optical/UV light curve of the Galactic HMXB Swift J0243.6+6124, shown in Figure~1 of \citealt{2024Alfonso_GiantOutburst}). 

Though the inefficient accretion rates suggested by some black hole-Be star HMXBs may drive X-ray quiescence in a progenitor system (and optical variability has been seen in some BeXRBs without coincident X-rays; \citealt{2023Coe_Outburst}), the X-ray luminosities observed in even the most luminous HMXB outbursts are still dimmer than the optical precursor observed in 2023zkd: our binned detections at $\sim$$10^{41}\;\mathrm{erg}\;\mathrm{s}^{-1}$ are more comparable to ULXs, e.g., M101-X1 (a system with a putative WR primary and $\sim$$10\;M_{\odot}$ black hole companion inferred from kinematic modeling; \citealt{2015MNRAS_Shen}) and MCG-03-34 \citep[whose X-ray luminosities are comparable to the optical luminosity of Precursor A][]{2006Miniutti_LuminousULX}. ULXs have recently been shown to exhibit long-lived variability at UV wavelengths \citep{2023Khan_ULXs}, and systematic optical searches for additional variability have recently begun \citep{2022Allak_ULXSearch}. 

There is substantial observational overlap between HMXBs/ULXs and SN~IIn regardless of whether a causal link exists: the optical spectra of ULXs might be able to mimic LBVs and WRs due to their fast, hydrogen-rich disk winds \citep{2015SupercriticalAccretion_ULX}, and the composite spectrum from a strong accretion phase may mimic that of a young SN~IIn \citep{2015Fabrika_SS433}. Nonetheless, a ULX-like progenitor was very recently proposed to explain periodic variations in the post-explosion light curve for the stripped-envelope SN 2020jli \citep{2024Chen}. There may be additional similarities between the physical environments of XRBs and the progenitors of re-brightening SNe~IIn; we consider this a promising avenue of future research.

\section{Conclusion}\label{sec:conclusion}
We have presented photometric and spectroscopic observations of the double-peaked and helium-rich type~IIn SN~2023zkd. We describe the most significant properties of the explosion from our analysis below: 

\begin{itemize}
    \item A host-galaxy with a mass at the lowest end of the distribution of observed SN~IIn hosts ($M_*\approx10^{8}\;M_{\odot}$), and with negligible ($\sim$$0.02\;M_{\odot}\;\mathrm{yr}^{-1}$) ongoing global star-formation;
    \item A double-peaked photometric evolution, with the event reaching an $r$-band maximum of $M_r\approx-18.7$~mag at the first peak and $M_r\approx-18.4$~mag at a second, concave-down peak $\sim$224~days later (in the rest frame), followed by a rapid ($\sim$60~day) dimming and associated reddening to $g-r>1$;
    \item A precursor phase characterized by persistent ($\sim$1,500~days) and luminous ($M_r\approx-15$~mag) emission, a final $\sim$400-day ramp-up to event discovery;
    \item Multi-component, asymmetric Balmer and He~I emission features that evolve minimally from the decline of the first light curve peak through the decline of the second; fits to the H$\alpha$ profiles of the dominant line-emitting region suggest CSM velocities starting at $\sim$$1,\!170\;\mathrm{km}\;\mathrm{s}^{-1}$ at +90d and reaching $\sim$$1,\!800\;\mathrm{km}\;\mathrm{s}^{-1}$ at the conclusion of the secondary brightening phase. Fits to the P-Cygni absorption of the H$\beta$ profiles indicate an SN-CSM shell velocity decreasing from $\sim$$4,\!700\;\mathrm{km}\;\mathrm{s}^{-1}$ at +90d to $\sim$$2,\!800\;\mathrm{km}\;\mathrm{s}^{-1}$ at the conclusion of the secondary peak; a slow-moving and persistent ($\sim$$400\;\mathrm{km}\;\mathrm{s}^{-1}$) CSM component is also detected that contributes to the observed luminosity of the secondary peak;
    \item A double-peaked emission profile for He~I lines in the +90d spectrum, with emission peaks at the CSM velocity; this strongly suggests an axisymmetric outflow whose interaction with the SN ejecta powers the first photometric peak; 
    \item A mass-loss history characterized by maximum mass ejection rates $\sim$1-2 and $\sim$3 years prior to discovery, with a total CSM mass of $5-6\;M_{\odot}$ ejected in comparable proportions during each mass-loss episode;
    \item A low Balmer decrement of $\sim$$2.0$ at +90d and 1.3-1.7 at all subsequent spectra, suggesting collisional excitation in an extremely high-density plasma ($n_e>10^{13}\;\mathrm{cm}^{-2}$) comparable to the material powering the 2012-B brightening of 2009ip;
    \item A H$\alpha$/He~I$\lambda$5876 ratio of 2-5, intermediate between He-rich SNe~IIn (1996al, 2021adou, 2016jbu) and the growing sample of transitional SNe~IIn/Ibn (2011hw, 2021foa, 2020bqj, 2005la, and iPTF15akq). 
\end{itemize}

The above properties are well-matched to a merger-driven explosion between a $M_{\mathrm{ZAMS}}>30\;M_{\odot}$ He core and a $\sim$$10\;M_{\odot}$ black hole companion following years of unstable mass transfer. Interestingly, our analysis suggests that the peak in the progenitor system's mass-loss rate does \textit{not} coincide with the more luminous precursor event (Precursor B), and instead may be associated with the lower-luminosity, longer-lived pre-explosion emission (Precursor A). This finding may naturally explain why two eruptive episodes are suggested by the mass-loss histories of multiple double-peaked SNe~IIn, but the majority of pre-explosion detections suggest only a single brightening phase in the progenitor's final year (this includes 2021qqp, \citealt{2024Hiramatsu_21qqp}; 2016jbu, \citealt{2022Brennan_2016jbu}; 2019zrk, \citealt{2022Fransson_2019zrk}; among others). As a result, caution should be taken in interpreting the months-long precursor emission detected in some SNe~IIn \citep[e.g.,][]{2021Strotjohann_MonthsLong}. Decade-long photometric baselines will allow us to confirm luminous emission at a secondary pre-explosion phase for the most nearby double-peaked SNe; if a secondary outburst is \textit{not} observed, it will force us to critically re-examine the photometric signposts of pre-explosion mass-loss. 

The shock models employed in this work are one-zone, and assume that the luminosity of the explosion at a given phase can be explained by CSM interaction at a single radius. As our spectral sequence for SN~2023zkd reveals, the reality can be significantly more complicated. Future studies will require more sensitive probes of CSM geometry, aided both by multi-zone models and theoretical studies of unstable mass-loss in single and binary systems, to be able to more reliably follow the thread of circumstellar interaction back to a likely progenitor system.

Our final conceptual picture for the progenitor system of 2023zkd is supported by the existence of an unresolved and low-velocity narrow component observed in hydrogen and helium at the decline of both light curve peaks. This feature has been observed in multiple other SNe~IIn. In cases where this component is completely blended with the other velocity components, it is reasonable to expect that only the higher-velocity CSM can be identified. Conversely, in cases where both a narrow and an intermediate-width component are seen, the intermediate-width component may be mistakenly attributed to radiative or collisional acceleration of a portion of the same slow-moving CSM. These dramatic differences in interpretation may be to blame for some of the scatter in CSM velocities reported for archival SNe~IIn \citep[which span $100-1,\!000\;\mathrm{km}\;\mathrm{s}^{-1}$;][]{2024Ransome_DiversityofSNeIIn}. Alternatively, given the peculiarities of the 2023zkd explosion, it may be the case that the multi-component geometry associated with 2023zkd is uncommon among SNe~IIn. 

If higher-resolution spectroscopy confirms the presence of equatorial CSM in an upcoming strongly-interacting SN~IIn, the structure of that profile over time can shed light on its evolution and fate (including whether it is fully destroyed or only partially disrupted by the SN ejecta). These data will allow us to answer more targeted questions about the explosion than have been possible in many other areas of SN science. The spectra of strongly-interacting SNe are substantially more complex than those of traditional SNe: electron scattering, asymmetric/clumpy CSM, dust formation, and differences in line intensities spanning wide dynamic ranges in optical depth, composition, and temperature may all produce ambiguous structure at small wavelength scales. High-resolution spectroscopy is critical for disentangling these physical effects.

The final photometry associated with 2023zkd suggests that the interaction shell had overrun the CSM whose interaction powered the second light curve peak. Nevertheless, if the SN ejecta or one of the polar shells is able to maintain its density structure and velocity, additional re-brightening phases may be possible. We advocate for continued multi-wavelength monitoring of this explosion to further constrain the properties of the CSM and confirm both a core-collapse explosion and a binary origin. Given the complex geometry and high inferred CSM masses, SN~2023zkd is also a prime target for late-time studies of dust formation at NIR wavelengths with \textit{JWST} \citep{2006Gardner_JWST} and the \textit{Nancy Grace Roman Space Telescope} \citep{2019Akeson_Roman} \citep[e.g.,][]{2011Fox_Dust,2021Fox}.

Samples of photometric SNe~IIn will dramatically increase in 2025, with the advent of the Vera C. Rubin Observatory's Legacy Survey for Space and Time expected to discover $\sim$$10^{5}\;$yr$^{-1}$ \citep{2024Ransome_DiversityofSNeIIn}. Algorithms designed to flag these long-lived and re-brightening transients will play a critical role in characterizing the full breadth of strongly-interacting events. Future studies of the phases and phenomenology of pre-explosion emission will be equally valuable for probing the progenitor system in its final years, although seasonal gaps combined with the relatively low single-band cadence will present additional challenges in reconstructing a physical picture. 

\textit{Software: Astropy \citep{astropy:2013, astropy:2018, astropy:2022}, BLAST \citep{2024Jones_BLAST}, corner \citep{corner_DFM}, dust-extinction \citep{2023Gordon_Extinction},  Matplotlib \citep{Hunter:2007_Matplotlib},  Prospector \citep{2021Johnson_Prospector}, Numpy \citep{harris2020array}, Pandas \citep{the_pandas_development_team_2024_10957263}, Scipy \citep{2020SciPy-NMeth}, Seaborn \citep{Waskom2021_Seaborn}, sfdmap2\footnote{\url{https://github.com/AmpelAstro/sfdmap2}}, SNANA \citep{2009Kessler_SNANA} YSE-PZ \citep{2022Coulter_YSEPZ,2023PASP_YSEPZ}} 

%%%%%%%%%%%%%%%%%%%%%%%%%%%%%%%%
%%%%%%%%%%%%%%%%%%%%%%%%%%%%%%%%
\section{Acknowledgments} 
\label{sec:acknowledgments}
We thank Edo Berger, Stan Woosley, Seán Brennan, Monica Gallegos-Garcia, Peter Blanchard, Harsh Kumar, Brian Metzger, and Kevin Burdge for helpful conversations that improved this manuscript. We additionally thank Yuri Beletsky and Dave Osip for obtaining the spectroscopic observations of SN~2023zkd with Magellan/LDSS-3. Finally, we are indebted to the National Institute of Standards and Technology (NIST); their H and He line lists have proven invaluable for a thorough analysis of the spectroscopic sequence of SN~2023zkd.

This work is supported by the National Science Foundation under Cooperative Agreement PHY-2019786 (The NSF AI Institute for Artificial Intelligence and Fundamental Interactions, http://iaifi.org/). A.G.\ also acknowledges support from the AI Institutes Virtual Organization (AIVO) and the MIT Libraries.
The Villar Astro Time Lab acknowledges support through the David and Lucile Packard Foundation, National Science Foundation under AST-2433718, AST-2407922 and AST-2406110, as well as an Aramont Fellowship for Emerging Science Research. 
C.R.A.\ is supported by the European Research Council (ERC) under the European Union’s Horizon 2020 research and innovation programme (grant agreement no. 948381)
D.O.J.\ acknowledges support from NSF grants AST-2407632 and AST-2429450, NASA grant 80NSSC24M0023, and HST/JWST grants HST-GO-17128.028, HST-GO-16269.012, and JWST-GO-05324.031, awarded by the Space Telescope Science Institute (STScI), which is operated by the Association of Universities for Research in Astronomy, Inc., for NASA, under contract NAS5-26555.
Parts of this research were supported by the Australian Research Council Centre of Excellence for Gravitational Wave Discovery (OzGrav), through project number CE230100016.
D.T.\ is supported by the Sherman Fairchild Postdoctoral Fellowship at Caltech.
K.A.B.\ is supported by an LSST-DA Catalyst Fellowship; this publication was thus made possible through the support of Grant 62192 from the John Templeton Foundation to LSST-DA.
D.F.\ is supported by a VILLUM FONDEN Young Investigator Grant (VIL25501).
The UCSC team is supported in part by a fellowship from the David and Lucile Packard Foundation to R.J.F.
C.G.\ is supported by research grants (VIL25501 and VIL69896) from VILLUM FONDEN. 
The LCO group is supported by NSF grants AST-1911225 and AST-1911151.  This work makes use of data from the Las Cumbres Observatory global telescope network.
M.M.\ gratefully acknowledges support from a Clay Postdoctoral Fellowship of the Smithsonian Astrophysical Observatory. 
G.N.\ gratefully acknowledges NSF support from a CAREER grant, AST-2239364, supported in-part by funding from Charles Simonyi, AST-2206195, OAC-2311355 and AST-2432428. G.N.\ and the NSF-Simons AI-Institute for the Sky (SkAI) are supported via NSF AST-2421845 and Simons Foundation MPS-AI-00010513. This work was partially performed at the NSF-Simons SkAI Institute. G.N.\ is also supported by the DOE through the Department of Physics at the University of Illinois, Urbana-Champaign (\#13771275), and support from the HST Guest Observer Program through HST-GO-16764 and HST-GO-17128 (PI: R.~Foley).
K.W.S.\ acknowledges funding from the Royal Society.
Q.W.\ is supported by the Sagol Weizmann-MIT Bridge Program.
This work is supported by research grants (VIL16599,VIL54489) from VILLUM FONDEN.

% YSE
The Young Supernova Experiment (YSE) and its research infrastructure is supported by the European Research Council under the European Union's Horizon 2020 research and innovation programme (ERC Grant Agreement 101002652, PI K.\ Mandel), the Heising-Simons Foundation (2018-0913, PI R.\ Foley; 2018-0911, PI R.\ Margutti), NASA (NNG17PX03C, PI R.\ Foley), NSF (AST--1720756, AST--1815935, PI R.\ Foley; AST--1909796, AST-1944985, PI R.\ Margutti), the David \& Lucille Packard Foundation (PI R.\ Foley), VILLUM FONDEN (project 16599, PI J.\ Hjorth), and the Center for AstroPhysical Surveys (CAPS) at the National Center for Supercomputing Applications (NCSA) and the University of Illinois Urbana-Champaign.

% Pan-STARRS
Pan-STARRS is a project of the Institute for Astronomy of the University of Hawaii, and is supported by the NASA SSO Near Earth Observation Program under grants 80NSSC18K0971, NNX14AM74G, NNX12AR65G, NNX13AQ47G, NNX08AR22G, 80NSSC21K1572, and by the State of Hawaii.  The Pan-STARRS1 Surveys (PS1) and the PS1 public science archive have been made possible through contributions by the Institute for Astronomy, the University of Hawaii, the Pan-STARRS Project Office, the Max-Planck Society and its participating institutes, the Max Planck Institute for Astronomy, Heidelberg and the Max Planck Institute for Extraterrestrial Physics, Garching, The Johns Hopkins University, Durham University, the University of Edinburgh, the Queen's University Belfast, the Harvard-Smithsonian Center for Astrophysics, the Las Cumbres Observatory Global Telescope Network Incorporated, the National Central University of Taiwan, STScI, NASA under grant NNX08AR22G issued through the Planetary Science Division of the NASA Science Mission Directorate, NSF grant AST-1238877, the University of Maryland, Eotvos Lorand University (ELTE), the Los Alamos National Laboratory, and the Gordon and Betty Moore Foundation.

% Gemini
Based on observations obtained at the international Gemini Observatory, a program of NSF NOIRLab, which is managed by the Association of Universities for Research in Astronomy (AURA) under a cooperative agreement with the U.S.\ National Science Foundation on behalf of the Gemini Observatory partnership: the U.S.\ National Science Foundation (United States), National Research Council (Canada), Agencia Nacional de Investigaci\'{o}n y Desarrollo (Chile), Ministerio de Ciencia, Tecnolog\'{i}a e Innovaci\'{o}n (Argentina), Minist\'{e}rio da Ci\^{e}ncia, Tecnologia, Inova\c{c}\~{o}es e Comunica\c{c}\~{o}es (Brazil), and Korea Astronomy and Space Science Institute (Republic of Korea). Data were taken by program GN-2024A-Q-136. This work was enabled by observations made from the Gemini North telescope, located within the Maunakea Science Reserve and adjacent to the summit of Maunakea. We are grateful for the privilege of observing the Universe from a place that is unique in both its astronomical quality and its cultural significance.

% Kast
A major upgrade of the Kast spectrograph on the Shane 3~m telescope at Lick Observatory was made possible through generous gifts from the Heising-Simons Foundation as well as William and Marina Kast. Research at Lick Observatory is partially supported by a generous gift from Google.

% LCO
This work makes use of data from the Las Cumbres Observatory global telescope network. The LCO group is supported by NSF grants AST-1911225 and AST-1911151.

% ATLAS
This work has made use of data from the Asteroid Terrestrial-impact Last Alert System (ATLAS) project. The Asteroid Terrestrial-impact Last Alert System (ATLAS) project is primarily funded to search for near earth asteroids through NASA grants NN12AR55G, 80NSSC18K0284, and 80NSSC18K1575; byproducts of the NEO search include images and catalogs from the survey area. This work was partially funded by Kepler/K2 grant J1944/80NSSC19K0112 and HST GO-15889, and STFC grants ST/T000198/1 and ST/S006109/1. The ATLAS science products have been made possible through the contributions of the University of Hawaii Institute for Astronomy, the Queen’s University Belfast, the Space Telescope Science Institute, the South African Astronomical Observatory, and The Millennium Institute of Astrophysics (MAS), Chile.

% ZTF
The ZTF forced-photometry service was funded under the Heising-Simons Foundation grant \#12540303 (PI: Graham).

% 2MASS
This publication makes use of data products from the Two Micron All Sky Survey, which is a joint project of the University of Massachusetts and the Infrared Processing and Analysis Center/California Institute of Technology, funded by the National Aeronautics and Space Administration and the National Science Foundation.

% SDSS
Funding for the Sloan Digital Sky  Survey IV has been provided by the  Alfred P.\ Sloan Foundation, the U.S.\  Department of Energy Office of Science, and the Participating  Institutions.
SDSS-IV acknowledges support and resources from the Center for High Performance Computing  at the University of Utah. The SDSS website is www.sdss4.org.
SDSS-IV is managed by the Astrophysical Research Consortium for the Participating Institutions of the SDSS Collaboration including the Brazilian Participation Group, the Carnegie Institution for Science, Carnegie Mellon University, Center for Astrophysics | Harvard \& Smithsonian, the Chilean Participation Group, the French Participation Group, Instituto de Astrof\'isica de Canarias, The Johns Hopkins University, Kavli Institute for the Physics and Mathematics of the Universe (IPMU) / University of Tokyo, the Korean Participation Group, Lawrence Berkeley National Laboratory, Leibniz Institut f\"ur Astrophysik Potsdam (AIP),  Max-Planck-Institut f\"ur Astronomie (MPIA Heidelberg), Max-Planck-Institut f\"ur Astrophysik (MPA Garching), Max-Planck-Institut f\"ur Extraterrestrische Physik (MPE), National Astronomical Observatories of  China, New Mexico State University, New York University, University of Notre Dame, Observat\'ario Nacional / MCTI, The Ohio State University, Pennsylvania State University, Shanghai Astronomical Observatory, United Kingdom Participation Group, Universidad Nacional Aut\'onoma de M\'exico, University of Arizona, University of Colorado Boulder, University of Oxford, University of Portsmouth, University of Utah, University of Virginia, University of Washington, University of Wisconsin, Vanderbilt University, and Yale University. Collaboration Overview Affiliate Institutions Key People in SDSS Collaboration Council Committee on Inclusiveness Architects SDSS-IV Survey Science Teams and Working Groups Code of Conduct SDSS-IV Publication Policy How to Cite SDSS External Collaborator Policy For SDSS-IV Collaboration Members.

% Legacy Surveys
The Legacy Surveys consist of three individual and complementary projects: the Dark Energy Camera Legacy Survey (DECaLS; Proposal ID \#2014B-0404; PIs: David Schlegel and Arjun Dey), the Beijing-Arizona Sky Survey (BASS; NOAO Prop.\ ID \#2015A-0801; PIs: Zhou Xu and Xiaohui Fan), and the Mayall z-band Legacy Survey (MzLS; Prop. ID \#2016A-0453; PI: Arjun Dey). DECaLS, BASS and MzLS together include data obtained, respectively, at the Blanco telescope, Cerro Tololo Inter-American Observatory, NSF’s NOIRLab; the Bok telescope, Steward Observatory, University of Arizona; and the Mayall telescope, Kitt Peak National Observatory, NOIRLab. Pipeline processing and analyses of the data were supported by NOIRLab and the Lawrence Berkeley National Laboratory (LBNL). The Legacy Surveys project is honored to be permitted to conduct astronomical research on Iolkam Du’ag (Kitt Peak), a mountain with particular significance to the Tohono O’odham Nation.
NOIRLab is operated by the Association of Universities for Research in Astronomy (AURA) under a cooperative agreement with the National Science Foundation. LBNL is managed by the Regents of the University of California under contract to the U.S.\ Department of Energy.
This project used data obtained with the Dark Energy Camera (DECam), which was constructed by the Dark Energy Survey (DES) collaboration. Funding for the DES Projects has been provided by the U.S.\ Department of Energy, the U.S.\ National Science Foundation, the Ministry of Science and Education of Spain, the Science and Technology Facilities Council of the United Kingdom, the Higher Education Funding Council for England, the National Center for Supercomputing Applications at the University of Illinois at Urbana-Champaign, the Kavli Institute of Cosmological Physics at the University of Chicago, Center for Cosmology and Astro-Particle Physics at the Ohio State University, the Mitchell Institute for Fundamental Physics and Astronomy at Texas A\&M University, Financiadora de Estudos e Projetos, Fundacao Carlos Chagas Filho de Amparo, Financiadora de Estudos e Projetos, Fundacao Carlos Chagas Filho de Amparo a Pesquisa do Estado do Rio de Janeiro, Conselho Nacional de Desenvolvimento Cientifico e Tecnologico and the Ministerio da Ciencia, Tecnologia e Inovacao, the Deutsche Forschungsgemeinschaft and the Collaborating Institutions in the Dark Energy Survey. The Collaborating Institutions are Argonne National Laboratory, the University of California at Santa Cruz, the University of Cambridge, Centro de Investigaciones Energeticas, Medioambientales y Tecnologicas-Madrid, the University of Chicago, University College London, the DES-Brazil Consortium, the University of Edinburgh, the Eidgenossische Technische Hochschule (ETH) Zurich, Fermi National Accelerator Laboratory, the University of Illinois at Urbana-Champaign, the Institut de Ciencies de l’Espai (IEEC/CSIC), the Institut de Fisica d’Altes Energies, Lawrence Berkeley National Laboratory, the Ludwig Maximilians Universitat Munchen and the associated Excellence Cluster Universe, the University of Michigan, NSF’s NOIRLab, the University of Nottingham, the Ohio State University, the University of Pennsylvania, the University of Portsmouth, SLAC National Accelerator Laboratory, Stanford University, the University of Sussex, and Texas A\&M University.

The Legacy Surveys imaging of the DESI footprint is supported by the Director, Office of Science, Office of High Energy Physics of the U.S.\ Department of Energy under Contract No.\ DE-AC02-05CH1123, by the National Energy Research Scientific Computing Center, a DOE Office of Science User Facility under the same contract; and by the U.S.\ National Science Foundation, Division of Astronomical Sciences under Contract No.\ AST-0950945 to NOAO.

The stellar population modeling in this work is based on data from the MILES library service developed by the Spanish Virtual Observatory in the framework of the IAU Commission G5 Working Group: Spectral Stellar Libraries.

YSE-PZ was developed by the UC Santa Cruz Transients Team with support from The UCSC team is supported in part by NASA grants NNG17PX03C, 80NSSC19K1386, and 80NSSC20K0953; NSF grants AST-1518052, AST-1815935, and AST-1911206; the Gordon \& Betty Moore Foundation; the Heising-Simons Foundation; a fellowship from the David and Lucile Packard Foundation to R.\ J.\ Foley; Gordon and Betty Moore Foundation postdoctoral fellowships and a NASA Einstein fellowship, as administered through the NASA Hubble Fellowship program and grant HST-HF2-51462.001, to D.~O.~Jones; and a National Science Foundation Graduate Research Fellowship, administered through grant No.\ DGE-1339067, to D.~A.~Coulter.

\bibliography{references}{}
\bibliographystyle{aasjournal}

\appendix 

\section{Confirming the ZTF Pre-Discovery Flux Excess}
\begin{figure}
    \centering
    \includegraphics[width=\linewidth]{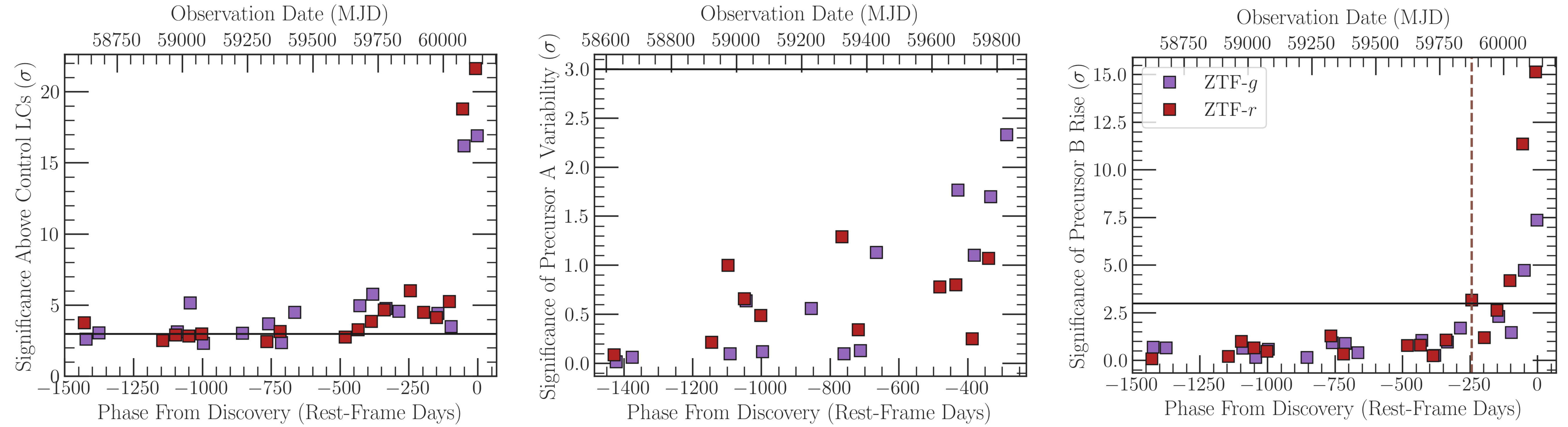}
    \caption{\textit{Left Panel:} Statistical significance of binned pre-explosion photometry from the Zwicky Transient Facility in ZTF-$g$ (purple) and ZTF-$r$ (red) above the variability of control light curves surrounding the explosion site. The majority of measurements are at $>$3$\sigma$ significance (horizontal black line) in both filters. \textit{Middle Panel:} Significance of variability about the mean during Precursor A (see text for details). No measurements in either filter reach $>$3$\sigma$ significance, and we are unable to distinguish between multiple eruptions and persistent enhanced emission during Precursor A. \textit{Right Panel:} Significance of flux measurements from Precursor B above the baseline flux for Precursor A (see text for details). Multiple observations starting $\sim$250~days from discovery (dashed vertical line) are statistically significant, suggesting a persistent rise in the final 250~days prior to event discovery.}
    \label{fig:ZTFtests}
\end{figure}

Here, we evaluate the probability that the binned pre-explosion ZTF detections coincident with the SN site do not arise from the SN progenitor. We define 10 evenly-spaced positions along an annulus of radius 5$\arcsec$ and centered on the SN position and query the ZTF Forced Photometry Service for all flux measurements at these positions (while ensuring that these positions lie in the same field as SN~2023zkd and were observed with the same CCD). For each control light curve, we undertake the same analysis as for the SN~2023zkd light curve and calculate uncertainty-weighted flux measurements in 50-day bins spanning the same pre-explosion epochs. In each bin, we calculate the mean and standard deviation of the flux measured across all 10 control light curves and use these values to convert the binned flux measurements for SN~2023zkd to z-scores (combining the uncertainties of our control light curves and our precursor detections in quadrature). We show the results of this experiment in the top left panel of Figure~\ref{fig:ZTFtests}.

74\% (25/34) of the reported pre-discovery detections for SN~2023zkd lie $>$3$\sigma$ above the distribution of binned background fluxes, and 24\% (8/34) are at $>$5$\sigma$. The earliest $>$5$\sigma$ detection occurs at $\mathrm{MJD}\approx59029.4$, 1044.6~days before discovery in the rest frame. This supports the marginal detection of excess flux in the case of Precursor A. We further note the potential presence of correlated structure in both filters in the 500-days prior to discovery, which we cannot clearly resolve in our final detections.

We can also calculate the probability that Precursor A is consistent with emission at a constant flux. We calculate the mean and standard deviation of binned flux values prior to $\mathrm{MJD}=59850$ in each filter and again convert the measured flux values to z-scores. The results are shown in the middle panel of Figure~\ref{fig:ZTFtests}. All flux measurements are $<$3$\sigma$ from the mean flux (the most significant deviation is 2.3$\sigma$ in ZTF-$g$ and 1.3$\sigma$ in ZTF-$r$), indicating that the photometric variability about the mean of Precursor~A is statistically insignificant.

Finally, we calculate the statistical significance of the brightening associated with Precursor B. We calculate z-scores for each precursor detection relative to the mean and standard deviation for the flux measurements of Precursor A. The results are shown in the right panel of Figure~\ref{fig:ZTFtests}. Two detections in ZTF-$g$ and four detections in ZTF-$r$ lie $>$3$\sigma$ above the mean flux of Precursor A; all occur in the 300~days prior to discovery and increase in significance leading up to the discovery date (the final binned detection is $\sim$15$\sigma$ above the precursor flux of Precursor A). The earliest statistically-significant deviation from Precursor A occurs at MJD=59873.4 in ZTF-$r$ (245.4~days prior to discovery in the rest frame); as a result, we confirm the brightening of the progenitor system in its final year and mark the division between Precursor A and B at $\mathrm{MJD}\approx59873$. 

\section{Identification of SN~2023zkd as a Photometric Anomaly with \texttt{LAISS}} \label{sec:laiss}
\texttt{LAISS} is a random forest classifier trained to distinguish between typical and atypical SNe on the basis of their photometric evolution and host-galaxy properties. We can interpret the predictions of the model using a force plot, which visualizes the approximate Shapley values associated with a candidate's features. Shapley values estimate the mean marginal contribution of each feature to a model output via additive feature attribution \citep[see][]{2017Lundberg_SHAP}. They were originally developed to in the context of cooperative game theory and are now widely used in studies of machine learning interpretability \citep{Hart1989_Shapley}.

We use Shapley values to predict which features are most significant in determining the final anomaly score predicted for 2023zkd. We use the SHAP\footnote{\url{https://shap.readthedocs.io/en/latest/index.html}} package to calculate the Shapley values associated with SN~2023zkd's properties at $\mathrm{MJD}=60327$, the epoch when the event was originally flagged. Light curve properties have been calculated with the \texttt{light-curve} package \citep{2021Malanchev_Lightcurve} and the host galaxy properties have been calculated from \texttt{GHOST} \citep{2021Gagliano_GHOST}. 

\begin{figure*}
    \centering
    \includegraphics[width=\linewidth]{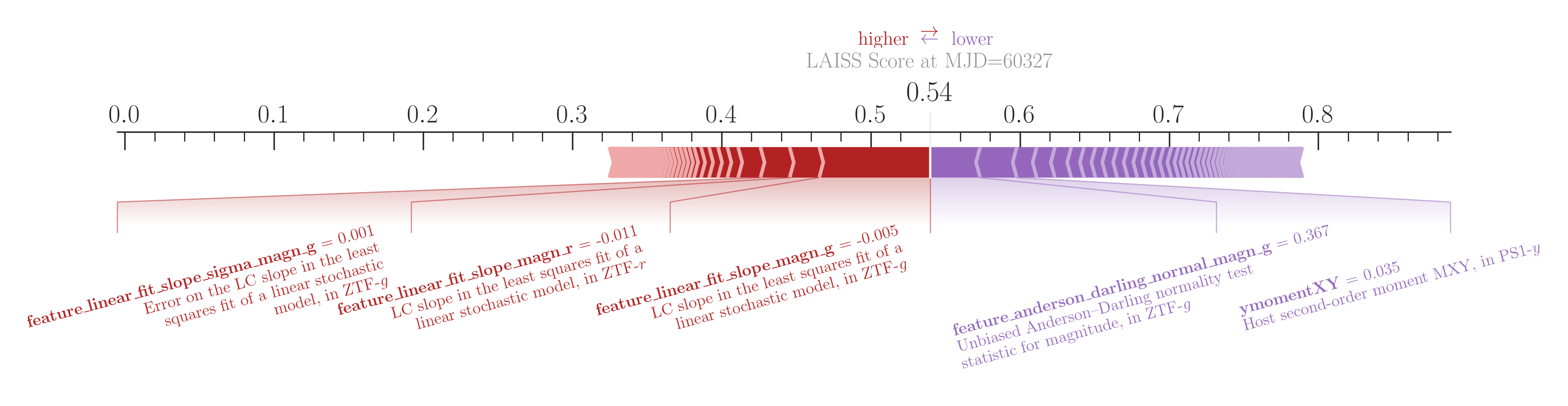}
    \caption{Force plot of SN~2023zkd for its properties on $\mathrm{MJD}\approx60327$, when the event was first flagged by \texttt{LAISS}. Light curve and host galaxy parameters with largest positive Shapley values (indicating positive mean marginal contribution to the predicted anomaly score) are shown in red, and parameters with largest negative values are shown in blue. Small negative linear slopes are measured in ZTF-$g$ and ZTF-$r$ as the top positive Shapley features, suggesting that an anomalous decline from first peak increased the anomaly score and led to its identification as an unusual transient.}
    \label{fig:shap_plot}
\end{figure*}

We visualize the results using a force plot for the features with largest Shapley values in Figure~\ref{fig:shap_plot}, which indicates that an anomaly label was assigned predominantly by the slope of the linear fit to the light curve evolution across filters. This parameter was also the single most important feature for distinguishing spectroscopic anomalies in \citet{2024Aleo_LAISS}, and was attributed to the unique color evolution of relatively blue \citep[e.g.,][]{2014Arcavi_TDEs} and red \citep[e.g., stripped-envelope SNe;][]{2018Taddia_SESNe,2024Khakpash_SESNtemplates} transients relative to the spectroscopically ``normal" SNe~Ia and SNe~II. In this case, the dimming of ZTF-$g$ after peak relative to ZTF-$r$ is due to the prominent H$\alpha$ emission feature that overlaps with ZTF-$r$, which dominates the brightness measurement in SNe~IIn as the continuum fades. We will further explore \texttt{LAISS}'s ability to identify long-lived SNe with ongoing photometric signatures of CSM interaction in an upcoming work.

\section{MOSFiT Corner Plot}
\begin{figure*}
    \centering
    \includegraphics[width=\linewidth]{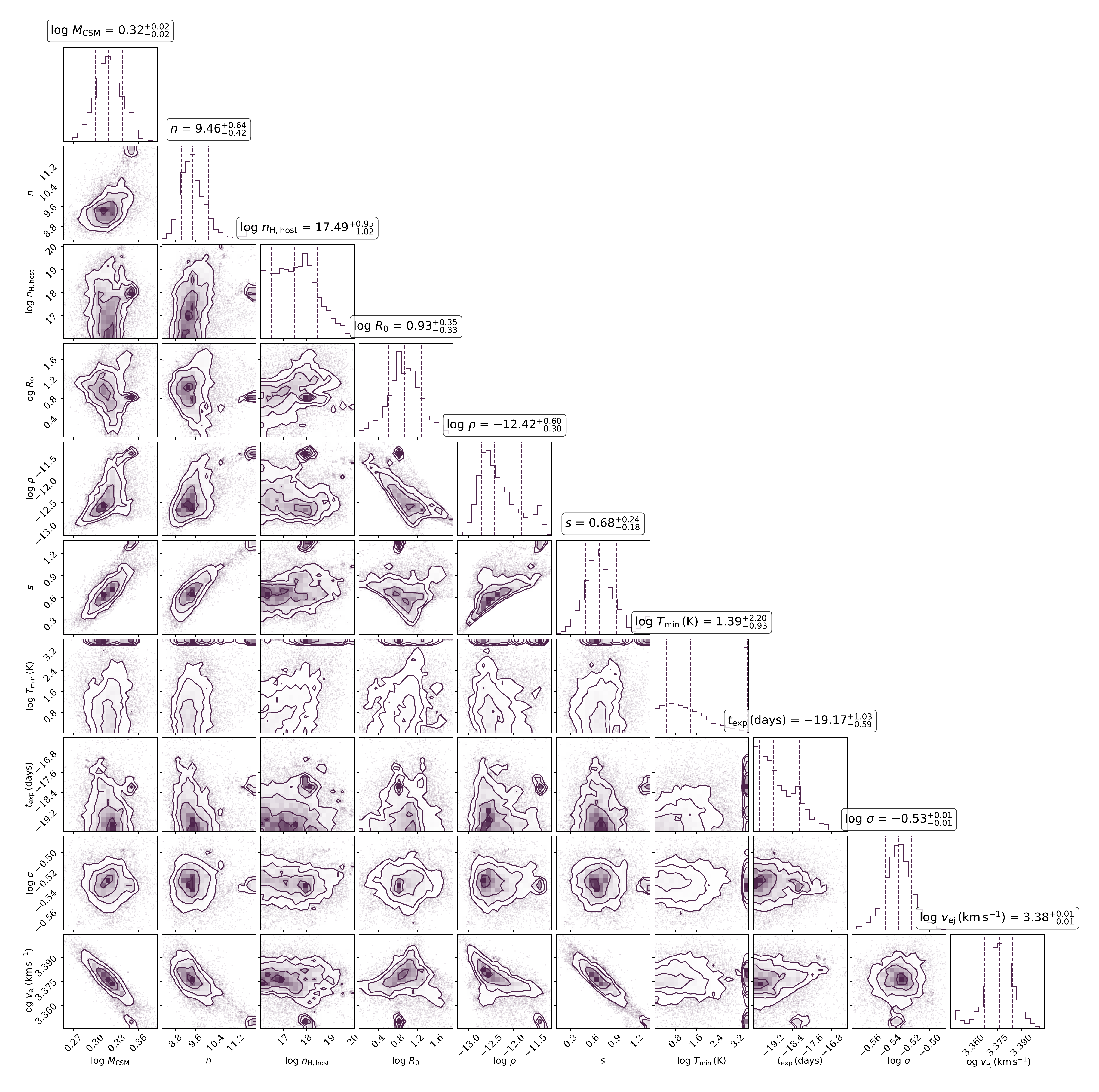}
    \caption{Posterior samples for the \texttt{MOSFiT} fit to the event photometry associated with the second peak. Marginal distributions are shown at top, with vertical lines corresponding to the median and 1$\sigma$ range of the posteriors.}
    \label{fig:mosfit_fit}
\end{figure*}

Figure~\ref{fig:mosfit_fit} shows the posterior samples after burn-in for the \texttt{MOSFiT} fit to the second light curve peak (described in \textsection\ref{subsec:mosfit}). The ejecta mass of the SN is fixed at $10\;M_{\odot}$, and the priors associated with each parameter are presented in Table~\ref{tab:mosfit_params}. 

\section{Prospector Corner Plot}
\begin{figure*}
    \centering
    \includegraphics[width=\linewidth]{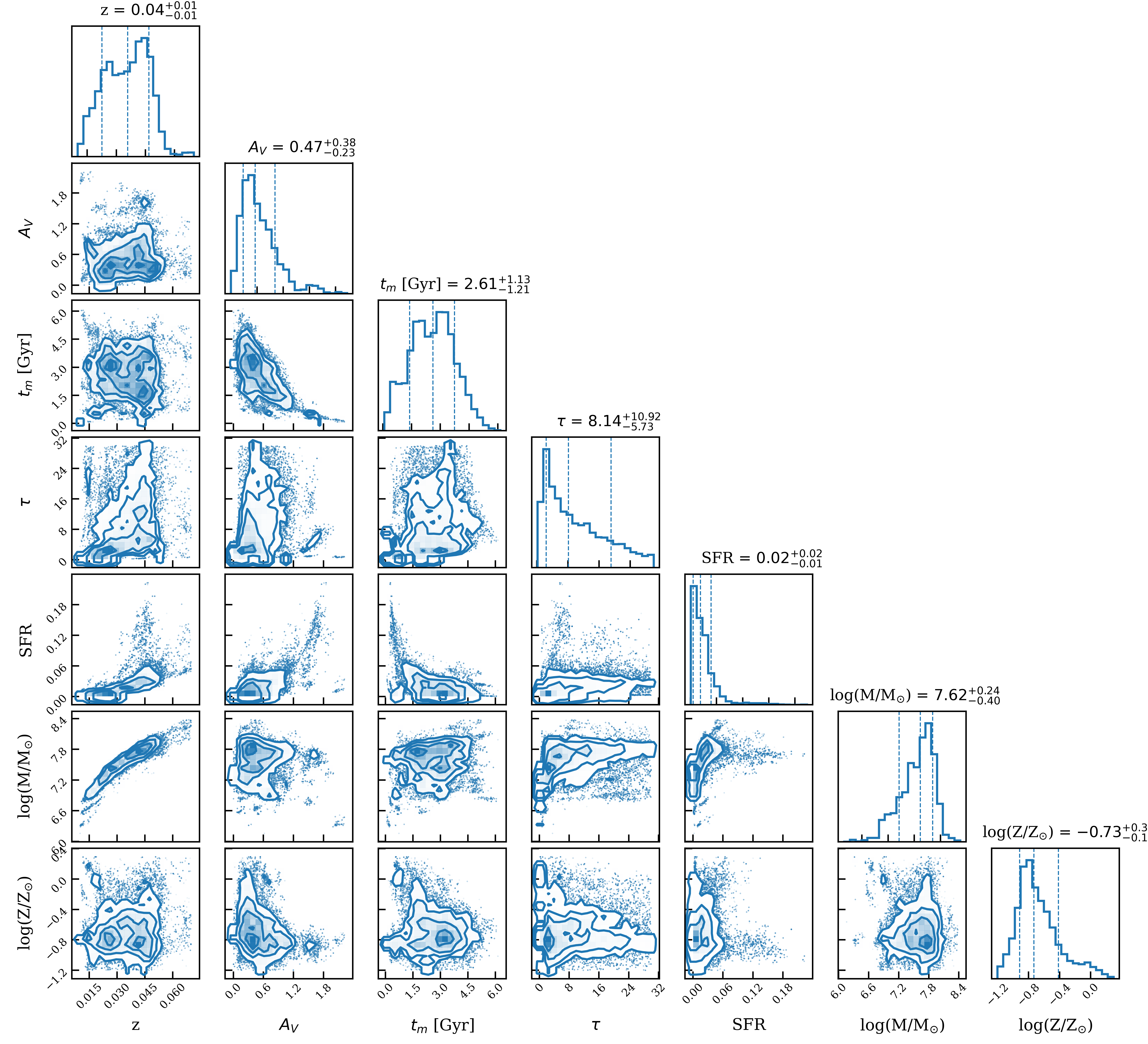}
    \caption{Posterior samples for Prospector fit to the host galaxy SED. Marginal distributions are shown at top, with vertical lines corresponding to the median and 1$\sigma$ range of the posteriors.}
    \label{fig:prospector_corner}
\end{figure*}

Figure~\ref{fig:prospector_corner} shows the posterior samples for the \texttt{Prospector} fit to the photometry of SN~2023zkd's host galaxy, described in \textsection\ref{sec:host}.

\section{Spectral Line Fitting Corner Plot}
\begin{figure*}
    \centering
    \includegraphics[width=\linewidth]{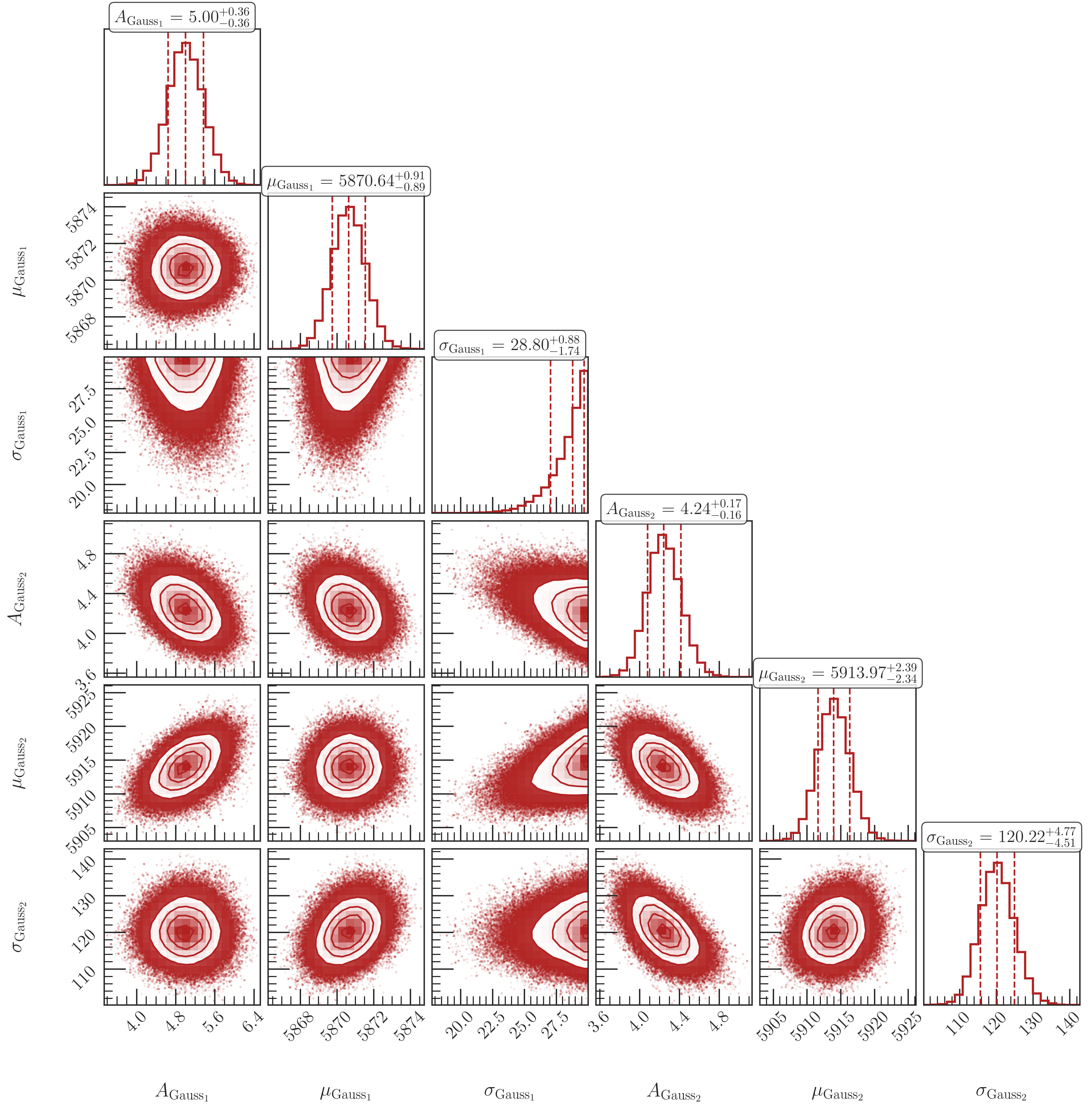}
    \caption{Corner plot showing the posteriors for the parameters corresponding to the continuum-subtracted He~I$\lambda5876$ profile of the Magellan/LDSS-3 spectrum at $\mathrm{MJD}=60519$ (+216d). The asymmetric profile is best fit by a narrow Gaussian component with amplitude $A_{\mathrm{Gauss}_1}$, center $\mu_{\mathrm{Gauss}_1}$, and standard deviation $\sigma_{\mathrm{Gauss}_1}$; and a broader Gaussian component with amplitude $A_{\mathrm{Gauss}_2}$, center $\mu_{\mathrm{Gauss}_2}$, and standard deviation $\sigma_{\mathrm{Gauss}_2}$. The median and 1$\sigma$ ranges of each parameter are shown as dashed lines.}
    \label{fig:corner_specfit}
\end{figure*}

Figure~\ref{fig:corner_specfit} shows the posteriors associated with our continuum-subtracted line fit to He~I$\lambda5876$ in the NOT/ALFOSC spectrum obtained on $\mathrm{MJD}=60519$ (+216d relative to the observed $r$-band maximum at $\mathrm{MJD}=60290.6$). The variables $\mu$ and $\sigma$ are in units of Angstroms for each Gaussian component, while the amplitude $A$ is in units of $10^{-17}\;\mathrm{erg}\;\mathrm{s}^{-1}\mathrm{cm}^{-2}\;\mathrm{Å}^{-1}$.

\end{document}